\documentclass[preprint,showpacs,preprintnumbers,amsmath,amssymb,showpacs,11pt]{revtex4}

\usepackage{graphicx}
\usepackage{dcolumn}
\usepackage{bm}
\usepackage{amssymb}

\def\bea{\begin{eqnarray}}
\def\ena{\end{eqnarray}}

\begin{document}

\title{Imprints of Relic Gravitational Waves in Cosmic Microwave Background
Radiation}

\author{D. Baskaran}
\affiliation{School of Physics and Astronomy, Cardiff University,
Cardiff CF24 3YB, UK}
\author{L. P. Grishchuk}
\affiliation{School of Physics and Astronomy, Cardiff University,
Cardiff CF24 3YB, UK ~and~ Sternberg Astronomical Institute,
Moscow State University, Moscow 119899, Russia}
\author{A. G. Polnarev}
\affiliation{ Astronomy Unit, School of Mathematical Sciences
Queen Mary, University of London, Mile End Road, London E1 4NS,
UK}

\date{\today}


\begin{abstract}
A strong variable gravitational field of the very early Universe
inevitably generates relic gravitational waves by amplifying their
zero-point quantum oscillations. We begin our discussion by
contrasting the concepts of relic gravitational waves and
inflationary `tensor modes'. We explain and summarize the
properties of relic gravitational waves that are needed to derive
their effects on CMB temperature and polarization anisotropies.
The radiation field is characterized by four invariants $I, V, E,
B$. We reduce the radiative transfer equations to a single
integral equation of Voltairre type and solve it analytically and
numerically. We formulate the correlation functions
$C^{XX'}_{\ell}$ for $X, X'= T, E, B$ and derive their amplitudes,
shapes and oscillatory features. Although all of our main
conclusions are supported by exact numerical calculations, we
obtain them, in effect, analytically by developing and using
accurate approximations. We show that the $TE$ correlation at
lower $\ell$'s must be negative (i.e. an anticorrelation), if it
is caused by gravitational waves, and positive if it is caused by
density perturbations. This difference in $TE$ correlation may be
a signature more valuable observationally than the lack or
presence of the $BB$ correlation, since the $TE$ signal is about
100 times stronger than the expected $BB$ signal. We discuss the
detection by WMAP of the $TE$ anticorrelation at $\ell \approx 30$
and show that such an anticorrelation is possible only in the
presence of a significant amount of relic gravitational waves
(within the framework of all other common assumptions). We propose
models containing considerable amounts of relic gravitational
waves that are consistent with the measured $TT$, $TE$ and $EE$
correlations.
\end{abstract}



\maketitle



\section{\label{sec:introduction}Introduction}

The detection of primordial gravitational waves is rightly
considered a highest priority task for the upcoming observational
missions \cite{Weiss2006}. Relic gravitational waves are
inevitably generated by strong variable gravitational field of the
very early Universe. The generating mechanism is the
superadiabatic (parametric) amplification of the waves' zero-point
quantum oscillations \cite{Grishchuk1974}. In contrast to other
known massless particles, the coupling of gravitational waves to
the `external' gravitational field is such that they could be
amplified or generated in a homogeneous isotropic FLRW universe.
This conclusion, at the time of its formulation, was on a
collision course with the dominating theoretical doctrine. At that
time, it was believed that the gravitational waves could not be
generated in a FLRW universe, and the possibility of their
generation required the early Universe to be strongly anisotropic
(see, for example, \cite{Zeldovich1971}).

The generating mechanism itself relies only on the validity of
general relativity and quantum mechanics. But the amount and
spectral content of relic gravitational waves depend on a specific
evolution of the cosmological scale factor (classical `pumping'
gravitational field) $a(\eta)$. The theory was applied to a
variety of $a(\eta)$, including those that are now called
inflationary models (for a sample of possible spectra of relic
gravitational waves, see Fig.4 in Ref.\ \cite{Grishchuk1980}). If
a unique $a(\eta)$ were known in advance from some fundamental
`theory of everything', we would have derived the properties of
the today's signal with no ambiguity. In the absence of such a
theory, we have to use the available partial information in order
to reduce the number of options. This allows us to evaluate the
level of the expected signals in various frequency intervals. The
prize is very high - the actual detection of a particular
background of relic gravitational waves will provide us with the
unique clue to the `birth' of the Universe and its very early
dynamical behaviour.

A crucial assumption that we make in this and previous studies is
that the observed large-scale cosmic microwave background (CMB)
anisotropies are caused by cosmological perturbations of
quantum-mechanical origin. If this is true, then general
relativity and quantum mechanics tell us that relic gravitational
waves should be a significant, if not a dominant, contributor to
the observed large-scale anisotropies. From the existing data on
the amplitude and spectrum of the CMB fluctuations we infer the
amplitude and spectral slope of the very long relic gravitational
waves. We then derive detailed predictions for indirect and direct
observations of relic gravitational waves in various frequency
bands.

At this point, it is important to clarify the difference between
the concepts of relic gravitational waves and what is now called
inflationary gravitational waves. The statements about
inflationary gravitational waves (`tensor modes') are based on the
inflation theory. This theory assumes that the evolution of the
very early Universe was driven by a scalar field, coupled to
gravity in a special manner. The theory does not deny the
correctness of the previously performed calculations for relic
gravitational waves. However, the inflationary theory proposes its
own way of calculating the generation of density perturbations
(`scalar modes'). The inflationary theory appeals exactly to the
same mechanism of superadiabatic (parametric) amplification of
quantum vacuum fluctuations, that is responsible for the
generation of relic gravitational waves, but enforces very
peculiar initial conditions in the `scalar modes' calculations.

According to the inflationary initial conditions, the amplitudes
of the `gauge-invariant' metric perturbations $\zeta$ associated
with the `scalar modes' (or, in other words, the amplitudes of the
curvature perturbations called $\zeta$ or $\cal{R}$) can be
arbitrarily large from the very beginning of their evolution.
Moreover, the theory demands that these amplitudes must be
infinitely large in the limit of the deSitter expansion law
$a(\eta) \propto |\eta|^{-1}$ which is responsible for the
generation of a flat (Harrison-Zeldovich-Peebles,
`scale-invariant') primordial spectrum, with the spectral index
${\rm n}=1$. At the same time, the amplitudes of the generated
gravitational waves are finite and small for all spectral indices,
including ${\rm n} =1$. Since both, gravitational waves and
density perturbations, produce CMB anisotropies and we see them
small today, the inflationary theory substitutes (for
`consistency') its prediction of infinitely large amplitudes of
density perturbations, in the limit ${\rm n}=1$, by the claim that
it is the amount of primordial gravitational waves, expressed in
terms of the `tensor/scalar ratio $r$', that should be zero,
$r=0$. For a detailed critical analysis of the inflationary
conclusions, see \cite{Grishchuk2005}; for arguments aimed at
defending those conclusions, see \cite{Lukash2006}.

The science motivations and CMB data analysis pipelines, designed
to evaluate the gravitational wave contribution, are usually based
on inflationary formulas \cite{Peiris2003, Spergel2006, Page2006,
PLANCK2006}. In particular, according to the inflation theory, the
primordial power spectrum of density perturbations has the form
(it follows, for example, from Eqs. (2.12a) and (2.12b) in
Ref.\cite{PLANCK2006} or from Eqs. (18) and (19) (or (A12) and (A13))
in Ref.\cite{Peiris2003}):
\begin{eqnarray}
P_S(k) =\frac{1}{4\pi^{3} M_{Pl}^{2}} \frac{1}{r} H^{2}|_{(k=aH)}.
\nonumber
\end{eqnarray}
Despite the fact that this spectrum diverges at $r=0$ (${\rm n}_s
=1$, ${\rm n}_t =0$) the CMB data analysts persistently claim that the
inflation theory is in spectacular agreement with observations and the
CMB data are perfectly well consistent with $r =0$ (the published
confidence level contours always include $r=0$ and are typically
centered at that point).

Our analysis in this paper, based on general relativity and
quantum mechanics, is aimed at showing that there is evidence of
signatures of relic gravitational waves in the already available
CMB data. We also make predictions for some future experiments and
observations.

The plan of the paper is as follows. In Sec.\ \ref{sec:gw} we
summarize the properties of a random background of relic
gravitational waves that are needed for CMB calculations. The
emphasis is on the gravitational wave (g.w.) mode functions, power
spectra, and statistical relations. In Sec.\ \ref{sec:rte} we
discuss the general equations of radiative transfer and explain
the existence of four invariants $I, V, E, B$ that fully
characterize the radiation field. We formulate the linearized
equations in the presence of a single Fourier mode of
gravitational waves. We prove that there exists a choice of
variables that reduces the problem of temperature and polarization
anisotropies to only two functions of time $\alpha(\eta)$ and
$\beta(\eta)$.

Section \ref{sec:integralequations} is devoted to further analysis
of the radiative transfer equations. The main result of this
section is the reduction of coupled integro-differential equations
to a single integral equation of Voltairre type. Essentially, the
entire problem of the CMB polarization is reduced to a single
integral equation. This allows us to use simple analytical
approximations and give transparent physical interpretation. In
Sec.\ \ref{sectionmultipoleexpansion} we generalize the analysis
to a superposition of random Fourier modes with arbitrary
wavevectors. We derive (and partially rederive the previously
known) expressions for multipole coefficients $a^X_{\ell m}$ of
the radiation field. We show that the statistical properties of
the multipole coefficients are fully determined by the statistical
properties of the underlying gravitational perturbations. This
section contains the expressions for general correlation and
cross-correlation functions $C^{XX'}_{\ell}$ for invariants of the
radiation field.

We work out astrophysical applications in Sec.\
\ref{sec:recombination} and Sec.\ \ref{reionera} where we discuss
the effects of recombination and reionization era, respectively.
Although all our main conclusions are supported by exact numerical
calculations, we show the origin of these conclusions and
essentially derive them by developing and using semi-analytical
approximations. The expected amplitudes, shapes, oscillatory
features, etc. of all correlation functions as functions of $\ell$
are under analytical control. The central point of this analysis
is the $TE$ correlation function. We show that, at lower
multipoles $\ell$, the $TE$ correlation function must be negative
(anticorrelation), if it is induced by gravitational waves, and
positive, if it is induced by density perturbations. We argue that
this difference in sign of $TE$ correlations can be a signature
more valuable observationally than the presence or absence of the
$BB$ correlations. This is because the $TE$ signal is about two
orders of magnitude larger than the expected $BB$ signal and is
much easier to measure. We summarize the competing effects of
density perturbations in Appendix \ref{app:densityperturbations}.

Theoretical findings are compared with observations in Sec.\
\ref{comparison}. In the context of relic gravitational waves it
is especially important that the WMAP team \cite{Page2006}
stresses (even if for a different reason) the actual detection of
the $TE$ anticorrelation near $\ell \approx 30$. We show that this
is possible only in the presence of a significant amount of relic
gravitational waves (within the framework of all other common
assumptions). We analyze the CMB data and suggest models with
significant amounts of gravitational waves that are consistent
with the measured $TT$, $TE$ and $EE$ correlation functions. Our
final conclusion is that there is evidence of the presence of
relic gravitational waves in the already available CMB data, and
further study of the $TE$ correlation at lower $\ell$'s has the
potential of a firm positive answer.


\section{\label{sec:gw} General properties of relic gravitational waves}

Here we summarize some properties of cosmological perturbations of
quantum-mechanical origin. We will need formulas from this summary
for our further calculations.

\subsection{\label{gwbasic} Basic definitions}

As usual (for more details, see, for example, \cite{Bose2002}), we
write the perturbed gravitational field of a flat FLRW universe in
the form
\begin{eqnarray}
ds^{2} = -c^{2} dt^{2} + a^{2}(t)(\delta_{ij} + h_{ij}) dx^idx^j=
a^{2}(\eta)\left[ - d\eta^{2} + (\delta_{ij} + h_{ij}) dx^idx^j
\right].
\label{FRWmetric}
\end{eqnarray}
We denote the present moment of time by $\eta=\eta_R$ and define
it by the observed quantities, for example, by today's value of
the Hubble parameter $H_0=H(\eta_R)$. In addition, we take the
present day value of the scale factor to be $a(\eta_R) = 2l_H$,
where $l_H = c/ H_0$.

The functions $h_{ij}\left(\eta,{\bf x}\right)$ are expanded over
spatial Fourier harmonics $e^{\pm i {\bf n\cdot x}}$, where ${\bf
n}$ is a dimensionless time independent wave-vector. The
wavenumber $n$ is $n=(\delta_{ij}n^in^j)^{1/2}$. The wavelength
$\lambda$, measured in units of laboratory standards, is related
to $n$ by $\lambda = 2\pi a /n$. The waves whose wavelengths today
are equal to today's Hubble radius carry the wavenumber $n_H=
4\pi$. Shorter waves have larger $n$'s and longer waves have
smaller $n$'s.

The often used dimensional wavenumber $k$, defined by $k= 2
\pi/\lambda(\eta_R)$ in terms of today's wavelength
$\lambda(\eta_R)$, is related to $n$ by a simple formula
\begin{eqnarray}
k = \frac{n}{2 l_H} \approx n\left(1.66 \times 10^{-4} ~h\right)
\textrm{Mpc}^{-1}.
\nonumber
\end{eqnarray}

The expansion of the field $h_{ij}\left(\eta,{\bf x}\right)$ over
Fourier components ${\bf n}$ requires a specification of
polarization tensors $\stackrel{s}{p}_{ij}({\bf n})$ (s = 1, 2).
They have different forms depending on whether the functions
$h_{ij}(\eta, {\bf x})$ represent gravitational waves, rotational
perturbations, or density perturbations.

In the case of gravitational waves, two independent linear
polarization states can be described by two real polarization
tensors
\begin{eqnarray}
\stackrel{1}{p}_{ij}({\bf n})=l_il_j - m_im_j,~~
\stackrel{2}{p}_{ij}({\bf n})=l_im_j+m_il_j,
\label{ptensors}
\end{eqnarray}
where spatial vectors $({\bf l},{\bf m},{\bf n}/n)$ are unit and
mutually orthogonal vectors. The polarization tensors
(\ref{ptensors}) satisfy the conditions
\begin{eqnarray}
\stackrel{s}{p}_{ij}\delta^{ij}=0,~~~\stackrel{s}{p}_{ij}n^{i}=0,~~~
\stackrel{s'}{p}_{ij}\stackrel{s}{p}{}^{ij}=2\delta_{s's}.
\label{orthpol}
\end{eqnarray}
The eigenvectors of $\stackrel{1}{p}_{ij}$ are $l_i$ and $m_i$,
whereas the eigenvectors of $\stackrel{2}{p}_{ij}$ are $l_i+m_i$
and $l_i-m_i$. In both cases, the first eigenvector has the
eigenvalue $+1$, whereas the second eigenvector has the eigenvalue
$-1$.

In terms of spherical coordinates $\theta,~ \phi$, we choose for
$({\bf l},{\bf m},{\bf n}/n)$ the right-handed triplet:
\begin{eqnarray}
\begin{array}{l}
{\bf l} = (\cos\theta \cos\phi,~ \cos\theta \sin\phi,~
-\sin\theta),\\
{\bf m}=(-\sin\phi,~ \cos\phi,~ 0),\\
{\bf n}/n =(\sin\theta \cos\phi,~ \sin\theta \sin\phi,~
\cos\theta).
\end{array}
\label{lmn}
\end{eqnarray}
The vector ${\bf l}$ points along a meridian in the direction of
increasing $\theta$, while the vector ${\bf m}$ points along a
parallel in the direction of increasing $\phi$. With this
specification, polarization tensors (\ref{ptensors}) will be
called the `+' and `${\times}$' polarisations. The eigenvectors of
`+' polarization correspond to north-south and east-west
directions, whereas the `${\times}$' polarization describes the
directions rotated by 45$^o$.

With a fixed ${\bf n}$, the choice of vectors ${\bf l}, {\bf m}$
given by Eq.\ (\ref{lmn}) is not unique. The vectors can be
subject to continuous and discrete transformations. The continuous
transformation is performed by a rotation of the pair ${\bf l},
{\bf m}$ in the plane orthogonal to ${\bf n}$:
\begin{eqnarray}
{\bf l'} = {\bf l} \cos \psi + {\bf m} \sin \psi, ~~~ {\bf m'} =
-{\bf l} \sin \psi + {\bf m} \cos\psi,
\label{rotlm}
\end{eqnarray}
where $\psi$ is an arbitrary angle. The discrete transformation is
described by the flips of the ${\bf l}, {\bf m}$ vectors:
\begin{eqnarray}
{\bf l'} = -{\bf l},~~ {\bf m'}= {\bf m}~~~~~~~{\rm or} ~~~~~~~
{\bf l'} = {\bf l},~~ {\bf m'}= -{\bf m}.
\label{flips}
\end{eqnarray}
When (\ref{rotlm}) is applied, polarization tensors
(\ref{ptensors}) transform as
\begin{eqnarray}
\begin{array}{l}
\stackrel{1}{p'}_{ij}({\bf n})=l'_il'_j - m'_im'_j =
\stackrel{1}{p}_{ij}({\bf n}) \cos {2\psi} +
\stackrel{2}{p}_{ij}({\bf n}) \sin {2\psi}, \\
\stackrel{2}{p'}_{ij}({\bf n})=l'_im'_j+m'_il'_j =
-\stackrel{1}{p}_{ij}({\bf n}) \sin {2\psi} +
\stackrel{2}{p}_{ij}({\bf n}) \cos {2\psi},
\end{array}
\label{ptensors2}
\end{eqnarray}
and when (\ref{flips}) is applied they transform as
\begin{eqnarray}
\stackrel{1}{p'}_{ij}({\bf n})= \stackrel{1}{p}_{ij}({\bf n}),
~~~~~~~~ \stackrel{2}{p'}_{ij}({\bf n})=
-\stackrel{2}{p}_{ij}({\bf n}).
\label{ptensors3}
\end{eqnarray}

Later in this section and in Appendix \ref{detailsgw}, we are
discussing the conditions under which the averaged observational
properties of a random field $h_{ij}({\bf n}, \eta)$ are symmetric
with respect to rotations around the axis ${\bf n}/n$ and with
respect to mirror reflections of the axes. Formally, this is
expressed as the requirement of symmetry of the g.w.\ field
correlation functions with respect to transformations
(\ref{rotlm}) and (\ref{flips}).

In this paper we will also be dealing with density perturbations.
In this case, the polarization tensors are
\begin{eqnarray}
\stackrel{1}{p}_{ij}=\sqrt{\frac{2}{3}} \delta_{ij},~~
\stackrel{2}{p}_{ij}= -\sqrt{3} \frac{n_in_j}{n^{2}}+
\frac{1}{\sqrt{3}}\delta_{ij}.
\label{ptensors4}
\end{eqnarray}
These polarization tensors satisfy the last of the conditions
(\ref{orthpol}).

In the rigorous quantum-mechanical version of the theory, the
functions $h_{ij}$ are quantum-mechanical operators. We write them
in the form:
\begin{eqnarray}
h_{ij}\left(\eta,{\bf x}\right) = \frac{\mathcal{
C}}{(2\pi)^{3/2}}\int\limits_{-\infty}^{+\infty}~d^{3}{\bf
n}\frac{1}{\sqrt{2n}} \sum_{s=1,2} \left[\stackrel{s}{p}_{ij}({\bf
n}) \stackrel{s}{h}_n\left(\eta\right)e^{i{\bf n\cdot x}}
\stackrel{s}{c}_{\bf n}+ {\stackrel{s}{p}_{ij}}^{*}({\bf n})
{\stackrel{s}{h}_n}^{*}(\eta)e^{-i{\bf n\cdot x}}
\stackrel{s}{c}_{\bf n}^{\dagger}\right],
\label{fourierh}
\end{eqnarray}
where the annihilation and creation operators,
$\stackrel{s}{c}_{\bf n}$ and $\stackrel{s}{c}_{\bf n}^{\dag}$,
satisfy the relationships
\begin{eqnarray}
[\stackrel{s'}{c}_{\bf n},~\stackrel{s}{c}_{\bf n'}^\dag] =
\delta_{s's}\delta^{(3)}({\bf n}-{\bf n}'), ~~
\stackrel{s}{c}_{\bf n}\left|0\right>=0.
\label{comrel}
\end{eqnarray}
The initial vacuum state $\left|0\right>$ of perturbations is
defined at some moment of time $\eta_0$ in the remote past, long
before the onset of the process of superadiabatic amplification.
This quantum state is maintained (in the Heisenberg picture) until
now. For gravitational waves, the normalization constant is
$\mathcal{C} =\sqrt{16\pi}l_{Pl}$.

The relationships (\ref{comrel}) determine the expectation values
and correlation functions of cosmological perturbations
themselves, and also of the CMB's temperature and polarization
anisotropies caused by these cosmological perturbations. In
particular, the variance of metric perturbations is given by
\begin{eqnarray}
\left<0\right|h_{ij}(\eta,{\bf x})h^{ij}(\eta,{\bf
x})\left|0\right> =
\frac{\mathcal{C}^{2}}{2\pi^{2}}\int\limits_{0}^{\infty}
~n^{2}\sum_{s=1,2}|\stackrel{s}{h}_n(\eta)|^{2}\frac{dn}{n}.
\label{meansq}
\end{eqnarray}
The quantity
\begin{eqnarray}
h^{2}(n,\eta) = \frac{\mathcal{
C}^{2}}{2\pi^{2}}n^{2}\sum_{s=1,2}|\stackrel{s}{h}_n(\eta)|^{2} =
\frac{1}{2}\sum_{s=1,2}|\stackrel{s}{h}(n,\eta)|^{2},
\label{gwpower}
\end{eqnarray}
is called the metric power spectrum. Note that we have introduced
\begin{eqnarray}
\stackrel{s}{h}(n,\eta) = \frac{\mathcal{
C}}{\pi}n\stackrel{s}{h}_n(\eta).
\label{gw_h_shortened}
\end{eqnarray}
The quantity (\ref{gwpower}) gives the mean-square value of the
gravitational field perturbations in a logarithmic interval of
$n$. The spectrum of the root-mean-square (rms) amplitude $h(n,
\eta)$ is determined by the square root of Eq.\ (\ref{gwpower}).
Having evolved the classical mode functions
$\stackrel{s}{h}_n(\eta)$ up to some arbitrary instant of time
$\eta$ (for instance, today $\eta=\eta_R$) one can find the power
spectrum $h(n, \eta)$ at that instant of time. For the today's
spectrum in terms of frequency $\nu$ measured in Hz, $\nu =n
H_0/4\pi$, we use the notation $h_{rms}(\nu)$.

In our further applications we will also need the power spectrum
of the first time-derivative of metric perturbations:
\begin{eqnarray}
\left<0\right|\frac{\partial h_{ij}(\eta,{\bf x})}{\partial \eta}
\frac{\partial h^{ij}(\eta,{\bf x})}{\partial \eta}\left|0\right>
= \frac{1}{2}\int\limits_{0}^{\infty} ~\sum_{s=1,2}\left|\frac{
d\stackrel{s}{h}(n,\eta)}{d\eta} \right|^{2} \frac{dn}{n}.
\label{powerder}
\end{eqnarray}

To simplify calculations, in the rest of the paper we will be
using a `classical' version of the theory, whereby the
quantum-mechanical operators $\stackrel{s}{c}_{\bf n}$ and
$\stackrel{s}{c}_{\bf n}^{\dag}$ are treated as classical random
complex numbers $\stackrel{s}{c}_{\bf n}$ and
$\stackrel{s}{c}_{\bf n}^{*}$. It is assumed that they satisfy the
relationships analogous to~(\ref{comrel}):
\begin{eqnarray}
\langle \stackrel{s}{c}_{\bf n}\rangle = \langle
\stackrel{s'}{c}^*_{\bf n'}\rangle = 0, ~~~ \langle
\stackrel{s}{c}_{\bf n} \stackrel{s'}{c}^*_{\bf n'}\rangle =
\langle \stackrel{s}{c}^*_{\bf n} \stackrel{s'}{c}_{\bf n'}\rangle
= \delta_{ss'}\delta^{(3)}({\bf n} - {\bf n'}), ~~~ \langle
\stackrel{s}{c}_{\bf n} \stackrel{s'}{c}_{\bf n'}\rangle = \langle
\stackrel{s}{c}^*_{\bf n} \stackrel{s'}{c}^*_{\bf n'}\rangle = 0,
\label{statCs}
\end{eqnarray}
where the averaging is performed over the ensemble of all possible
realizations of the random field (\ref{fourierh}).

The relationships (\ref{statCs}) are the only statistical
assumptions that we make. They fully determine all the
expectations values and correlation functions that we will
calculate, both for cosmological perturbations and for the induced
CMB fluctuations. For example, the metric power spectrum
(\ref{gwpower}) follows now from the calculation:
\begin{eqnarray}
\frac{1}{2} \left< h_{ij}(\eta,{\bf x})h^{ij}(\eta,{\bf x})\right>
= \frac{\mathcal{C}^{2}}{2\pi^{2}}\int\limits_{0}^{\infty}
~n^{2}\sum_{s=1,2} |\stackrel{s}{h}_n(\eta)|^{2}\frac{dn}{n}.
\label{gwpow}
\end{eqnarray}

The quantities $|\stackrel{s}{h}_n(\eta)|^{2}$ are responsible for
the magnitude of the mean-square fluctuations of the field in the
corresponding polarization states $s$. In general, the assumption
of statistical independence of two linear polarization components
in one polarization basis is not equivalent to this assumption in
another polarization basis. As we show in Appendix
\ref{detailsgw}, statistical properties are independent of the
basis (i.e. independent of $\psi$, Eq.\ (\ref{ptensors2})), if the
condition
\begin{eqnarray}
|\stackrel{+}{h}_n(\eta)|^{2} = |\stackrel{\times}{h}_n(\eta)|^{2}
\label{eqval}
\end{eqnarray}
is satisfied. As for the discrete transformations
(\ref{ptensors3}), they leave the g.w.\ field correlation
functions unchanged.

In our further discussion of the CMB polarization it will be
convenient to use also the expansion of $h_{ij}$ over circular,
rather than linear, polarization states. In terms of definitions
(\ref{ptensors}), the left $(s=L)$ and right $(s=R)$ circular
polarization states are described by the complex polarization
tensors
\begin{eqnarray}
\begin{array}{c}
\stackrel{L}{p}_{ij}=\frac{1}{\sqrt 2}\left(\stackrel{1}{p}_{ij}+
i \stackrel{2}{p}_{ij}\right), ~~~~~
\stackrel{R}{p}_{ij}=\frac{1}{\sqrt 2}\left(\stackrel{1}{p}_{ij}-
i \stackrel{2}{p}_{ij}\right), \\
{\stackrel{L}{p}_{ij}}^{*}=\stackrel{R}{p}_{ij}, ~~~~~
{\stackrel{R}{p}_{ij}}^{*}=\stackrel{L}{p}_{ij},
\end{array}
\label{pctensors}
\end{eqnarray}
satisfying the conditions (for $s = L, R$)
\begin{eqnarray}
\stackrel{s}{p}_{ij}\delta^{ij}=0,~~\stackrel{s}{p}_{ij}n^{i}=0,~~
\stackrel{s'}{p}_{ij}{\stackrel{s}{p}{}^{ij}}^{*}=2\delta_{s's}.
\label{orth2pol}
\end{eqnarray}

A continuous transformation (\ref{rotlm}) brings the tensors
(\ref{pctensors}) to the form
\begin{eqnarray}
\stackrel{L}{p}{'}_{ij}=\stackrel{L}{p}_{ij} e^{-i2\psi},~~~~
\stackrel{R}{p}{'}_{ij}=\stackrel{R}{p}_{ij} e^{i2\psi}.
\label{pctensors2}
\end{eqnarray}
Functions transforming according to the rule (\ref{pctensors2})
are called the spin-weighted functions of spin +2 and spin -2,
respectively \cite{Newman1966, Goldberg1967, Thorne1980}. A
discrete transformation (\ref{flips}) applied to (\ref{pctensors})
interchanges the left and right polarisation states:
\begin{eqnarray}
\stackrel{L}{p}{'}_{ij}=\stackrel{R}{p}_{ij},~~~~~~~
\stackrel{R}{p}{'}_{ij}=\stackrel{L}{p}_{ij}.
\label{pctensors3}
\end{eqnarray}

The assumption of statistical independence of two linear
polarisation states is, in general, not equivalent to the
assumption of statistical independence of two circular
polarization states. Moreover, it is shown in Appendix
\ref{detailsgw} that symmetry between left and right is violated,
unless
\begin{eqnarray}
|\stackrel{L}{h}_n(\eta)|^{2} = |\stackrel{R}{h}_n(\eta)|^{2}.
\label{eqval2}
\end{eqnarray}
However, if conditions (\ref{eqval}), (\ref{eqval2}) are
satisfied, statistical properties of the random g.w.\ field remain
unchanged under transformations from one basis to another,
including the transitions between linear and circular
polarizations. The summation over $s$ in the power spectra such as
(\ref{gwpow}) can be replaced by the multiplicative factor 2.
Moreover, the mode functions $\stackrel{s}{h}_n (\eta)$ for two
independent polarization states will be equal up to a constant
complex factor $e^{i \alpha}$. This factor can be incorporated in
the redefinition of random coefficients $\stackrel{s}{c}_{\bf n}$
without violating the statistical assumptions (\ref{statCs}).
After this, the index $s$ over the mode functions can be dropped:
\begin{eqnarray}
\stackrel{s}{h}_n(\eta) = h_n (\eta).
\label{eqval3}
\end{eqnarray}

There is no special reason for the quantum-mechanical generating
mechanism to prefer one polarization state over another. It is
natural to assume that the conditions (\ref{eqval}),
(\ref{eqval2}) hold true for relic gravitational waves, but in
general they could be violated. In calculations below we often use
these equalities, but we do not enforce them without warning.

\subsection{Mode functions and power spectra \label{sec:modefunctions}}

The perturbed Einstein equations give rise to the g.w.\ equation
for the mode functions $\stackrel{s}{h}_{n}(\eta)$. This equation
can be transformed to the equation for a parametrically disturbed
oscillator \cite{Grishchuk1974}:
\begin{eqnarray}
\stackrel{s}{\mu}^{''}_{n} + \stackrel{s}{\mu}_{n}\left[n^{2} -
\frac{a''}{a}\right] =0,
\label{meq}
\end{eqnarray}
where $\stackrel{s}{\mu}_{n}(\eta)\equiv
a(\eta)\stackrel{s}{h}_{n}(\eta)$ and $'=d/d\eta = (a/c)d/dt$. (In
this paper, we ignore anisotropic stresses. For the most recent
account of this subject, which includes earlier references, see
\cite{Watanabe2006}.) Clearly, the behaviour of the mode functions
depend on the gravitational `pumping' field $a(\eta)$, regardless
of the physical nature of the matter sources driving the
cosmological scale factor $a(\eta)$. Observational data about
relic gravitational waves allow us to make direct inferences about
$H(\eta)$ and $a(\eta)$ \cite{Grishchuk1991}, and it is only
through extra assumptions that we can make inferences about such
things as, say, the scalar field potential (if it is relevant at
all).

Previous analytical calculations (see, for example,
\cite{Grishchuk2001a,Bose2002} and references there) were based on
models where $a(\eta)$ consists of pieces of power-law evolution
\begin{eqnarray}
a(\eta) = l_o |\eta| ^{1+\beta},
\label{scalef}
\end{eqnarray}
where $l_o$ and $\beta$ are constants. The functions $a(\eta)$,
$a^{\prime}(\eta)$, $\stackrel{s}{h}_{n}(\eta)$,
$\stackrel{s}{h^{\prime}}_{n}(\eta)$ were continuously joined at
the transition points between various power-law eras.

It is often claimed in the literature that this method of joining
the solutions is unreliable, unless the wavelength is ``much
longer than the time taken for the transition to take place".
Specifically, it is claimed that the joining procedure leads to
huge errors in the g.w.\ power spectrum for short waves. It is
important to show that these claims are incorrect. As an example,
we will consider the transition between the radiation-dominated
and matter-dominated eras.

The exact scale factor, which accounts for the simultaneous
presence of matter, $\rho_m \propto a(\eta)^{-3}$, and radiation,
$\rho_{\gamma} \propto a(\eta)^{-4}$, has the form
\begin{eqnarray}
a(\eta) = 2l_H \left(\frac{1+z_{eq}}{2+z_{eq}}\right)~\eta~
\left(\eta + \frac{2 \sqrt{2+z_{eq}}}{1+z_{eq}} \right),
\label{scalemr}
\end{eqnarray}
where $z_{eq}$ is the redshift of the era of equality of energy
densities in matter and radiation $\rho_m(z_{eq}) =
\rho_{\gamma}(z_{eq})$,
\begin{eqnarray}
1+z_{eq} = \frac{a(\eta_R)}{a(\eta_{eq})}
=\frac{\Omega_m}{\Omega_{\gamma}}.
\nonumber
\end{eqnarray}
For this model, the values of parameter $\eta$ at equality and
today are given by the expressions
\begin{eqnarray}
\eta_{eq} = \left(\sqrt{2} -1\right)
\frac{\sqrt{2+z_{eq}}}{1+z_{eq}}, ~~~~~\eta_R = 1 -
\frac{\sqrt{2+z_{eq}}}{1+z_{eq}} + \frac{1}{1+z_{eq}}.
\nonumber
\end{eqnarray}
The current observations favour the value $z_{eq} \approx 3 \times
10^{3}$ \cite{LAMBDA}.

The piece-wise approximation to the scale factor (\ref{scalemr})
has the form
\begin{eqnarray}
a(\eta) = \frac{4l_H}{\sqrt{1+z_{eq}}} \eta, ~~\eta\leq\eta_{eq};
~~~~~ a(\eta) = 2l_H \left(\eta
+\eta_{eq}\right)^{2},~~\eta\geq\eta_{eq} ,
\label{jointsf}
\end{eqnarray}
where $\eta_{eq}$ and $\eta_R$ for this joined $a(\eta)$ are given
by $\eta_{eq} = 1/2 \sqrt{1+z_{eq}}$ and $\eta_R = 1 -
1/2\sqrt{1+z_{eq}} $. It can be seen from (\ref{scalemr}) and
(\ref{jointsf}) that the relative difference between the two scale
factors is very small in the deep radiation-dominated era, $\eta
\ll 1/\sqrt{z_{eq}}$, and in the deep matter-dominated era, $\eta
\gg 1/\sqrt{z_{eq}}$. But the difference reaches about $25\%$ at
times near equality.

The initial conditions for the g.w.\ equation (\ref{meq}) are the
same in the models (\ref{scalemr}), (\ref{jointsf}) and they are
determined by quantum-mechanical assumptions at the stage (which
we call the $i$-stage) preceding the radiation epoch. The
$i$-stage has finished, and the radiation-dominated stage has
started, at some $\eta_i$ with a redshift $z_i$. The numerical
value of $z_i$ should be somewhere near $10^{29}$ (see below).

The initial conditions at the radiation-dominated stage can be
specified at that early time $\eta_i$ or, in practice, for
numerical calculations, at much latter time, as long as the
appropriate g.w.\ solution is taken as \cite{Bose2002}
\begin{eqnarray}
\stackrel{s}{\mu}_n(\eta) = -2iB\sin{n\eta}, ~~~
\stackrel{s}{\mu}_n^{'}(\eta) = -2inB\cos{n\eta},
\label{initcond}
\end{eqnarray}
where
\begin{eqnarray}
B =F(\beta)\left(\frac{n\sqrt{1+z_{eq}}}{1+z_i}\right)^\beta,
\nonumber
\end{eqnarray}
and $F(\beta)$ is a slowly varying function of the constant
parameter $\beta$, $|F(-2)|=2$. Parameter $\beta$ describes the
power-law evolution at the $i$-stage and determines the primordial
spectral index ${\rm n}$, ${\rm n} = 2\beta +5$. In particular,
$\beta=-2$ corresponds to the flat (scale-invariant) primordial
spectrum ${\rm n} =1$.

For numerical calculations we use the constant $B$ in the form:
\begin{eqnarray}
B=2 \left(\frac{4 \pi \sqrt{1+z_{eq}}}{1+z_i}\right)^{\beta}
\left(\frac{n}{n_H}\right)^{\beta}.
\label{Bnorm}
\end{eqnarray}

\begin{figure}
\begin{center}
\includegraphics[width=10cm]{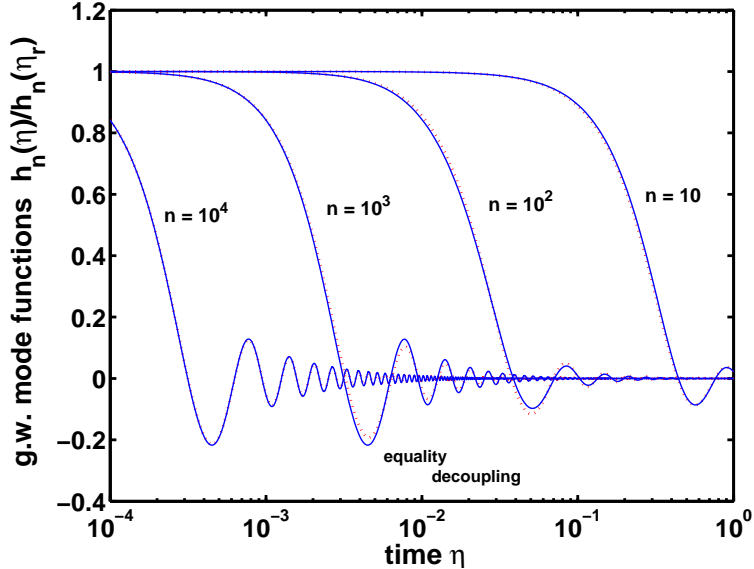}
\end{center}
\caption{ The g.w.\ mode functions $ h_n(\eta)/h_n(\eta_r)$ in a
matter-radiation universe. The solid curves are numerically
calculated solutions on the scale factor (\ref{scalemr}), while
the dotted curves are analytical solutions on the scale factor
(\ref{jointsf}).} \label{hfigure}
\end{figure}

The wave-equation (\ref{meq}) with the scale factor
(\ref{scalemr}) can not be solved in elementary functions.
However, it can be solved numerically using the initial data
(\ref{initcond}). Concretely, we have imposed the initial data
(\ref{initcond}) at $\eta_r = 10^{-6}$, which corresponds to the
redshift $z \approx 3\times 10^{7}$, and have chosen $z_{eq} =
6\times 10^{3}$ for illustration. Numerical solutions for
$h_n(\eta)/h_n(\eta_r)$ characterized by different wavenumbers $n$
are shown by solid lines in Fig.\ \ref{hfigure}. We should compare
these solutions with the joined solutions found on the joined
evolution (\ref{jointsf}) for the same wavenumbers.

The piece-wise scale factor (\ref{jointsf}) allows one to write
down the piece-wise analytical solutions to the g.w.\ equation
(\ref{meq}):
\begin{eqnarray}
\mu_n(\eta) = \left\{\begin{array}{l}
-2iB\sin{n\eta},~~~~~~\eta\leq\eta_{eq}\\
\sqrt{n(\eta + \eta_{eq})}\left[A_nJ_{3/2} \left(\frac{}{}n(\eta +
\eta_{eq})\right)-iB_nJ_{-3/2} \left(\frac{}{}n(\eta +
\eta_{eq})\right)\right],~~~\eta\geq\eta_{eq},
\end{array}\right.
\label{hm-r}
\end{eqnarray}
where $J_{3/2}, J_{-3/2}$ are Bessel functions. The coefficients
$A_n$ and $B_n$ are calculated from the condition of continuous
joining of $\mu_n$ and $\mu'_n$ at $\eta_{eq}$ \cite{Bose2002}:
\begin{eqnarray}
\begin{array}{l}
A_n = -i\sqrt{\frac{\pi}{2}}\frac{B}{4y_2^{2}}
\left[(8y_2^{2}-1)\sin{y_2}+4y_2\cos{y_2}+\sin{3y_2}\right],\\
B_n = -\sqrt{\frac{\pi}{2}}\frac{B}{4y_2^{2}}
\left[(8y_2^{2}-1)\cos{y_2}-4y_2\sin{y_2}+\cos{3y_2}\right],
\end{array}
\nonumber
\end{eqnarray}
where $y_2 =n\eta_{eq}$. In Fig.\ \ref{hfigure} we show the joined
solutions (\ref{hm-r}) by dotted curves.

It is clear from Fig.\ \ref{hfigure} that the g.w.\ solutions as
functions of $\eta$ are pretty much similar in the two models. The
solid and dotted curves are slightly different near equality
(where the relative difference between the scale factors is
noticeably large) and only for modes that entered the Hubble
radius around or before equality. Moreover, the g.w.\ amplitudes
of the modes that entered the Hubble radius before equality
gradually equalize in course of later evolution. There is nothing
like a huge underestimation or overestimation of the
high-frequency g.w.\ power that was alleged to happen in the
joined model.

It is easy to understand these features. Let us start from waves
that entered the Hubble radius well after equality, i.e. waves
with wavenumbers ${n}/{n_H}\ll {\sqrt{1+z_{eq}}}$. As long as
these waves are outside the Hubble radius, their amplitudes remain
constant and equal in the two models, despite the fact that the
scale factors are somewhat different near equality. The waves
start oscillating in the regime where the relative difference
between (\ref{scalemr}) and (\ref{jointsf}) is small, and
therefore the mode functions, as functions of $\eta$, coincide.

The waves with wavenumbers ${n}/{n_H}\gg {\sqrt{1+z_{eq}}}$ enter
the Hubble radius well before the equality. They oscillate in the
WKB regime according to the law $h_n(\eta) \propto
e^{-in\eta}/a(\eta)$, having started with equal amplitudes in the
two models. Near equality, the mode functions are different in the
two models, as much as the scale factors are different. But the
relative difference between (\ref{scalemr}) and (\ref{jointsf})
decreases with time, and therefore the mode functions in the two
models gradually equalize. The amplitudes of these mode functions,
as well as the scale factors (\ref{scalemr}), (\ref{jointsf}), are
exactly equal today. The only difference between these
high-frequency mode functions is in phase, that is, in different
numbers of cycles that they experienced by today. This is because
the moment of time defined as `today' in the two models is given,
in terms of the common parameter $\eta$, by slightly different
values of $\eta_R$.

Finally, for intermediate wavenumbers ${n}/{n_H}\approx
{\sqrt{1+z_{eq}}}$, the modes enter the Hubble radius when the
scale factors differ the most. Therefore, they start oscillating
with somewhat different amplitudes. The difference between these
mode functions is noticeable by the redshift of decoupling
$z_{dec}$ (characterized by somewhat different values of
$\eta_{dec}$), as shown in Fig.\ \ref{hfigure}. The difference
survives until today, making the graph for the g.w.\ power
spectrum (\ref{gwpower}) a little smoother (in comparison with
that derived from the joined model) in the region of frequencies
$10^{-16}$ Hz that correspond to the era of equality.

Having demonstrated that the use of joined analytical solutions is
well justified, and the previously plotted graphs for
$h_{rms}(\nu)$ and $\Omega_{gw}(\nu)$ are essentially correct, we
shall now exhibit the more accurate graphs based on numerical
calculations with the initial conditions (\ref{initcond}),
(\ref{Bnorm}). We adopt $z_{eq}=3\times 10^{3}$ and $H_0 = 75
\frac{km}{s}/Mpc$ \cite{LAMBDA}. We adjust the remaining free
constant $z_i$ in such a way that the temperature correlation
function $\ell( \ell + 1)C_{\ell}/2 \pi$ at $\ell = 2$ is equal to
211 $~\mu \textrm{K} ^{2}$ \cite{Hinshaw2006}. This requires us to
take $z_i = 1.0 \times 10^{29}$ for $\beta =-2$ and $z_i = 2.4
\times 10^{30}$ for $\beta =-1.9$.

\begin{figure}
\begin{center}
\includegraphics[width=9cm]{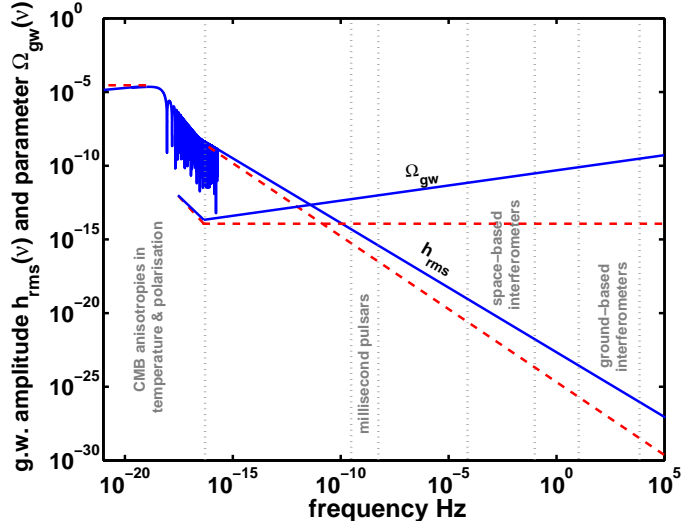}
\end{center}
\caption{The present-day spectra for $h_{rms}(\nu)$ and
$\Omega_{gw}(\nu)$. The solid lines correspond to the primordial
spectral index $\beta =-1.9$, i.e. ${\rm n} =1.2$, while the
dashed lines are for $\beta= -2$, i.e. ${\rm n} =1$.}
\label{figure_gwSpectrum}
\end{figure}

The graphs for today's spectra, normalised as stated above, are
shown in Fig.\ \ref{figure_gwSpectrum}. In order to keep the
figure uncluttered, the oscillations of $h_{rms}(\nu)$ are shown
only at low frequencies, while the oscillations of
$\Omega_{gw}(\nu)$ are not shown at all. We have also indicated
possible detection techniques in various frequency bands.

The function $\Omega_{gw}(\nu)$ is the spectral value of the
cosmological parameter $\Omega_{gw}$ \cite{Grishchuk2001a, Grishchuk1988}:
\begin{eqnarray}
\Omega_{gw}(\nu_1, \nu_2) = \frac{\rho_{gw}(\nu_1, \nu_2)}{\rho_c}
= \frac{1}{\rho_c}\int_{\nu_1}^{\nu_2} \rho_{gw}(\nu) \frac{d
\nu}{\nu}= \int_{\nu_1}^{\nu_2} \Omega_{gw} (\nu) \frac{d
\nu}{\nu},
\nonumber
\end{eqnarray}
that is,
\begin{eqnarray}
\Omega_{gw}(\nu) = \frac{1}{\rho_c} \rho_{gw}(\nu).
\nonumber
\end{eqnarray}
Using the high-frequency definition of $\rho_{gw}(\nu)$ (valid
only for waves which are significantly shorter than $l_H$) we
derive \cite{Grishchuk2001a}:
\begin{eqnarray}
\Omega_{gw}(\nu) = \frac{\pi^{2}}{3} h_{rms}^{2}(\nu) \left(
\frac{\nu}{\nu_H}\right)^{2}. \label{omegagw}
\end{eqnarray}
It is this definition of $\Omega_{gw}(\nu)$ that is used for
drawing the curves in Fig.\ \ref{figure_gwSpectrum}.

We have to warn the reader that a great deal of literature on
stochastic g.w.\ backgrounds uses the incorrect definition
\begin{eqnarray}
\Omega_{gw}(\nu) = \frac{1}{\rho_c} \frac{d \rho_{gw}(\nu)}{d ~ln
\nu},\nonumber
\end{eqnarray}
which suggests that the $\Omega_{gw}(\nu)$ parameter is zero if
the g.w.\ spectral energy density $\rho_{gw}(\nu)$ is
frequency-independent, regardless of the numerical value of
$\rho_{gw}(\nu)$. Then, from this incorrect definition, a formula
similar to Eq.\ (\ref{omegagw}) is often being derived by making
further compensating errors.

The higher-frequency part of $h_{rms}(\nu)$ is relevant for direct
searches for relic gravitational waves, while the lower-frequency
part is relevant to the CMB calculations that we turn to in the
next section. The direct and indirect methods of detecting relic
gravitational waves are considered in a large number of papers
(see, for example, references
\cite{Grishchuk1988,Allen1999,Corbin2006,Boyle2005,Cooray2005,Smith2006,
SPC2006,Chongchitnan2005, Mandic2006},\cite{Grishchuk2005} even though we
disagree with some of them).

It is important to keep in mind that according to the inflationary
theory the dashed lines for $\Omega_{gw}(\nu)$ and $h_{rms}(\nu)$
in Fig.\ \ref{figure_gwSpectrum} should be at a zero level, because
they describe the g.w.\ background with a flat primordial spectrum
$\beta = -2$, ${\rm n}=1$. The `consistency relation' of the
inflationary theory demands $r=0$, i.e. vanishingly small g.w.\
background, in the limit ${\rm n} =1$.


\section{\label{sec:rte} Equations of radiative transfer}

\subsection{Characterization of a radiation field}

The radiation field is usually characterized by the four Stokes
parameters $(I,Q,U,V)$, \cite{Chandrasekhar1960,Landau1975}. The
parameter $I$ is the total intensity of radiation, $Q$ and $U$
describe the magnitude and direction of linear polarization, and
$V$ is the circular polarization. The Stokes parameters can be
viewed as functions of photons' coordinates and momenta
$(x^{\alpha},p^{\alpha})$. Since photons propagate with the speed
of light $c$, the momenta satisfy the condition
$p_{\alpha}p^{\alpha}=0$. Thus, the Stokes parameters are
functions of $(t,x^{i},\nu,e^{i})$, where $\nu$ is the photon's
frequency, and $e^{i}$ is a unit vector in the direction of
observation (opposite to the photon's propagation).

In a given space-time point $(t,x^i)$, the Stokes parameters are
functions of $\nu, \theta, \phi$, where $\theta, \phi$ are
coordinates on a unit sphere:
\begin{eqnarray}
d \sigma^{2} = g_{ab} dx^a dx^b = d \theta^{2} +\sin^{2} \theta d
\phi^{2}.
\label{sphmetr}
\end{eqnarray}
The radial direction is the direction of observation.

The Stokes parameters are components of the polarization tensor
$P_{ab}$ \cite{Landau1975} which can be written as
\begin{eqnarray}
P_{ab}(\theta,\phi) = \frac{1}{2}
\left(\begin{array}{c}~~ I + Q ~~~~~~~~~~ -(U - iV)\sin{\theta} \\
-(U + iV)\sin{\theta} ~~~~~ (I - Q)\sin^{2}{\theta}
\end{array}\right).
\label{PolarizationtensorPab}
\end{eqnarray}
(We do not indicate the dependence of Stokes parameters on $\nu$.)

The symmetry of $P_{ab}$ with respect to rotations around the
direction of observation requires the vanishing of linear
polarization $Q=0$, $U=0$, but the circular polarization $V$ can
be present. The symmetry of $P_{ab}$ with respect to coordinate
reflections in the observation plane requires $V=0$, but the
linear polarization can be present. The first symmetry means that
the readings of a linear polarimeter are the same when it is
rotated in the observation plane. The second symmetry means that
the readings of the left-handed and right-handed circular
polarimeters are the same. (Compare with the discussion on
gravitational waves in Sec.\ \ref{gwbasic}.)

Under arbitrary transformations of $\theta, \phi$, the components
of $P_{ab}(\theta,\phi)$ transform as components of a tensor, but
some quantities remain invariant. We want to build linear
invariants from $P_{ab}$ and its derivatives, using the metric
tensor $g_{ab}(\theta, \phi)$ and a completely antisymmetric
pseudo-tensor $\epsilon^{ab}(\theta, \phi)$,
\begin{eqnarray}
\epsilon^{ab} = \left(\begin{array}{c} ~~ 0 ~~~~~~ -\sin^{-1}{\theta} \\
\sin^{-1}{\theta} ~~~~~~~~ 0 \end{array}\right).
\nonumber
\end{eqnarray}
The first two invariants are easy to build:
\begin{eqnarray}
I(\theta,\phi) = g^{ab}(\theta,\phi)P_{ab}(\theta,\phi), ~~~
V(\theta,\phi) = i\epsilon^{ab}(\theta,\phi)P_{ab}(\theta,\phi).
\label{IandV}
\end{eqnarray}
Then, it is convenient to single out the trace and antisymmetric
parts of $P_{ab}$, and introduce the symmetric trace-free (STF)
part $P_{ab}^{STF}$:
\begin{eqnarray}
P_{ab}(\theta,\phi) = \frac{1}{2}Ig_{ab} -
\frac{i}{2}V\epsilon_{ab} + P_{ab}^{STF},
\nonumber
\end{eqnarray}
\begin{eqnarray}
P_{ab}^{STF}= \frac{1}{2}\left(\begin{array}{c}~ Q ~~ -U\sin{\theta} \\
-U \sin{\theta} ~ -Q\sin^{2}{\theta}
\end{array}\right).
\nonumber
\end{eqnarray}

Clearly, the construction of other linear invariants requires the
use of covariant derivatives of the tensor $P_{ab}^{STF}$. There
is no invariants that can be built from the first derivatives
$P_{ab;c}^{STF}$, so we need to go to the second derivatives. One
can check that there are only two linearly independent invariants
that can be built from the second derivatives:
\begin{eqnarray}
E\left(\theta,\phi\right)= -2\left(P^{STF}_{ab}\right)^{;a;b},~~~
B\left(\theta,\phi\right)=-2\left(P^{STF}_{ab}\right)^{;b;d}
\epsilon^{a}_{~d},
\label{EandB}
\end{eqnarray}
The quantities $I$ and $E$ are scalars, while $V$ and $B$ are
pseudoscalars. $V$ and $B$ change sign under flips of directions
(coordinate transformations with negative determinants). This is
also seen from the fact that their construction involves the
pseudo-tensor $\epsilon_{ab}$.

The invariant quantities $(I,V,E,B)$, as functions of $\theta,
\phi$, can be expanded over ordinary spherical harmonics $Y_{\ell
m}(\theta,\phi)$, $Y_{\ell m}^* = (-1)^m Y_{\ell , -m}$:
\begin{subequations}
\label{multipolecoefficients}
\begin{eqnarray}
I(\theta,\phi) &=& \sum_{\ell
=0}^{\infty}\sum_{m=-\ell}^{\ell}a_{\ell m}^TY_{\ell
m}(\theta,\phi),
\label{multipolecoefficientsI} \\
V(\theta,\phi) &=& \sum_{\ell =0}^{\infty}\sum_{m=-\ell }^{\ell
}a_{\ell m}^VY_{\ell m}(\theta,\phi),
\label{multipolecoefficientsV}\\
E(\theta,\phi) &=& \sum_{\ell =2}^{\infty}\sum_{m=-\ell }^{\ell }
\left[\frac{(\ell +2)!}{(\ell -2)!}\right]^{\frac{1}{2}} a_{\ell
m}^EY_{\ell m}(\theta,\phi),
\label{multipolecoefficientsE} \\
B(\theta,\phi) &=& \sum_{\ell =2}^{\infty}\sum_{m=-\ell }^{\ell }
\left[\frac{(\ell +2)!}{(\ell -2)!}\right]^{\frac{1}{2}} a_{\ell
m}^BY_{\ell m}(\theta,\phi).
\label{multipolecoefficientsB}
\end{eqnarray}
\end{subequations}
The set of multipole coefficients $(a_{\ell m}^T, a_{\ell m}^V,
a_{\ell m}^E, a_{\ell m}^B)$ completely characterizes the
radiation field. We will use these quantities in our further
discussion.

To make contact with previous work, we note that the multipole
coefficients $a_{\ell m}^E, a_{\ell m}^B$ can also be expressed in
terms of the tensor $P_{ab}$ itself, rather than its derivatives.
This is possible because one can interchange the order of
differentiation under the integrals that define $a_{\ell m}^E,
a_{\ell m}^B$ in terms of the right hand side (r.h.s.) of Eq.\
(\ref{EandB}). This leads to the appearance of the spin-weighted
spherical harmonics or tensor spherical harmonics
\cite{Newman1966,Goldberg1967,Thorne1980,Zaldarriaga1997,Kamionkowski1997}.
For example, the tensor $P_{ab}$ can be written as
\begin{eqnarray}
P_{ab} &=& \frac{1}{2}\sum_{\ell =0}^{\infty}\sum_{m=-\ell }^{\ell
}\left(g_{ab} a_{\ell m}^T
- i\epsilon_{ab} a_{\ell m}^V\right)Y_{\ell m}(\theta,\phi)\nonumber \\
&& + \frac{1}{\sqrt{2}}\sum_{\ell=2}^{\infty}\sum_{m=-l}^{l}
\left( -a_{\ell m}^E Y_{(\ell m)ab}^G(\theta,\phi) + a_{\ell
m}^BY_{(\ell m)ab}^C(\theta,\phi)\right),
\nonumber
\end{eqnarray}
where $Y_{(\ell m)ab}^G(\theta,\phi)$ and $Y_{(\ell
m)ab}^C(\theta,\phi)$ are the ``gradient" and ``curl" tensor
spherical harmonics forming a set of orthonormal functions for STF
tensors \cite{Kamionkowski1997}. The invariants $E$ and $B$ can
also be written in terms of the spin raising and lowering
operators $\eth$ and $\bar{\eth}$ \cite{Zaldarriaga1997}:
\begin{eqnarray}
E = -\frac{1}{2}\left[ \bar{\eth}^{2}\left( Q+iU\right) +
{\eth}^{2}\left( Q-iU\right) \right],~~~~ B = \frac{i}{2}\left[
\bar{\eth}^{2}\left( Q+iU\right) - {\eth}^{2}\left( Q-iU\right)
\right].
\nonumber
\end{eqnarray}

The quantities $E$ and $B$ are called the $E$ (or ``gradient") and
$B$ (or ``curl") modes of polarization. The $\ell $-dependent
numerical coefficients in (\ref{multipolecoefficientsE}) and
(\ref{multipolecoefficientsB}) were introduced in order to make
the definitions of this paper fully consistent with the previous
literature \cite{Zaldarriaga1997,Kamionkowski1997}.

\subsection{Radiative transfer in a perturbed universe}

We need to work out the radiative transfer equation in a slightly
perturbed universe described by Eq.\ (\ref{FRWmetric}). The
Thompson scattering of initially unpolarized light cannot generate
circular polarization, so we shall not consider the $V$ Stokes
parameter. Following
\cite{Chandrasekhar1960,Basko1980,Polnarev1985}, we shall write
the radiative transfer equation in terms of a 3-component quantity
(symbolic vector) ${\bf \hat{n}}(x^{\alpha}, p^{\alpha})$. The
components $(\hat{n}_1, \hat{n}_2, \hat{n}_3)$ are related to the
Stokes parameters by
\begin{eqnarray}
{\bf \hat{n}} = \left(\begin{array}{c} \hat{n}_1 \\ \hat{n}_2 \\
\hat{n}_3
\end{array}\right) =
\frac{1}{2}\frac{c^{2}}{h\nu^{3}} \left(\begin{array}{c}
I+Q\\I-Q\\-2U
\end{array}\right),
\label{symbolicvectorstokesparameters}
\end{eqnarray}
where $h$ is the Planck constant. The quantities $\hat{n}_1,
\hat{n}_2, (\hat{n}_1+\hat{n}_2+\hat{n}_3)/2$ are the numbers of
photons of frequency $\nu$ coming from the direction $z$ and
passing through a slit oriented, respectively, in the directions
$x$, $y$, and the bisecting direction between $x$ and $y$.

The equation of radiative transfer can be treated as a Boltzmann
equation in a phase space. The general form of this equation is as
follows \cite{Lindquist1966}
\begin{eqnarray}
\frac{D{\bf \hat{n}}}{ds} = {\bf \hat{C}}\left[ {\bf \hat{n}}
\right],
\label{Liuville}
\end{eqnarray}
where $s$ is a parameter along the world-line of a photon,
$\frac{D}{ds}$ is a total derivative along this world-line, and
${\bf \hat{C}}$ is a collision term. We shall explain each term of
this equation separately.

The total derivative in Eq.\ (\ref{Liuville}) reads:
\begin{eqnarray}
\frac{D{\bf \hat{n}}}{ds} =
\left[\frac{dx^\alpha}{ds}\frac{\partial }{\partial x^{\alpha}} +
\frac{dp^{\alpha}}{ds}\frac{\partial}{\partial p^{\alpha}}\right]
{\bf \hat{n}},
\label{totalderivative}
\end{eqnarray}
where $dx^{\alpha}/ds$ and $dp^{\alpha}/ds$ are determined by the
light-like geodesic world-line,
\begin{eqnarray}
\frac{dx^{\alpha}}{ds} = p^{\alpha},~~~\frac{dp^{\alpha}}{ds} =
-\Gamma^{\alpha}_{\beta\gamma}p^{\beta}p^{\gamma},~~~
g_{\alpha\beta}p^{\alpha}p^{\beta} = 0.
\label{generalgeodesicequations}
\end{eqnarray}
Strictly speaking, the square bracket in Eq.\
(\ref{totalderivative}) should also include the additive matrix
term ${\bf \hat{R}}$. This term is responsible for the rotation of
polarization axes that may take place in course of parallel
transport along the photon's geodesic line \cite{Caderni1978}. In
the perturbation theory that we are working with, this matrix does
not enter the equations in the zeroth and first order
approximations \cite{Basko1980}, and therefore we neglect ${\bf
\hat{R}}$.

In our problem, the collision term ${\bf \hat{C}}$ describes the
Thompson scattering of light on free (not combined in atoms)
electrons \cite{Chandrasekhar1960}. We assume that the electrons
are at rest with respect to one of synchronous coordinate systems
(\ref{FRWmetric}). We work with this coordinate system, so that it
is not only synchronous but also `comoving' with the electrons.
(This choice is always possible when the functions $h_{ij}$ in
(\ref{FRWmetric}) are gravitational waves. Certain complications
in the case of density perturbations will be considered later,
Appendix \ref{app:densityperturbations}.) Thus, the collision term
${\bf \hat{C}}$ is given by the expression
\begin{eqnarray}
{\bf \hat{C}}\left[ {\bf \hat{n}}\right] =
-\sigma_TN_e(x^{\alpha})\left(\frac{c dt}{ds}\right)\left[ {\bf
\hat{n}}(t,x^i,\nu,\theta,\phi) - \frac{1}{4\pi}\int d\Omega' {\bf
\hat{P}}(\theta,\phi ; \theta',\phi'){\bf
\hat{n}}(t,x^i,\nu,\theta',\phi') \right],
\label{collisionterm}
\end{eqnarray}
where $\sigma_T = 6.65\cdot10^{-24}~\textrm{cm}^{2}$ is the
Thompson cross section, $N_e$ is the density of free electrons,
and ${\bf \hat{P}}(\theta,\phi ; \theta',\phi')$ is the
Chandrasekhar scattering matrix. (The explicit form of the
scattering matrix is discussed in Appendix \ref{app:poln}.) The
factor $cdt/ds$ arises because of our use of the element $ds$,
instead of $cdt$, in the left hand side (l.h.s.) of Eq.\
(\ref{Liuville}). In accord with the meaning of the scattering
term, the quantity $\sigma_T N_e(x^{\alpha})(c dt/ds)$ is the
averaged number of electrons that could participate in the
scattering process when the photon traversed the element $ds$
along its path.

Let us now write down the equations of radiative transfer in the
presence of the gravitational field (\ref{FRWmetric}). First, we
write down the equations for the light-like geodesic line
$x^{\alpha}(s) =(\eta(s), x^i(s))$:
\begin{eqnarray}
p^{0} =\frac{d\eta}{ds} = \frac{\nu}{ca}, ~~~ p^i =\frac{dx^i}{ds}
= \frac{\nu}{ca}e^i, ~~~ \frac{d\nu}{ds} = -\nu\left[
\frac{1}{a}\frac{da}{d\eta} + \frac{1}{2}e^ie^j\frac{\partial
h_{ij}}{\partial \eta} \right]\frac{d\eta}{ds},
\label{geodesicequations}
\end{eqnarray}
\begin{eqnarray}
\left(\delta_{ij} + h_{ij}\right)e^i e^j =1.
\nonumber
\end{eqnarray}
We do not need the expression for $de^i/ds$, because it is a
first-order (in terms of metric perturbations) quantity, and this
quantity enters the equations of radiative transfer only in
products with other first-order terms. We neglect such
second-order corrections.

Second, we write for the `vector' ${\bf \hat{n}}$:
\begin{eqnarray}
{\bf \hat{n}} = {\bf \hat{n}}^{(0)} + {\bf \hat{n}}^{(1)},
\end{eqnarray}
where ${\bf \hat{n}}^{(0)}$ is the zeroth order solution, and
${\bf \hat{n}}^{(1)}$ is the first order correction arising
because of the presence of metric perturbations. We shall now
formulate the equation for ${\bf \hat{n}}^{(1)}$, taking into
account the zero-order solution to Eq.\ (\ref{Liuville}).

In the zero-order approximation we assume that $h_{ij}=0$ and that
the radiation field is fully homogeneous, isotropic, and
unpolarized. Therefore,
\begin{eqnarray}
{\bf \hat{n}}^{(0)} = n^{(0)}(\eta,\nu){\bf \hat{u}},
\label{zerothordersolution}
\end{eqnarray}
where
\begin{eqnarray}
{\bf \hat{u}} = \left( \begin{array}{c}1 \\ 1 \\0
\end{array} \right).
\nonumber
\end{eqnarray}
Since the scattering matrix ${\bf \hat{P}}$ does not couple to the
radiation field if it has no quadrupole anisotropy, the collision
term (\ref{collisionterm}) vanishes, ${\bf
\hat{C}}\left[{\bf\hat{n}}^{(0)}\right] = 0$, and the equation for
$n^{(0)}(\eta,\nu)$ reads
\begin{eqnarray}
\frac{\partial n^{(0)}}{\partial \eta} -
\frac{\nu}{a}\frac{da}{d\eta}\frac{\partial n^{(0)}}{\partial \nu}
= 0.
\nonumber
\end{eqnarray}
The general solution to this equation is $n^{(0)} = n_0\left(\nu
a(\eta)\right)$, which makes it convenient to use a new variable
\begin{eqnarray}
\tilde{\nu} = \nu a(\eta).
\nonumber
\end{eqnarray}
In the zero-order approximation, $\nu = const/a(\eta)$ and
therefore $\tilde{\nu}$ is a constant along the light ray.

We are now in a position to write down the first-order
approximation to the Boltzmann equation (\ref{Liuville}). We take
$(\eta,x^i,\tilde{\nu},e^i)$ as independent variables, i.e.
$n^{(0)} = n_0\left(\tilde{\nu}\right)$, ${\bf \hat{n}}^{(1)}=
{\bf \hat{n}}^{(1)}(\eta,x^i,\tilde{\nu},e^i)$, and use the
identity
\begin{eqnarray}
\frac{dp^{\alpha}}{ds}\frac{\partial}{\partial p^{\alpha}}=
\frac{d \tilde{\nu}}{ds}\frac{\partial}{\partial \tilde{\nu}}+
\frac{d e^i}{ds}\frac{\partial}{\partial e^i}
\nonumber
\end{eqnarray}
in the first-order approximation to (\ref{totalderivative}).
Taking also into account the geodesic equation
(\ref{geodesicequations}) we arrive at the equation
\begin{eqnarray}
\left[\frac{\partial {\bf \hat{n}}^{(1)} }{\partial \eta} +
e^i\frac{\partial {\bf \hat{n}}^{(1)} }{\partial x^i}
-\frac{1}{2}\tilde{\nu}e^ie^j\frac{\partial h_{ij}}{\partial
\eta}\frac{\partial {\bf \hat{n}}^{(0)} }{\partial
\tilde{\nu}}\right]\frac{d\eta}{ds} = {\bf \hat{C}}\left[ {\bf
\hat{n}}^{(1)} \right].
\nonumber
\end{eqnarray}

Introducing new notations $q(\eta) = \sigma_T a(\eta) N_e(\eta)$
and $f(\tilde{\nu}) = {\partial {\ln{n}_{0}} }/{\partial
\ln\tilde{\nu}}$ \cite{Basko1980} (the astrophysical meaning and
numerical values of the functions $q(\eta)$ and $f(\tilde{\nu})$
are discussed in Appendix {\ref{app:Astrophysics}}) we write down
the final form of the transfer equation:
\begin{eqnarray}
&&\left[ \frac{\partial }{\partial \eta} + q(\eta)+
e^i\frac{\partial}{\partial x^i} \right]
{\bf \hat{n}}^{(1)}(\eta,x^i,\tilde{\nu},e^i) = \nonumber \\
&& = ~
\frac{f(\tilde{\nu})n_0(\tilde{\nu})}{2}e^ie^j\frac{\partial
h_{ij}}{\partial \eta}{\bf \hat{u}} + q(\eta) \frac{1}{4\pi}\int
d\Omega' ~ {\bf \hat{P}}(e^i ; {e'^j}){\bf
\hat{n}}^{(1)}(\eta,x^i,\tilde{\nu},e'^j).\nonumber \\
\label{firstorderradiativetransfer}
\end{eqnarray}

It is seen from Eq.\ (\ref{firstorderradiativetransfer}) that the
`source' for the generation of ${\bf \hat{n}}^{(1)}$ consists of
two terms on the r.h.s. of this equation. First, it is the
gravitational field perturbation $h_{ij}$, participating in the
combination $e^ie^j{\partial h_{ij}}/{\partial \eta}$. It directly
generates a structure proportional to ${\bf \hat{u}}$, i.e. a
variation in the $I$ Stokes parameter and a temperature
anisotropy. In this process, a quadrupole component of the
temperature anisotropy necessarily arises due to the presence of
the term $e^i {\partial}/{\partial x^i}$, even if the
above-mentioned combination itself does not have angular
dependence. The second term on the r.h.s. of Eq.\
(\ref{firstorderradiativetransfer}) generates polarization, i.e. a
structure different from ${\bf \hat{u}}$. This happens because of
the mixing of different components of ${\bf \hat{n}}^{(1)}$,
including those proportional to ${\bf \hat{u}}$, in the product
term ${\bf \hat{P}}{\bf \hat{n}}^{(1)}$. In other words,
polarization is generated by the scattering of anisotropic
radiation field. Clearly, polarization is generated only in the
intervals of time when $q(\eta) \neq 0$, i.e. when free electrons
are available for the Thompson scattering (see, for example,
\cite{Naselsky2003}).

\subsection{The radiative transfer equations for a single
gravitational wave \label{sec:fourier}}

We work with a random gravitational field $h_{ij}$ expanded over
spatial Fourier components (\ref{fourierh}). It is convenient to
make similar expansion for the quantities ${\bf \hat{n}}^{(1)}
(\eta,x^i,\tilde{\nu},e^i)$. Since Eq.\
(\ref{firstorderradiativetransfer}) is linear, the Fourier
components of ${\bf \hat{n}}^{(1)}$ inherit the same random
coefficients $\stackrel{s}{c}_{\bf n}$ that enter Eq.\
(\ref{fourierh}):
\begin{eqnarray}
{\bf \hat{n}}^{(1)}(\eta,x^i,\tilde{\nu},e^i) = \frac{
\mathcal{C}}{(2\pi)^{3/2}}\int\limits_{-\infty}^{+\infty}~\frac{d^{3}{\bf
n}}{\sqrt{2n}}\sum_{s=1,2}\left[{\bf {\hat{n}}}^{(1)}_{{\bf
n},s}(\eta,\tilde{\nu},e^i)e^{i{\bf n \cdot
x}}\stackrel{s}{c}_{\bf n}+ {{\bf {\hat{n}}}^{(1)*}_{{\bf
n},s}}(\eta,\tilde{\nu},e^i)
e^{-i{\bf n \cdot x}}\stackrel{s}{c}_{\bf n}^*\right].\nonumber \\
\label{Fouriern}
\end{eqnarray}
Equation (\ref{firstorderradiativetransfer}) for a particular
Fourier component takes the form:
\begin{eqnarray}
&&\left[ \frac{\partial }{\partial \eta} + q(\eta) + ie^in_i
\right]
{\bf {\hat{n}}}^{(1)}_{{\bf n},s}(\eta,\tilde{\nu},e^i) = \nonumber \\
&& \qquad\qquad =
\frac{f(\tilde{\nu})n_0(\tilde{\nu})}{2}e^ie^j\stackrel{s}{p}_{ij}({\bf
n})\frac{d\stackrel{s}{h}_{n}(\eta)}{d\eta}{\bf \hat{u}} +
\frac{q(\eta)}{4\pi}\int d\Omega' ~ {\bf \hat{P}}(e^i ;
{e'^j}){\bf {\hat{n}}}^{(1)}_{{\bf n},s}(\eta,\tilde{\nu},e'^j).
\label{fourierfirstorderradiativetransfer}
\end{eqnarray}

To simplify technical details, we start with a single
gravitational wave propagating exactly in the direction of $z$ in
terms of definitions (\ref{ptensors}), (\ref{lmn}),
(\ref{pctensors}). The coordinate system, associated with the
wave, is specified by $\theta =0$, $\phi =0$. This simplifies the
polarization tensors (\ref{pctensors}) and makes them constant
matrices. At the same time, the observation direction is arbitrary
and is defined by $e^i =(\sin\theta \cos\phi,~ \sin\theta
\sin\phi,~\cos\theta)$. We consider circularly polarized states
with $s=1=L$, $s=2=R$. Then, we find
\begin{eqnarray}
e^ie^j\stackrel{s}{p}_{ij}({\bf n}) = (1-\mu^{2})e^{\pm2i\phi},
\label{eep}
\end{eqnarray}
where $\mu = \cos{\theta}$ and the $\pm$ signs correspond to $s=L,
R$, respectively. This simplification of the angular dependence is
possible only for one Fourier component, but not for all of them
together. We shall still need the results for a wave propagating
in an arbitrary direction. The necessary generalization will be
done in Sec.\ \ref{arbgw}.

The $\pm2\phi$ angular dependence of the source term in Eq.\
(\ref{fourierfirstorderradiativetransfer}) generates the
$\pm2\phi$ angular dependence in the solution \cite{Basko1980,
Polnarev1985}. We show in Appendix \ref{app:poln} that the terms
in ${\bf {\hat{n}}^{(1)}}_{{\bf n},s}(\eta,\tilde{\nu}, \mu,
\phi)$ with any other $\phi$-dependence satisfy homogeneous
differential equations, and therefore they vanish if they were not
present initially (which we always assume). Similarly, the
$\tilde{\nu}$ dependence of the solution can be factored out.
Finally, we show in Appendix \ref{app:poln} that one linear
combination of the three components of ${\bf
{\hat{n}}}^{(1)}_{{\bf n}}$ always satisfies a homogeneous
equation and therefore vanishes at zero initial data. Thus, the
problem of solving Eq.\ (\ref{fourierfirstorderradiativetransfer})
reduces to the problem of finding two functions of the arguments
$\eta, \mu$.

Explicitly, we can now write
\begin{eqnarray}
{\bf {\hat{n}}}^{(1)}_{n,s}(\eta,\tilde{\nu},\mu,\phi) =
\frac{f(\tilde{\nu}) n_0(\tilde{\nu})}{2}\left
[\alpha_{n,s}(\eta,\mu)(1-\mu^{2})\left(
\begin{array}{c}1 \\ 1 \\0
\end{array} \right)
+ \beta_{n,s}(\eta,\mu) \left( \begin{array}{c}(1+\mu^{2}) \\
-(1+\mu^{2}) \\ \mp4i\mu
\end{array} \right)\right ]e^{\pm2i\phi}. \nonumber \\
\label{nintermsofalphabeta}
\end{eqnarray}
Clearly, function $\alpha$ is responsible for temperature
anisotropy ($I$ Stokes parameter), while function $\beta$ is
responsible for polarization ($Q$ and $U$ Stokes parameters).

Temporarily dropping out the labels $n, s$ and introducing the
auxiliary function $\xi(\eta,\mu) = \alpha(\eta,\mu)+
\beta(\eta,\mu)$, we get from Eqs.\
(\ref{fourierfirstorderradiativetransfer}),(\ref{nintermsofalphabeta})
a pair of coupled equations \cite{Polnarev1985}:
\begin{eqnarray}
\frac{{\partial\beta}(\eta,\mu)}{\partial\eta}+\left(q(\eta)+in\mu\right)
\beta(\eta,\mu)=\frac{3}{16}q(\eta)\mathcal{I}(\eta),
\label{eqbeta}
\end{eqnarray}
\begin{eqnarray}
\frac{{\partial\xi}(\eta,\mu)}{\partial\eta}+\left(q(\eta)+in\mu\right)
\xi(\eta,\mu)=\frac{d h(\eta)}{d \eta},
\label{eqxi}
\end{eqnarray}
where
\begin{eqnarray}
\mathcal{I}(\eta)=\int\limits^{1}_{-1}d\mu{'}\left
[(1+\mu'^{2})^{2}\beta(\eta,\mu{'})-\frac{1}{2}(1-\mu'^{2})^{2}
\xi(\eta,\mu{'})\right].
\label{I}
\end{eqnarray}


\section{Radiative transfer equations as a single
integral equation \label{sec:integralequations}}

In some previous studies
\cite{Crittenden1993,Zaldarriaga1997,Seljak1997}, equations
(\ref{eqbeta}), (\ref{eqxi}) are being solved by first expanding
the $\mu$-dependence of $\beta$ and $\xi$ in terms of Legendre polynomials.
This generates an infinite series of coupled ordinary differential
equations. Then, the series is being truncated at some order.

We go by a different road. We demonstrate that the problem can be
reduced to a single mathematically consistent integral equation.
There are technical and interpretational advantages in this
approach. The integral equation enables us to derive physically
transparent analytical solutions and make reliable estimates of
the generated polarization. The numerical implementation of the
integral equation considerably saves time and allows
simple control of accuracy.

\subsection{Derivation of the integral equation}

In order to show that the solutions of Eqs.\ (\ref{eqbeta}),
(\ref{eqxi}) for $\alpha(\eta, \mu)$ and $\beta(\eta, \mu)$ are
completely determined by a single integral equation, we first
introduce new quantities \cite{Frewin1994}
\begin{eqnarray}
\Phi(\eta)=\frac{3}{16}g(\eta)\mathcal{I}(\eta),
\label{Phi}
\end{eqnarray}
\begin{eqnarray}
H(\eta) = e^{-\tau(\eta)}\frac{dh(\eta) }{d\eta}.
\label{H}
\end{eqnarray}
Solutions to Eqs.\ (\ref{eqbeta}), (\ref{eqxi}) can be written as
\begin{eqnarray}
\beta(\eta,\mu)= e^{\tau(\eta)-in\mu\eta}\int\limits^{\eta}_{0}
d\eta'~\Phi(\eta')e^{in\mu \eta'},
\label{formalsolbeta} \\
\xi(\eta,\mu)=e^{\tau(\eta)-in\mu \eta}\int\limits^{\eta}_{0}
d\eta'~H(\eta')e^{in\mu \eta'}
\label{formalsolxi} ,
\end{eqnarray}
Expression (\ref{formalsolbeta}) is a formal solution to Eq.\
(\ref{eqbeta}) in the sense that $\beta(\eta, \mu)$ is expressed
in terms of $\Phi(\eta)$ which itself depends on $\beta(\eta,
\mu)$ (see (\ref{I}) and (\ref{Phi})).

We now put (\ref{formalsolbeta}) and (\ref{formalsolxi}) into Eq.\
(\ref{I}) to get a new formulation for $\mathcal{I}(\eta)$:
\begin{eqnarray}
\mathcal{I}(\eta) =
e^{\tau(\eta)}\int\limits_{-1}^{1}\int\limits_{0}^{\eta}d\mu
d\eta'\left[ \left(1+\mu^{2}\right)^{2}\Phi(\eta') -
\frac{1}{2}\left(1-\mu^{2}\right)^{2}H(\eta')
\right]e^{in\mu(\eta'-\eta)}.
\label{integralequation-prilem}
\end{eqnarray}
Using the kernels $K_{\pm}(\eta-\eta')$,
\begin{eqnarray}
K_{\pm}(\eta-\eta')=\int\limits^{1}_{-1}d\mu
(1\pm\mu^{2})^{2}e^{in\mu(\eta-\eta')},
\label{kernal}
\end{eqnarray}
Eq.\ (\ref{integralequation-prilem}) can be rewritten as
\begin{eqnarray}
\mathcal{I}(\eta) = e^{\tau(\eta)}\int\limits_{0}^{\eta}
d\eta'\left[ K_+(\eta-\eta')\Phi(\eta') -
\frac{1}{2}K_-(\eta-\eta')H(\eta') \right].
\label{integralequation-prilem2}
\end{eqnarray}
Multiplying both sides of this equality by
$(3/16)q(\eta)e^{-\tau(\eta)}$ and recalling the definition
(\ref{Phi}) we arrive at a closed form equation for $\Phi(\eta)$:
\begin{eqnarray}
\Phi(\eta)= \frac{3}{16}q(\eta)\int \limits^{\eta}_{0}
d\eta'\Phi(\eta')K_{+}(\eta-\eta') + F(\eta),
\label{integralequation}
\end{eqnarray}
where $F(\eta)$ is the known gravitational-field term given by the
metric perturbations,
\begin{eqnarray}
F(\eta)=-\frac{3}{32}q(\eta)\int\limits^{\eta}_{0}d\eta'
H(\eta')K_{-}(\eta-\eta'),
\label{phi_0}
\end{eqnarray}

The derived equation (\ref{integralequation}) for $\Phi(\eta)$ is
the integral equation of Voltairre type. As soon as $\Phi(\eta)$
is found from this equation, we can find $\beta(\eta, \mu)$ from
Eq.\ (\ref{formalsolbeta}). Then, Eqs.\ (\ref{formalsolxi}) and
(\ref{formalsolbeta}) completely determine all the components of
${\bf \hat{n}}^{(1)}$ according to Eq.\
(\ref{nintermsofalphabeta}).

Clearly, we are mainly interested in temperature and polarization
anisotropies seen at the present time $\eta = \eta_R$. Introducing
$\zeta = n(\eta_R - \eta)$ and restoring the indices $n$ and $s$,
we obtain the present-day values of $\alpha$ and $\beta$:
\begin{subequations}
\label{alphabetapresent}
\begin{eqnarray}
&&\alpha_{n,s}(\mu) \equiv \alpha_{n,s}(\eta_R,\mu)=
\int\limits^{\eta_R}_{0}d\eta\left(H_{n,s}(\eta) - \frac{}{}
\Phi_{n,s}(\eta) \right)e^{-i\mu\zeta}, \label{alphapresent} \\
&&\beta_{n,s}(\mu) \equiv\beta_{n,s}(\eta_R,\mu) =
\int\limits^{\eta_R}_{0}d\eta~\Phi_{n,s}(\eta)e^{-i\mu \zeta}.
\label{betapresent}
\end{eqnarray}
\end{subequations}
The integrals can safely be taken from $\eta = 0$ as the optical
depth $\tau$ quickly becomes very large in the early Universe, and
the `source' functions $H_{n,s}(\eta)$ and $\Phi_{n,s}(\eta)$
quickly vanish there. We will work with expressions
(\ref{alphabetapresent}) in our further calculations.

\subsection{Analytical solution to the integral equation
\label{sec:analyticalsolutioninteq}}

The integral equation (\ref{integralequation}) can be solved
analytically in the form of a series expansion. Although our
graphs and physical conclusions in this paper are based on the
exact numerical solution to Eq.\ (\ref{integralequation}), it is
important to have a simple analytical approximation to the exact
numerical solution. We will show below why the infinite series can
be accurately approximated by its first term and how this
simplification helps in physical understanding of the derived
numerical results.

We start with the transformation of kernels (\ref{kernal}) of the
integral equation (\ref{integralequation}). Using the identity
$\mu^ke^{ix\mu} = (d/idx)^k e^{ix\mu}$, the kernels can be written
as
\begin{eqnarray}
K_{\pm}(\eta - \eta')= \int\limits_{-1}^{+1}d\mu
~(1\pm\mu^{2})^{2} e^{in\mu(\eta - \eta')}~ = 2\left(1 \mp
\frac{d^{2}}{dx^{2}}\right)^{2}\frac{\sin{x}}{x},
\nonumber
\end{eqnarray}
where $x=n(\eta - \eta')$. Now, taking into account the expansion
\begin{eqnarray}
\frac{\sin{x}}{x} =
\sum_{m=0}^{\infty}\frac{(-1)^m}{(2m+1)!}x^{2m},
\nonumber
\end{eqnarray}
the r.h.s. of Eq.\ (\ref{integralequation}) can be presented in
the form of a series:
\begin{eqnarray}
\Phi(\eta) =
\frac{3}{2}q(\eta)\sum_{m=0}^{\infty}n^{2m}\int\limits_{0}^{\eta}d\eta'~\left(\eta
- \eta'\right)^{2m}\left[ \frac{}{}\lambda_+(m)\Phi(\eta') -
\lambda_-(m)H(\eta')\right],
\label{seriesexpansionkernel}
\end{eqnarray}
where
\begin{eqnarray}
\lambda_+(m) = \frac{(-1)^m}{(2m+1)!}\left[1 -
4\frac{(m+2)}{(2m+3)(2m+5)}\right],~\lambda_-(m) =
\frac{(-1)^m}{(2m+1)!(2m+3)(2m+5)}.
\nonumber
\end{eqnarray}

Since the r.h.s. of Eq.\ (\ref{seriesexpansionkernel}) is a series
in even powers of the wavenumber $n$, the l.h.s. of the same
equation can also be expanded in powers of $n^{2m}$:
\begin{eqnarray}
\Phi(\eta) = \sum_{m=0}^{\infty}\Phi^{(m)}(\eta)n^{2m}.
\label{assymtoticexpansion}
\end{eqnarray}
Using expansion (\ref{assymtoticexpansion}) in both sides of Eq.\
(\ref{seriesexpansionkernel}) we transform this equation to
\begin{eqnarray}
\sum_{m=0}^{\infty}\Phi^{(m)}(\eta) n^{2m} =&&
\frac{3}{2}q(\eta)\left[\sum_{m=0}^{\infty}\sum_{j=0}^{\infty}
{\lambda_+(m)} n^{2(m+j)}\int\limits_{0}^{\eta}d\eta'
\Phi^{(j)}(\eta')(\eta-\eta')^{2m} \right.\nonumber\\
&&~~~\left. -\sum_{m=0}^{\infty}\lambda_-(m)n^{2m}
\int\limits_{0}^{\eta}d\eta'H(\eta')(\eta-\eta')^{2m}\right].
\label{47}
\end{eqnarray}
The left side and the second term in the right side of Eq.\
(\ref{47}) are series in $n^{2m}$, but the first sum on the r.h.s.
of Eq.\ (\ref{47}) is still a mixture of different powers. This
sum can be rearranged to be manifestly a series in $n^{2m}$:
\begin{eqnarray}
\sum_{m=0}^{\infty}\sum_{j=0}^{\infty}{\lambda_{+}(m)}n^{2(m+j)}
\int\limits_{0}^{\eta}d\eta'\Phi^{(j)}(\eta')(\eta-\eta')^{2m} =
\qquad\qquad\qquad\nonumber \\
\qquad\qquad\qquad = \sum_{m=0}^{\infty}n^{2m}
\left[\sum_{k=0}^{m}{\lambda_{+}(k)}\int\limits_{0}^{\eta}d\eta'
\Phi^{(m-k)}(\eta')(\eta-\eta')^{2k}\right].
\label{48}
\end{eqnarray}

According to Eq.\ (\ref{47}) we have to make equal the
coefficients of terms with the same power $n^{2m}$ in both sides
of the equation. This produces a set of integral equations
\begin{eqnarray}
\Phi^{(m)}(\eta) = q(\eta)S^{(m)}(\eta) +
\frac{7}{10}q(\eta)\int\limits_{0}^{\eta}d\eta'~\Phi^{(m)}(\eta'),
\label{50}
\end{eqnarray}
where functions $S^{(m)}(\eta)$ depend only on the known function
$H(\eta)$ and functions $\Phi^{(m-k)}(\eta)$ presumed to be found
from equations of previous orders:
\begin{eqnarray}
S^{(m)}(\eta) &=&
-\frac{3}{2}{\lambda_-(m)}\int\limits_{0}^{\eta}d\eta'~H(\eta')
(\eta-\eta')^{2m}
\nonumber\\
&& +
\frac{3}{2}\sum_{k=1}^{m}{\lambda_+(k)}\int\limits_{0}^{\eta}d\eta'~
\Phi^{(m-k)}(\eta')(\eta-\eta')^{2k}.
\label{51}
\end{eqnarray}
The important advantage of the performed expansion in powers of
$n^{2m}$ is that Eq.\ (\ref{50}) of any order $m$ is now a
self-contained analytically solvable integral equation.

Exact solution to the integral equation (\ref{50}) is given by the
formula
\begin{eqnarray}
\Phi^{(m)}(\eta) = q(\eta)\int\limits_{0}^{\eta}d\eta'
\frac{d{S}^{(m)}(\eta')}{d\eta'}e^{\frac{7}{10}\tau(\eta,\eta')}.
\label{result1}
\end{eqnarray}
We can further simplify this formula. Taking an $\eta$-derivative
of expression (\ref{51}) we find
\begin{eqnarray}
\frac{d{S}^{(m)}(\eta)}{d\eta} = -\frac{1}{10}H(\eta)\delta_{0m} -
\omega_-(m)\int\limits_{0}^{\eta}d\eta'H(\eta')(\eta-\eta')^{2m-1}\nonumber
\\ ~~+
\sum_{k=1}^{m}\omega_+(k)
\int\limits_{0}^{\eta}d\eta'\Phi^{(m-k)}(\eta')(\eta-\eta')^{2k-1},
\nonumber
\end{eqnarray}
where $ \omega_{\pm}(m) = {3m}\lambda_{\pm}(k)$. Substituting this
expression into Eq.\ (\ref{result1}), we arrive at the final
result
\begin{eqnarray}
\Phi^{(m)}(\eta) =
q(\eta)e^{-\frac{7}{10}\tau(\eta)}\int\limits_{0}^{\eta}d\eta'
e^{\frac{7}{10}\tau(\eta')}\left[-H(\eta')\left(
\frac{1}{10}\delta_{0m}+\omega_-(m)\Psi_{(m)}(\eta,\eta')\right)
\right.\nonumber
\\ \left. ~~ +
\sum_{k=1}^{m}\omega_+(k)\Phi^{(m-k)}(\eta')\Psi_{(k)}(\eta,\eta')
\right],
\label{assymtoticsolution}
\end{eqnarray}
where
\begin{eqnarray}
\Psi_{(k)}(\eta,\eta') = \int\limits_{\eta'}^{\eta}d\eta''~
e^{-\frac{7}{10}\tau(\eta'',\eta')}(\eta''- \eta')^{2k-1}.
\label{Psi_n}
\end{eqnarray}

Functions $\Psi_{(k)}(\eta, \eta')$ depend only on the ionization
history of the background cosmological model described by
$q(\eta)$. These functions can be computed in advance. Complete
determination of the function $\Phi(\eta)$, Eq.\
(\ref{assymtoticexpansion}), requires only one integration by
$\eta$ at each level $m$ in Eq.\ (\ref{assymtoticsolution}),
starting from $m=0$. The zero-order term $\Phi^{(0)}(\eta)$ does
not depend on functions $\Psi_{(k)}$ and is determined exclusively
by $H(\eta)$, Eq.\ (\ref{H}). The zero-order term can be presented
as
\begin{eqnarray}
\Phi_n^{(0)}(\eta) = -\frac{1}{10}g(\eta)\int\limits^{\eta}_0
d\eta' \frac{dh_n(\eta')}{d\eta}e^{-\frac{3}{10}\tau(\eta,\eta')}
\label{zerothorderapproximation1}.
\end{eqnarray}

It is crucial to remember that the function $\Phi(\eta)$, Eq.\
(\ref{Phi}), always contains the narrow visibility function
$g(\eta)$ (see Appendix \ref{app:Astrophysics}). In particular,
function $\Phi^{(0)}(\eta)$ is nonzero only for $\eta$ within the
width of $g(\eta)$, and is proportional to this width. In the era
of decoupling, we denote the characteristic width of $g(\eta)$ by
$\Delta \eta_{dec}$. With $\Delta \eta_{dec}$ we associate the
characteristic wavenumber $n_*$:
\begin{eqnarray}
n_* = \frac{2 \pi}{\Delta \eta_{dec}}.
\nonumber
\end{eqnarray}
Numerically, $\Delta \eta_{dec} \approx 3\times 10^{-3}$ and $n_*
\approx 2 \times 10^{3}$. In what follows, we will be interested
in CMB multipoles $\ell \lesssim 10^{3}$. They are mostly
generated by perturbations with wavenumbers $n \lesssim 10^{3}$.
Therefore, for wavenumbers of interest, we regard $n/n_*$ as a
small parameter.

We shall now show that $\Phi^{(0)}(\eta)$ is the dominant term of
the series (\ref{assymtoticexpansion}). The next term,
$\Phi^{(1)}(\eta) n^{2}$, is at least a factor $(n/n_*)^{2}$
smaller than $\Phi^{(0)}(\eta)$, and so on. The explicit
expression for $\Phi^{(1)}$ is as follows
\begin{eqnarray}
\Phi^{(1)}(\eta) = -g(\eta)\int\limits_{0}^{\eta}d\eta'
\left[\omega_-(1)\frac{dh(\eta')}{d\eta'}
-\omega_+(1)e^{\tau(\eta')}\Phi^{(0)}(\eta')
\right]e^{-\frac{3}{10}\tau(\eta,\eta')}\Psi_{(1)}(\eta,\eta').
\nonumber
\end{eqnarray}
Effectively, function $\Phi^{(1)}(\eta)$, in comparison with
$\Phi^{(0)}(\eta)$, contains an extra factor
$\Psi_{(1)}(\eta,\eta')$. Taking into account the fact that the
functions $g(\eta)$ and $e^{-\frac{7}{10}\tau(\eta'',\eta')}$ are
localized in the interval of arguments not larger than $\Delta
\eta_{dec}$, this factor evaluates to a number not larger than
$(\Delta \eta_{dec})^{2}$. Therefore, the $m=1$ term in Eq.\
(\ref{assymtoticexpansion}) is at least a factor $(n/n_*)^{2}$
smaller than the $m=0$ term.

\begin{figure}
\begin{center}
\includegraphics[width=7cm]{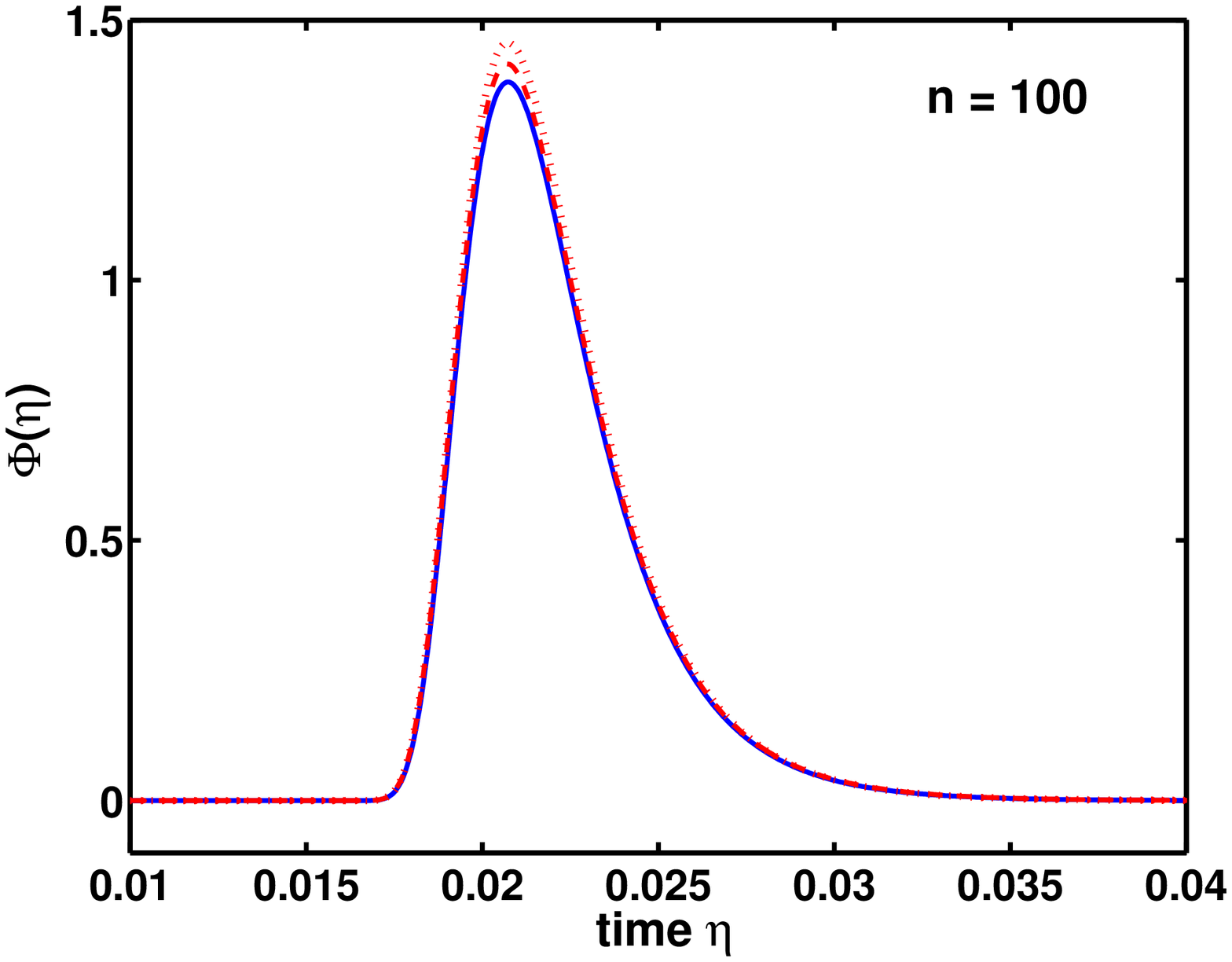}\qquad
\includegraphics[width=7cm]{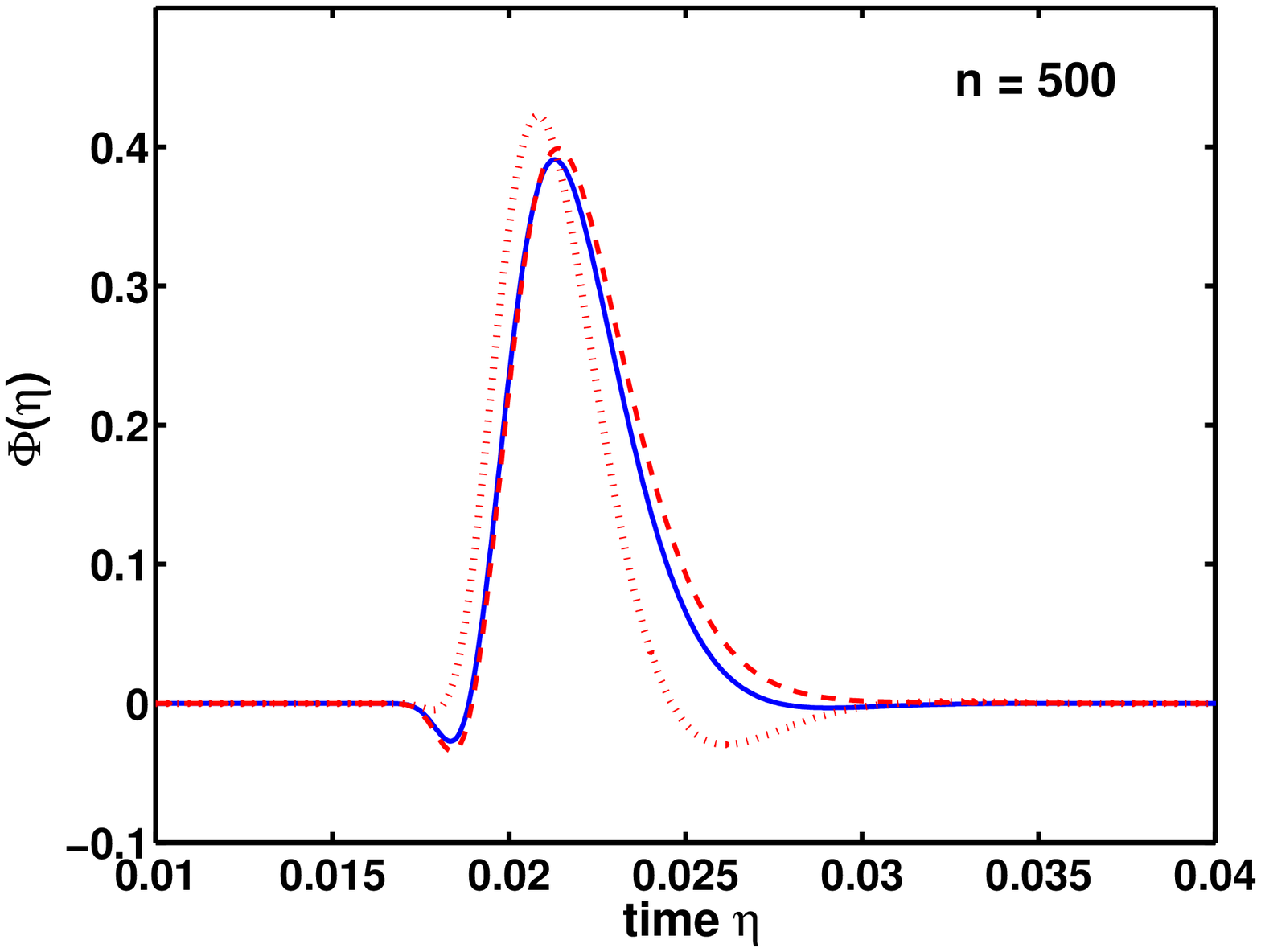}\\
\includegraphics[width=7cm]{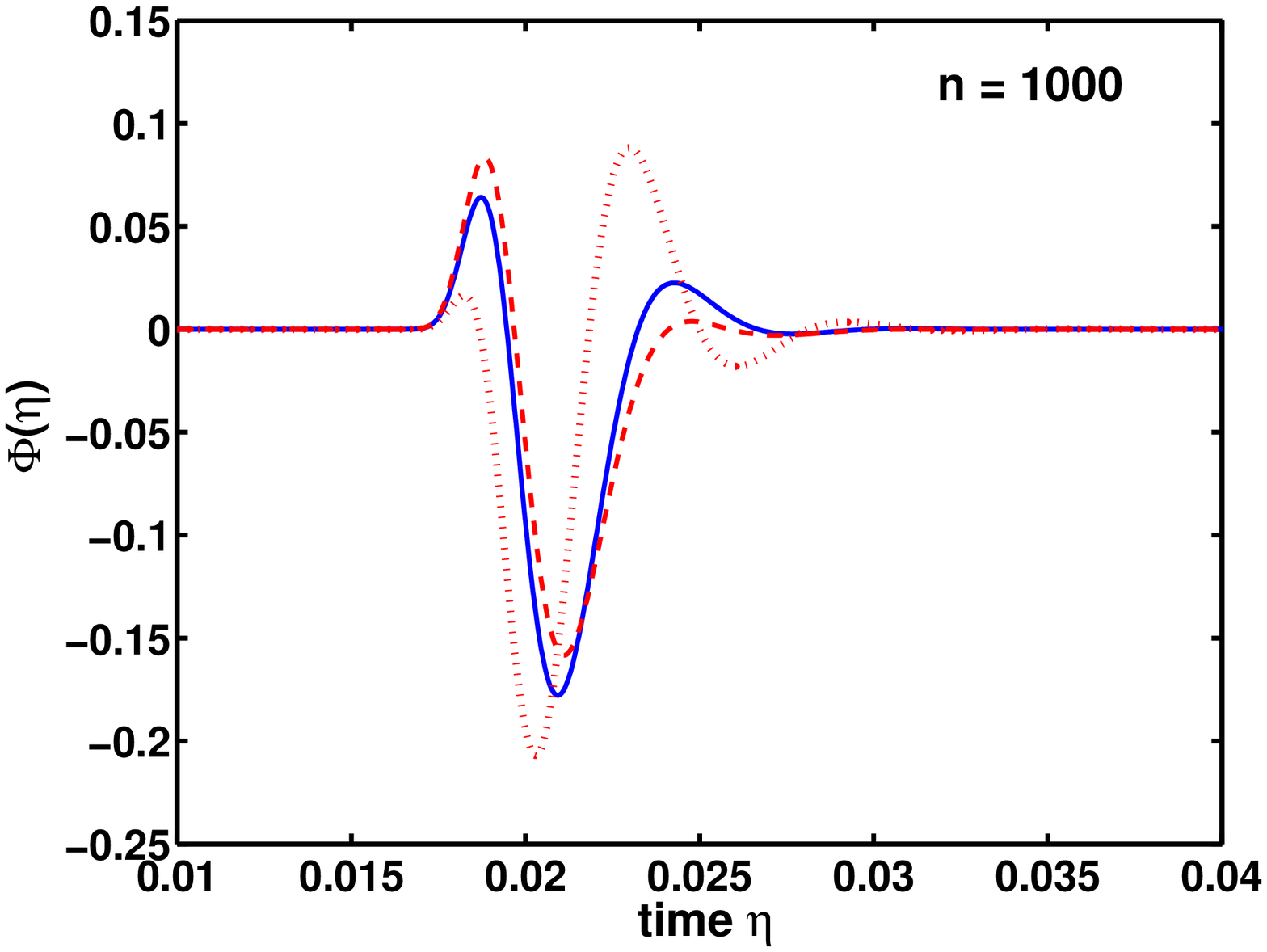}\qquad
\includegraphics[width=7cm]{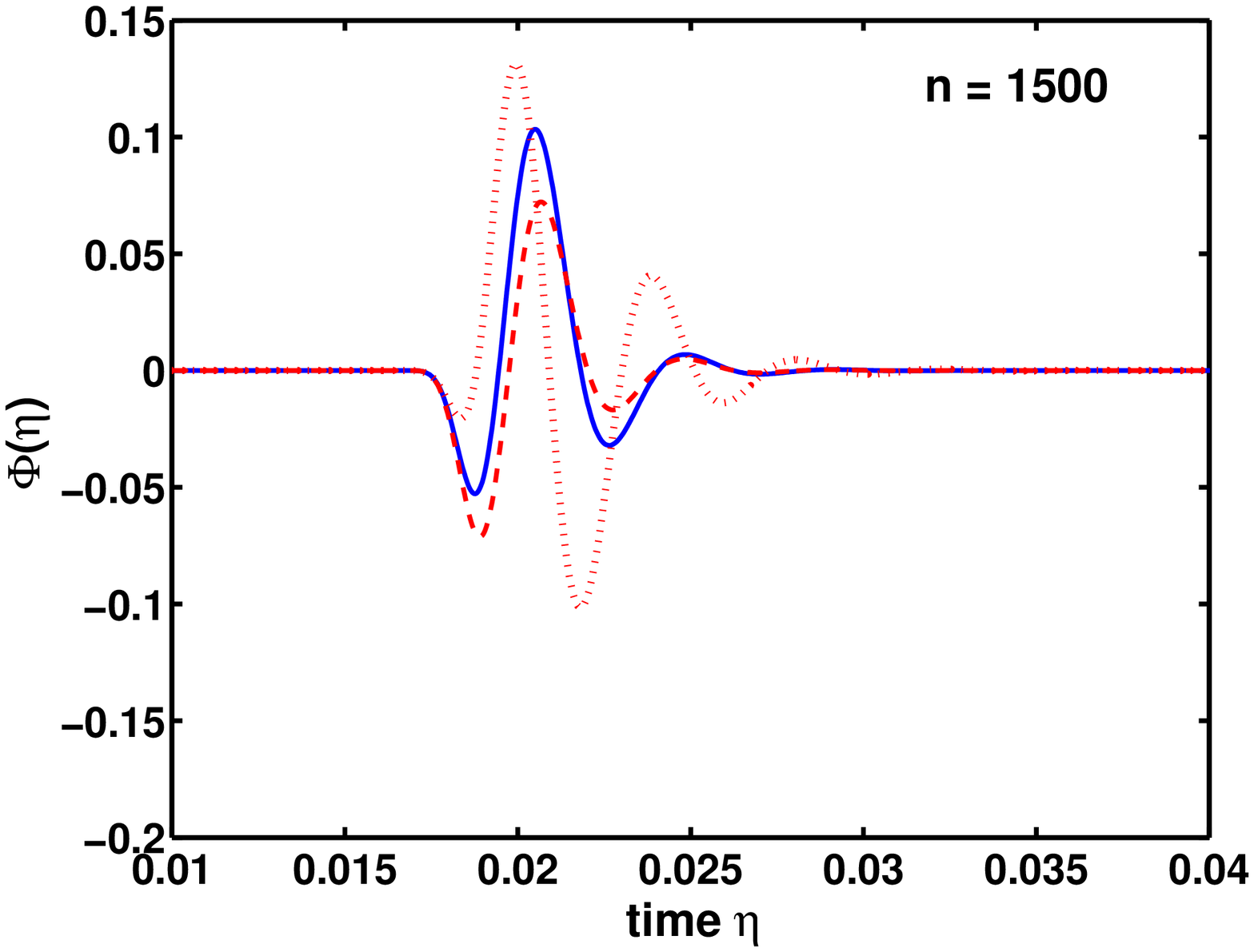}
\end{center}
\caption{Function $\Phi_n(\eta)$ for different values of $n$. The
solid line is exact numerical solution to Eq.\
(\ref{integralequation}). The dashed line is the zero order
approximation (\ref{zerothorderapproximation1}). The dotted line
shows the approximation (\ref{zerothorderapproximation2}) (see
below). The g.w.\ mode functions are normalized such that
$h_n(\eta_r)=1$.}
\label{figure_recombination2}
\end{figure}

These analytical evaluations are confirmed by numerical analysis
as shown in Fig.\ \ref{figure_recombination2}. The solid line
shows the exact numerical solution found from Eq.\
(\ref{integralequation}) for $h_n(\eta)$ and $q(\eta)$ described
in Sec.\ \ref{sec:modefunctions} and Appendix
\ref{app:Astrophysics} respectively. The dashed line is plotted
according to formula (\ref{zerothorderapproximation1}), with the
same $h_n(\eta)$ and $q(\eta)$. It is seen from Fig.\
\ref{figure_recombination2} that the zero-order term
$\Phi^{(0)}(\eta)$ is a good approximation. The deviations are
significant, they reach (20-25)\%, only for the largest
wavenumbers $n$ in the domain of our interest.

\section{Multipole Expansion and Power Spectra of the Radiation Field
\label{sectionmultipoleexpansion}}

\subsection{Multipole coefficients \label{multcoeff}}

Having found $\Phi_{n,s}(\eta)$ for a single gravitational wave
specified by Eq.\ (\ref{eep}) one can find $\alpha$ and $\beta$
functions according to Eqs.\ (\ref{alphapresent}),
(\ref{betapresent}). Then, using Eqs.\
(\ref{nintermsofalphabeta}),
(\ref{symbolicvectorstokesparameters}),
(\ref{PolarizationtensorPab}), (\ref{IandV}) and (\ref{EandB}),
one can find the multipole coefficients $a_{\ell m}^X$ ($X= I, E,
B$) participating in the decompositions
(\ref{multipolecoefficients}). Although this route has been
traversed before \cite{Zaldarriaga1997}, we have made independent
calculations in a more general arrangement. Formulas derived in
this subsection are effectively a confirmation of the correctness
of calculations in Ref.\ \cite{Zaldarriaga1997}.

First, we integrate over photon frequencies (fot the definition of
$\gamma$ see Eq.\ (\ref{gamma})) and arrive at the following
expressions
\begin{subequations}
\label{teb}
\begin{eqnarray}
I_{n,s}\left(\mu,\phi\right)&=& \gamma\left[\frac{}{}
\left(1-\mu^{2}\right)\alpha_{n,s}\left(\mu\right)e^{\pm2i\phi}\right]\label{tebI},
\\ E_{n,s}\left(\mu,\phi\right)&=& -\gamma
\left[\left(1-\mu^{2}\right)
\left(\left(1+\mu^{2}\right)\frac{d^{2}}{d\mu^{2}}+8\mu\frac{d}{d\mu}+12\right)\beta_{n,s}\left(\mu\right)e^{\pm2i\phi}\right]
\label{tebE},
\\ B_{n,s}\left(\mu,\phi\right)&=& \mp\gamma
\left[2\left(1-\mu^{2}\right)
\left(i\mu\frac{d^{2}}{d\mu^{2}}+4i\frac{d}{d\mu}\right)\beta_{n,s}\left(\mu\right)e^{\pm2i\phi}\right],
\label{tebB}
\end{eqnarray}
\end{subequations}
where, as before, the upper and lower signs correspond to $s=1=L$
and $s=2=R$, respectively. The $\pm2\phi$ dependence in Eqs.\
(\ref{teb}) implies that only the $m=\pm2$ multipoles are nonzero.

Then, we integrate Eqs.\ (\ref{teb}) over angular variables in
order to find $a_{\ell m}^X$ according to Eq.\
(\ref{multipolecoefficients}). Using the notations
$T_{\ell}(\zeta)$, $E_{\ell}(\zeta)$, $B_{\ell}(\zeta)$ for the
functions arising in course of calculations (and called multipole
projection functions)
\begin{subequations}
\label{TEB}
\begin{eqnarray}
T_{{\ell}}(\zeta)&=&
\sqrt{\frac{({\ell}+2)!}{({\ell}-2)!}}~\frac{j_{\ell}(\zeta)}{\zeta^{2}},\label{TEBT}\\
E_{{\ell}}(\zeta)&=&
\left[\left(2-\frac{l(l-1)}{\zeta^{2}}\right)j_{\ell}(\zeta)-\frac{2}{\zeta}j_{{\ell}-1}
(\zeta)\right],\label{TEBE}\\
B_{{\ell}}(\zeta)&=&
2\left[-\frac{({\ell}-1)}{\zeta}j_{\ell}(\zeta)+j_{{\ell}-1}(\zeta)\right],\label{TEBB}
\end{eqnarray}
\end{subequations}
and replacing $\alpha(\mu)$ and $\beta(\mu)$ by their expressions
(\ref{alphabetapresent}), we finally arrive at
\begin{subequations}
\label{a_lm}
\begin{eqnarray}
a^T_{{\ell} m}(n,s)&=& (-i)^{\ell-2}
\left(\delta_{2,m}\delta_{1,s}\frac{}{}
+\delta_{-2,m}\delta_{2,s}\right)a^T_{\ell}(n,s),\label{a_lmT}\\
a^E_{{\ell} m}(n,s)&=& (-i)^{\ell-2}
\left(\delta_{2,m}\delta_{1,s}\frac{}{}
+\delta_{-2,m}\delta_{2,s}\right)a^E_{\ell}(n,s) ,
\label{a_lmE}\\
a^B_{{\ell} m}(n,s)&=&(-i)^{\ell-2}
\left(\delta_{2,m}\delta_{1,s}\frac{}{}
-\delta_{-2,m}\delta_{2,s}\right)a^B_{\ell}(n,s)\label{a_lmB},
\end{eqnarray}
\end{subequations}
where
\begin{subequations}
\label{a_l}
\begin{eqnarray}
a^T_{{\ell}}(n,s)&=&{\gamma}\sqrt{4\pi(2{\ell}+1)}
\int\limits_{0}^{\eta_R}d\eta~ \left(H_{n,s}(\eta)\frac{}{}-
\Phi_{n,s}\left(\eta\right)\right) T_{{\ell}}(\zeta),
\label{a_lT}\\
a^E_{{\ell}}(n,s)&=&{\gamma}\sqrt{4\pi(2{\ell}+1)}
\int\limits_{0}^{\eta_R}d\eta~\Phi_{n,s}(\eta) E_{{\ell}}(\zeta) ,
\label{a_lE} \\
a^B_{{\ell}}(n,s)&=&{\gamma}\sqrt{4\pi(2{\ell}+1)}
\int\limits_{0}^{\eta_R}d\eta~\Phi_{n,s}(\eta) B_{{\ell}}(\zeta).
\label{a_lB}
\end{eqnarray}
\end{subequations}

\subsection{Superposition of gravitational waves with arbitrary wavevectors
\label{arbgw}}

It is important to remember that the result (\ref{a_lm}) is valid
only for a special wave, with the wavevector ${\bf n}$ oriented
exactly along the coordinate axis $z$. Since the perturbed
gravitational field is a random collection of waves with all
possible wavevectors ${\bf n}$, and we are interested in their
summarized effect as seen in some fixed observational direction
$\theta, \phi$, we have to find the generalization of Eq.\
(\ref{a_lm}) to an arbitrary wave, and then to sum them up.

To find the effect of an arbitrary wave, there is no need to do
new calculations. It is convenient to treat calculations in Sec.\
\ref{multcoeff} as done in a (primed) coordinate system specially
adjusted to a given wave in such a manner that the wave propagates
along $z'$, ${\bf n}'= (0, 0, n)$. The observational direction
$e^i$ is characterized by $\theta', \phi'$. The quantities $X=I,
E, B$ calculated in Sec.\ \ref{multcoeff} are functions of
$\theta', \phi'$ expanded over $Y_{\ell m}(\theta', \phi')$,
\begin{eqnarray}
X_{{\bf n}',s}(\theta,\phi') =
\sum_{{\ell}=0}^{\infty}\sum_{m=-\ell}^{\ell}\textsf{a}^X_{{\ell}m}(n,s)
Y_{\ell m}(\theta',\phi'), \label{Xold}
\end{eqnarray}
where $\textsf{a}^X_{{\ell}m}$ is a set of coefficients (compare
with Eq.\ (\ref{multipolecoefficients})): $a^T_{\ell m}$,
$\left[(\ell+2)!/(\ell-2)!\right]^{1/2} a^E_{{\ell}m}$,
$\left[(\ell+2)!/(\ell-2)!\right]^{1/2} a^B_{{\ell}m}$.

Now, imagine that this special (primed) coordinate system is
rotated with respect to the observer's (unprimed) coordinate
system by some Euler angles
\begin{eqnarray}
\alpha = \phi_{\bf n}, ~~~ \beta = \theta_{\bf n}, ~~~ \gamma =0
\nonumber
\end{eqnarray}
(see, for example \cite{Baskaran2004}). The same observational
direction $e^i$ is now characterized by $\theta, \phi$, and the
same wavevector ${\bf n}'$ is now characterized by the unit vector
\begin{eqnarray}
{\bf \tilde{n}} = {\bf n}/n = (\sin {\theta_{\bf n}} \cos
{\phi_{\bf n}}, \sin {\theta_{\bf n}} \sin {\phi_{\bf n}}, \cos
{\theta_{\bf n}}).
\nonumber
\end{eqnarray}
Obviously, the already calculated numerical values of the
invariants $X(\theta', \phi')$ do not depend on the rotation of
the coordinate system. Being expressed in terms of $\theta, \phi$,
the invariants describe the effect produced by a wave with a given
(arbitrary) unit wavevector ${\bf \tilde{n}}$, as seen in the
direction $\theta, \phi$.

The transformation between coordinate systems $(\theta', \phi')$
and $(\theta, \phi)$ is accompanied by the transformation of
spherical harmonics,
\begin{eqnarray}
Y_{\ell m}(\theta',\phi') =
\sum_{m'=-\ell}^{\ell}D^{\ell}_{m',m}\left({\bf \tilde{n}}\right)
Y_{\ell m'}(\theta,\phi),
\nonumber
\end{eqnarray}
where
\begin{eqnarray}
D^{\ell}_{m',m}\left({\bf \tilde{n}}\right) \equiv
D^{\ell}_{m',m}\left(\phi_{\bf n}, \theta_{\bf n}, 0 \right)
\nonumber
\end{eqnarray}
are the Wigner symbols \cite{Varshalovich1988}. Later, we will
need their orthogonality relationship
\begin{eqnarray}
\int d{\Omega}D^{\ell}_{m,p}({\bf
\tilde{n}})D^{\ell'*}_{m',p}({\bf \tilde{n}}) = \frac{4\pi}{2\ell
+ 1} \delta_{\ell\ell'}\delta_{mm'},
\label{B4a}
\end{eqnarray}
where $d \Omega = \sin \theta_{\bf n} d{\theta_{\bf n}}
d{\phi_{\bf n}}$.

We can now rewrite Eq.\ (\ref{Xold}) in terms of $\theta, \phi$,
and thus find the contribution of a single arbitrary Fourier
component,
\begin{eqnarray}
X_{{\bf n},s}(\theta,\phi) =
\sum_{{\ell}=0}^{\infty}\sum_{m=-\ell}^{\ell}\left(\sum_{m'=-\ell}^{\ell}
\textsf{a}^X_{{\ell} m'}(n,s)
D^{\ell}_{mm'}({\bf\tilde{n}})\right) Y_{\ell m}(\theta,\phi).
\label{individualXinobscoord}
\end{eqnarray}
The superposition of all Fourier components of the perturbed
gravitational field gives, at the observer's position ${\bf x} =
0$ and at $\eta=\eta_R$ (see (\ref{Fouriern})), the final result:
\begin{eqnarray}
X(\theta,\phi) = \frac{\mathcal{
C}}{(2\pi)^{3/2}}\int\limits_{-\infty}^{+\infty}~\frac{d^{3}{\bf
n}}{\sqrt{2n}}\sum_{s=1,2}\left[X_{{\bf
n},s}(\theta,\phi)\stackrel{s}{c}_{\bf n}+ X_{{\bf
n},s}^*(\theta,\phi)\stackrel{s}{c}_{\bf n}^*\right].
\label{Xforallfouriermodes}
\end{eqnarray}

 From this expression, combined with Eq.\
(\ref{individualXinobscoord}), one can read off the random
multipole coefficients $a^X_{\ell m}$ that participate in the
expansions (\ref{multipolecoefficients}):
\begin{eqnarray}
a^{X}_{\ell m} = \frac{\mathcal{C}}{(2\pi)^{3/2}} \int
\limits_{-\infty}^{+\infty} ~\frac{d^{3}{\bf n}}{\sqrt{2n}}
\sum_{s=1,2}\sum_{m'=-\ell}^{\ell}
\left[a^X_{{\ell}m'}(n,s)D^{\ell}_{m,m'}\left({\bf
\tilde{n}}\right)\stackrel{s}{c}_{\bf n} +
(-1)^{m}a^{X*}_{{\ell}-m'}(n,s)D^{\ell*}_{m,-m'}\left({\bf
\tilde{n}} \right)\stackrel{s}{c}_{\bf n}^*\right]. \nonumber \\
\label{multipoleXcompl}
\end{eqnarray}
Since $X$ is a real field, the multipole coefficients $a^{X}_{\ell
m}$ obey the reality conditions
\begin{eqnarray}
a^{X*}_{\ell m} = (-1)^ma^{X}_{\ell, -m},
\label{realcond}
\end{eqnarray}
as is seen directly from (\ref{multipoleXcompl}). The
gravitational-wave nature of metric perturbations is encoded in
concrete values of the coefficients (\ref{a_l}),
(\ref{multipolecoefficients}). But in all other aspects the
argumentation presented here is general.

\subsection{Angular power spectra for temperature and polarization
anisotropies
\label{powsp}}

It follows from Eq.\ (\ref{multipoleXcompl}) that the statistical
properties of the multipole coefficients $a^{X}_{\ell m}$ are
fully determined by the statistical properties of the
gravitational field perturbations represented by the random
coefficients $\stackrel{s}{c}_{\bf n}$. A particular realization
of $\stackrel{s}{c}_{\bf n}$ is responsible for the particular
realization of $a^{X}_{\ell m}$ actually observed in the sky.
Having derived the distribution function for $\stackrel{s}{c}_{\bf
n}$ from some fundamental considerations (for example, from the
assumption of the initial quantum-mechanical vacuum state of
perturbations) we could estimate the probability of the observed
set $a^{X}_{\ell m}$ within the ensemble of all possible sets. We
could also evaluate the inevitable uncertainty in the
observational determination of the parameters of the underlying
random process. This uncertainty is associated with the inherent
absence of ergodicity of any random process on a 2-sphere (i.e.
sky) \cite{Grishchuk1997}. In this paper, however, we adopt a
minimalistic approach; we postulate only the relationships
(\ref{statCs}) and calculate only the quadratic correlation
functions for $a^{X}_{\ell m}$.

Clearly, the mean values of the multipole coefficients are zeros,
\begin{eqnarray}
\langle a^X_{\ell m} \rangle = \langle {a^{X}_{\ell m}}^*\rangle =
0.
\nonumber
\end{eqnarray}
To calculate the variances and cross-correlation functions, we
have to form the products $a^{X*}_{\ell m}a^{X'}_{\ell' m'}$ and
then take their statistical averages. First, we find
\begin{eqnarray}
\langle a^{X*}_{\ell m}a^{X'}_{\ell' m'} \rangle = &&
\frac{\mathcal{C}^{2}}{(2\pi)^{3}}\int\frac{n^{2}dnd{\Omega}}{2n}
\sum_{s=1,2}\sum_{m_1=-\ell}^{\ell}\sum_{m_1'=-\ell}^{\ell}
\nonumber \\ &&
\left[a^{X*}_{{\ell}m_1}(n,s)a^{X'}_{{\ell}'
m_1'}(n,s)D^{\ell*}_{m m_1}({\bf \tilde{n}})D^{\ell'}_{m'
m_1'}({\bf \tilde{n}})\right. \nonumber \\ && \left.+
a^{X}_{{\ell} m_1}(n,s)a^{X'*}_{{\ell}'
m_1'}(n,s)D^{\ell}_{-m,m_1}({\bf
\tilde{n}})D^{\ell'*}_{-m',m_1'}({\bf \tilde{n}})\right].
\label{prodX}
\end{eqnarray}

We now take into account the fact (compare with Eq.\ (\ref{a_lm}))
that
\begin{eqnarray}
a^{X}_{{\ell} m_1}(n,s)a^{X'}_{{\ell}' m_1'}(n,s) \propto
\delta_{m_1 m_1'}.
\nonumber
\end{eqnarray}
This property allows us to get rid of summation over $m_1'$ in
Eq.\ (\ref{prodX}). Then, we perform integration over $d{\Omega}$
and use the orthogonality relationships (\ref{B4a}). We finally
arrive at
\begin{eqnarray}
\langle a^{X*}_{\ell m}a^{X'}_{\ell' m'} \rangle =
C_{\ell}^{XX'}\delta_{\ell\ell'}\delta_{mm'},
\label{B5}
\end{eqnarray}
where
\begin{eqnarray}
C_{\ell}^{XX'} = \frac{\mathcal{C}^{2}}{4\pi^{2}(2\ell+1)}\int
~ndn \sum_{s=1,2}\sum_{m=-\ell}^{\ell} \left( a^{X}_{\ell
m}(n,s)a^{X'*}_{\ell m}(n,s) + a^{X*}_{\ell m}(n,s)a^{X'}_{\ell
m}(n,s) \right).
\label{Clxx'}
\end{eqnarray}
Other quadratic averages, such as $\langle a^{X}_{\ell
m}a^{X'}_{\ell' m'} \rangle$, $\langle a^{X*}_{\ell m}
a^{X'*}_{\ell' m'} \rangle$, follow from (\ref{B5}) and the
reality condition (\ref{realcond}).

The angular correlation and cross-correlation functions of the
fields $I, E, B$ are directly expressible in terms of Eq.\
(\ref{Clxx'}). For example,
\begin{eqnarray}
\langle
I\left(\theta_1,\phi_1\right)I\left(\theta_2,\phi_2\right)\rangle
= \Gamma(\delta) =
\sum_{\ell=0}^{\infty}\frac{2{\ell}+1}{4\pi}C^{TT}_{\ell}P_{\ell}(\cos{\delta}),
\nonumber
\end{eqnarray}
where $\delta$ is the angular separation between the directions
$(\theta_1,\phi_1)$ and $(\theta_2,\phi_2)$ on the sky.
If the actually measured values of $a^{X}_{\ell m}$
represent a particular realization of a gaussian random process,
quantities $C_{\ell}^{XX'}$, constructed from the measured
$a^{X}_{\ell m}$ according to the r.h.s. of Eq.\ (\ref{Clxx'}),
are the best unbiased estimates \cite{Grishchuk1997}.

One can note that the final result (\ref{B5}), (\ref{Clxx'}) is
the integral of individual contributions (\ref{a_lm}) from single
gravitational waves given in a special frame discussed in Sec.\
\ref{multcoeff}. However, one cannot jump directly from
(\ref{a_lm}) to (\ref{Clxx'}) (which is often done in the
literature). In general, Eq.\ (\ref{Clxx'}) does not follow from
Eq.\ (\ref{a_lm}). By calculations given in this subsection we
have rigorously shown that Eq.\ (\ref{Clxx'}) is justified only if
special statistical assumptions (\ref{statCs}) are adopted.

One can also note that the correlation functions containing the
label $B$ once, i.e. $C^{TB}_{\ell}$ and $C^{EB}_{\ell}$, vanish
if the extra assumptions (\ref{eqval}), (\ref{eqval2}) are made.
Indeed, under these assumptions one can use one and the same mode
function for both polarization states $s$, Eq.\ (\ref{eqval3}).
Then, the coefficients $a^B_{{\ell} m}(n,s)$, Eq.\ (\ref{a_lmB}),
differ essentially only in sign for two different $s$, i.e.
$a^B_{{\ell} 2}(n, L) = -a^B_{{\ell},-2}(n, R)$. Therefore, their
contributions will cancel out in expressions (\ref{Clxx'}) for
$C^{TB}_{\ell}$ and $C^{EB}_{\ell}$. This statement is in
agreement with Ref.\ \cite{Kamionkowski1999}.

Without having access to a `theory of everything' which could
predict one unique distribution of the CMB radiation field over
the sky, we have to rely on the calculated statistical averages
(\ref{Clxx'}). We can also hope that our universe is a `typical'
one, so that the observed values of the correlation functions
should not deviate too much from the statistical mean values.


\section{Effects of Recombination Era \label{sec:recombination}}

All our final graphs and physical conclusions in this paper are
based on exact formulas and numerical calculations, starting from
numerical representation of the key functions $H(\eta)$ and
$\Phi(\eta)$, Eqs.\ (\ref{H}), (\ref{integralequation}). However,
we derive and explain all our results by developing manageable and
accurate analytical approximations. At every level of calculations
we compare exact numerical results with analytical ones.

\subsection{Temperature anisotropy angular power spectrum}

\begin{figure}
\begin{center}
\includegraphics[width=6cm]{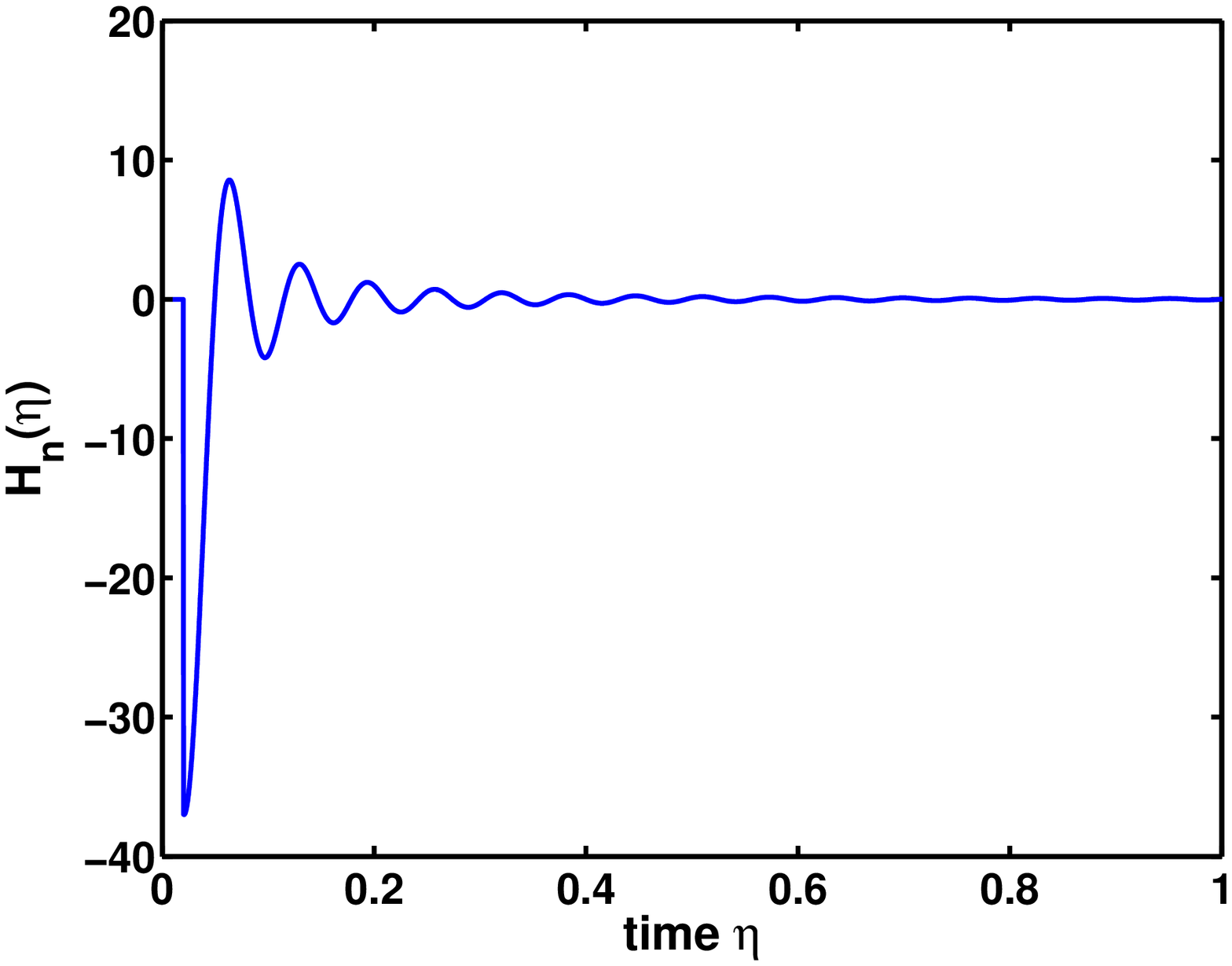}\qquad
\includegraphics[width=6cm]{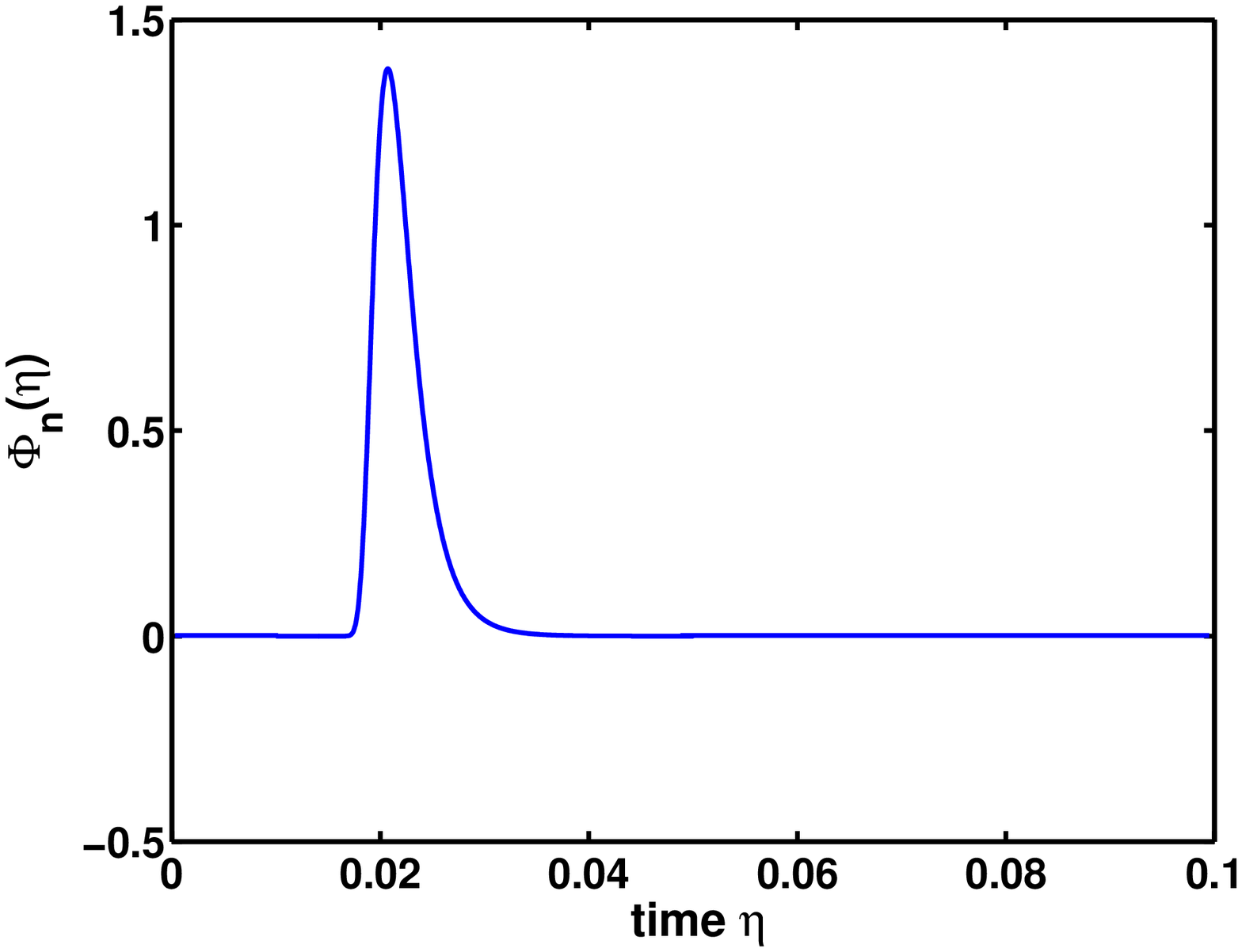}
\end{center}
\caption{The `source' functions $H_{n,s}(\eta)$ and
$\Phi_{n,s}(\eta)$ ($n=100$) of temperature and polarization
anisotropies (the normalization is chosen such that $h_n(\eta_r) =
1$). }
\label{figure_recombination3}
\end{figure}

The temperature anisotropy power spectrum $C_{\ell}^{TT}$ is
determined by the multipole coefficients $a_{\ell}^T(n,s)$, Eqs.\
(\ref{Clxx'}), (\ref{a_lmT}), (\ref{a_lT}). The typical graphs for
the functions $H_{n,s}(\eta)$ and $\Phi_{n,s}(\eta)$ are shown in
Fig.\ \ref{figure_recombination3}.

Since the visibility function $g(\eta)$ is a narrow function, a
convenient analytical approximation is the limit of an
instanteneous recombination. The function $e^{-\tau}$ is replaced
by a step function changing from 0 to 1 at $\eta =\eta_{dec}$,
$e^{-\tau}=h(\eta-\eta_{dec})$, and the function $g(\eta)$ is
replaced by a delta-function, $g(\eta)=\delta(\eta-\eta_{dec})$.
In this limit, the contribution to $a_{\ell}^T(n,s)$ from the
scattering term $\Phi_{n,s}^{(0)}(\eta)$ is proportional to
$\Delta\eta_{dec}$. It can be neglected in comparison with the
contribution from the gravitational term $H_{n,s}(\eta)$. The
ratio of these contributions is of the order of
$n\Delta\eta_{rec}$, and it tends to zero in the limit of
instanteneous recombination, $\Delta\eta_{dec}\rightarrow 0$.

\begin{figure}
\begin{center}
\includegraphics[width=6cm]{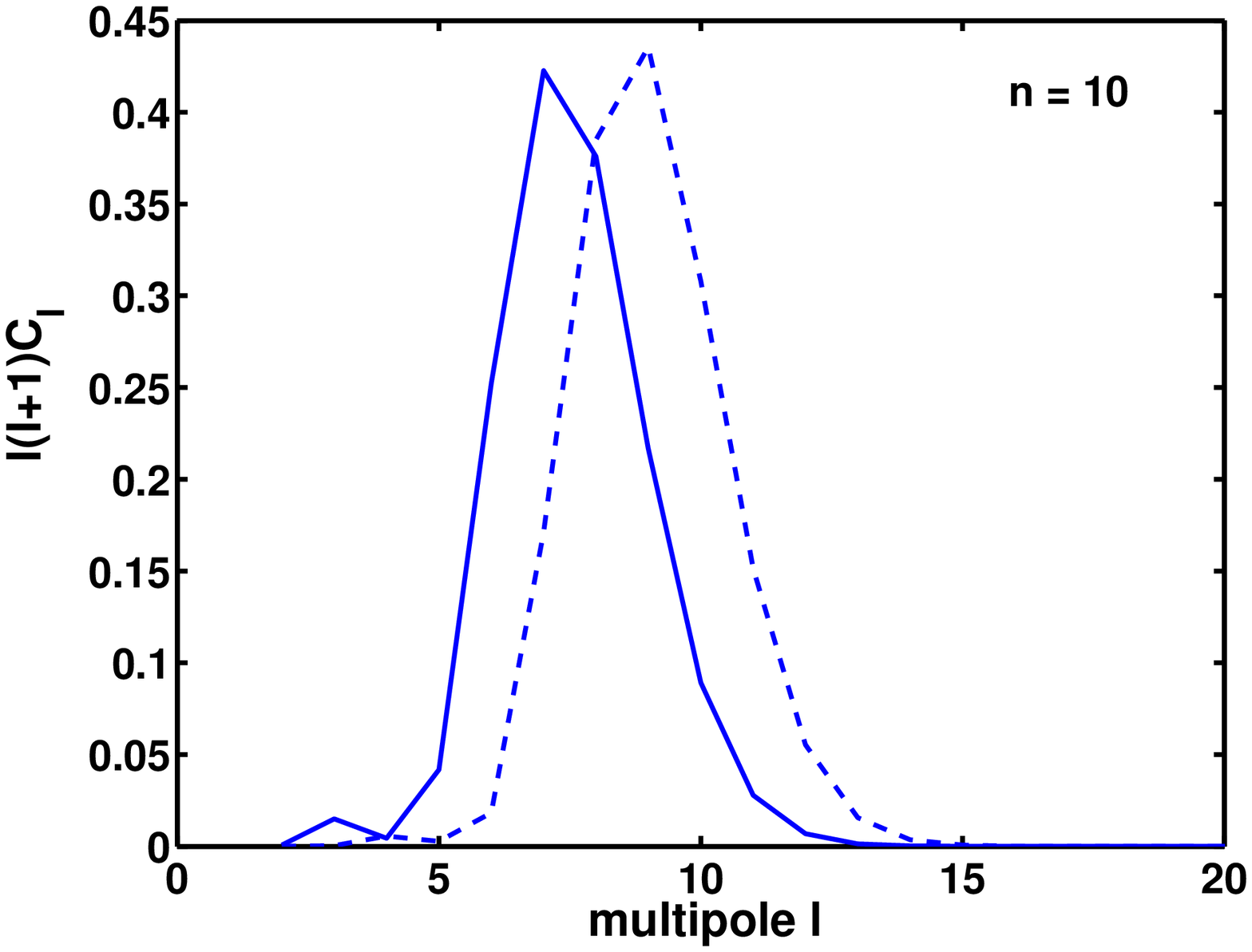}\qquad
\includegraphics[width=6cm]{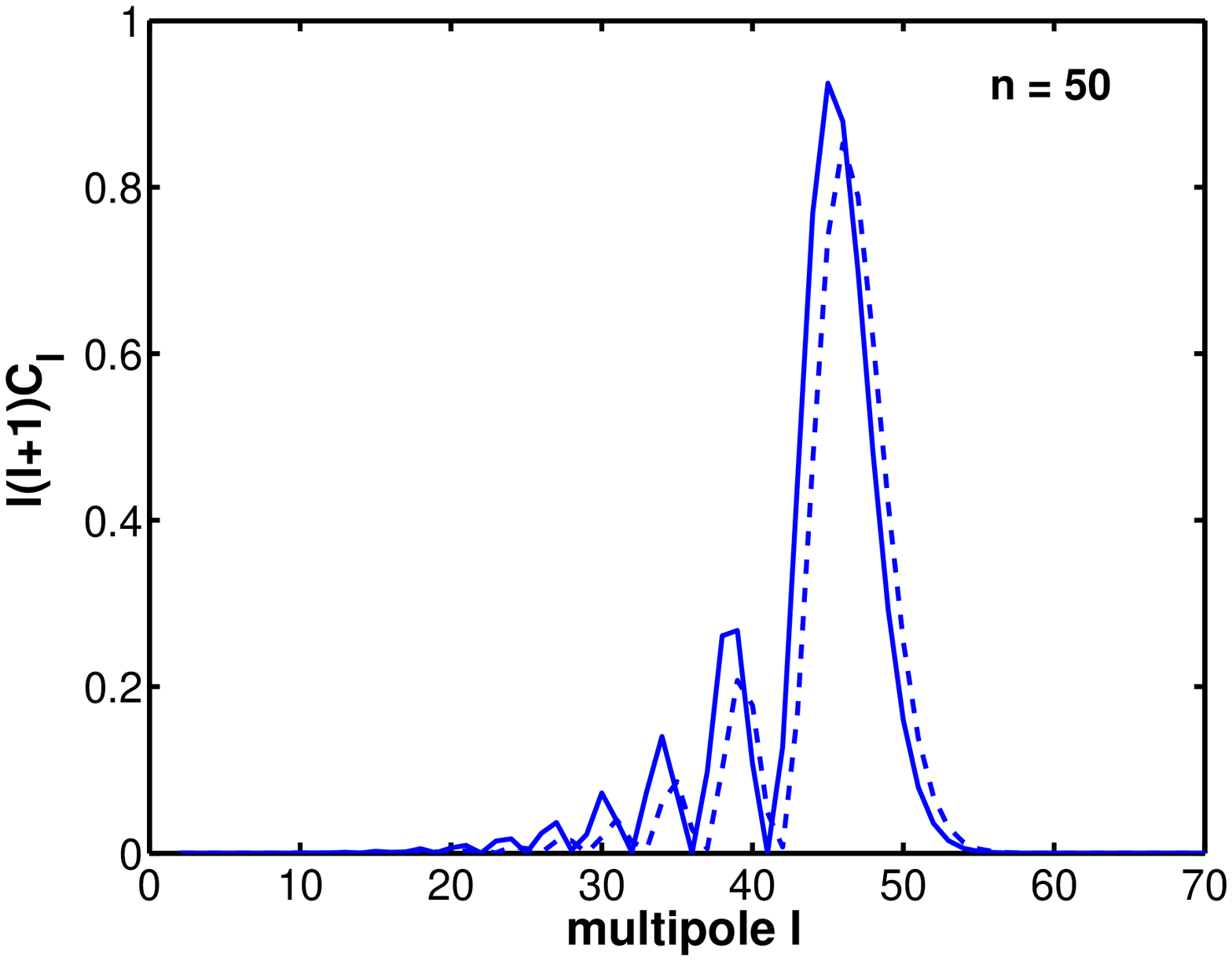}
\\
\includegraphics[width=6cm]{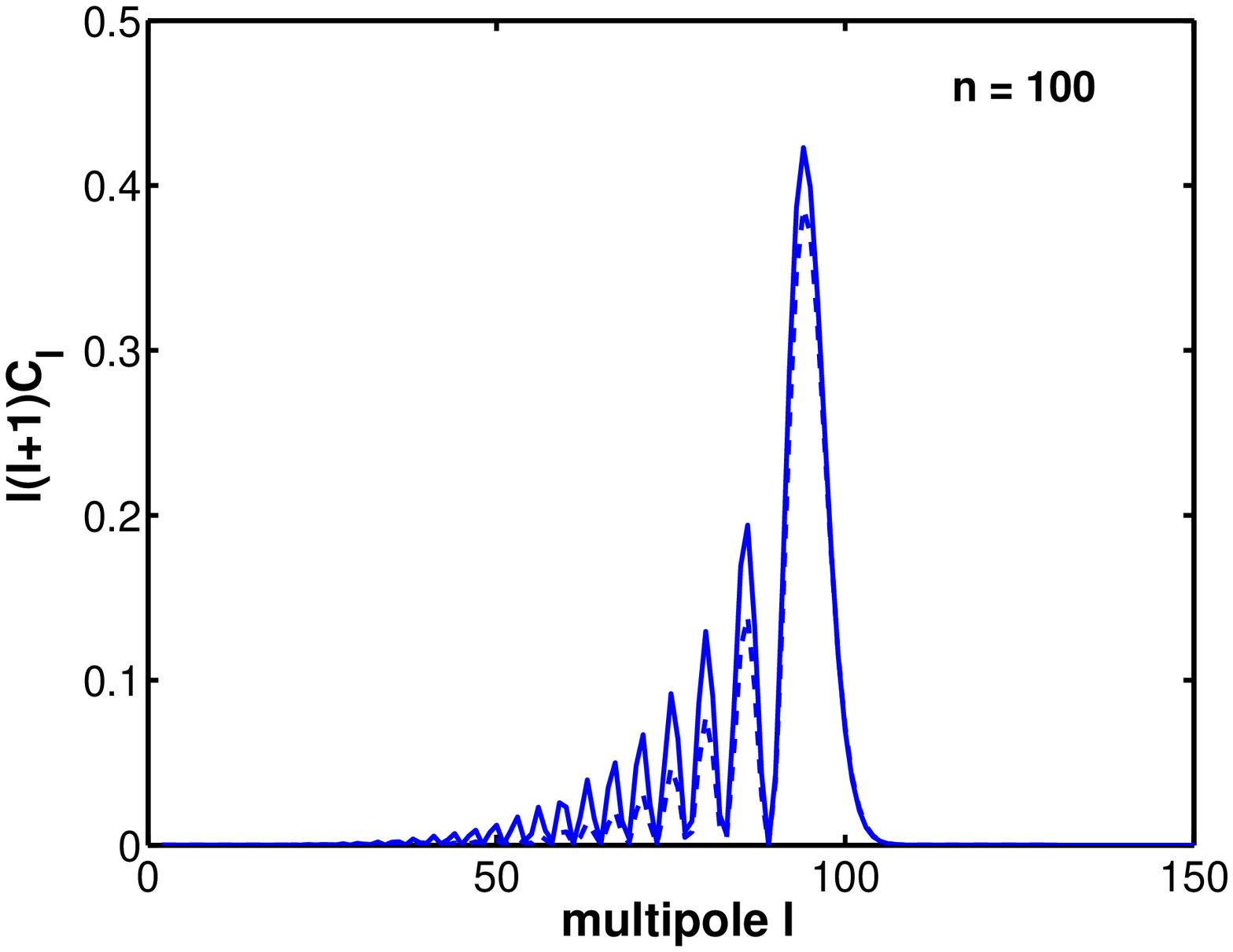}\qquad
\includegraphics[width=6cm]{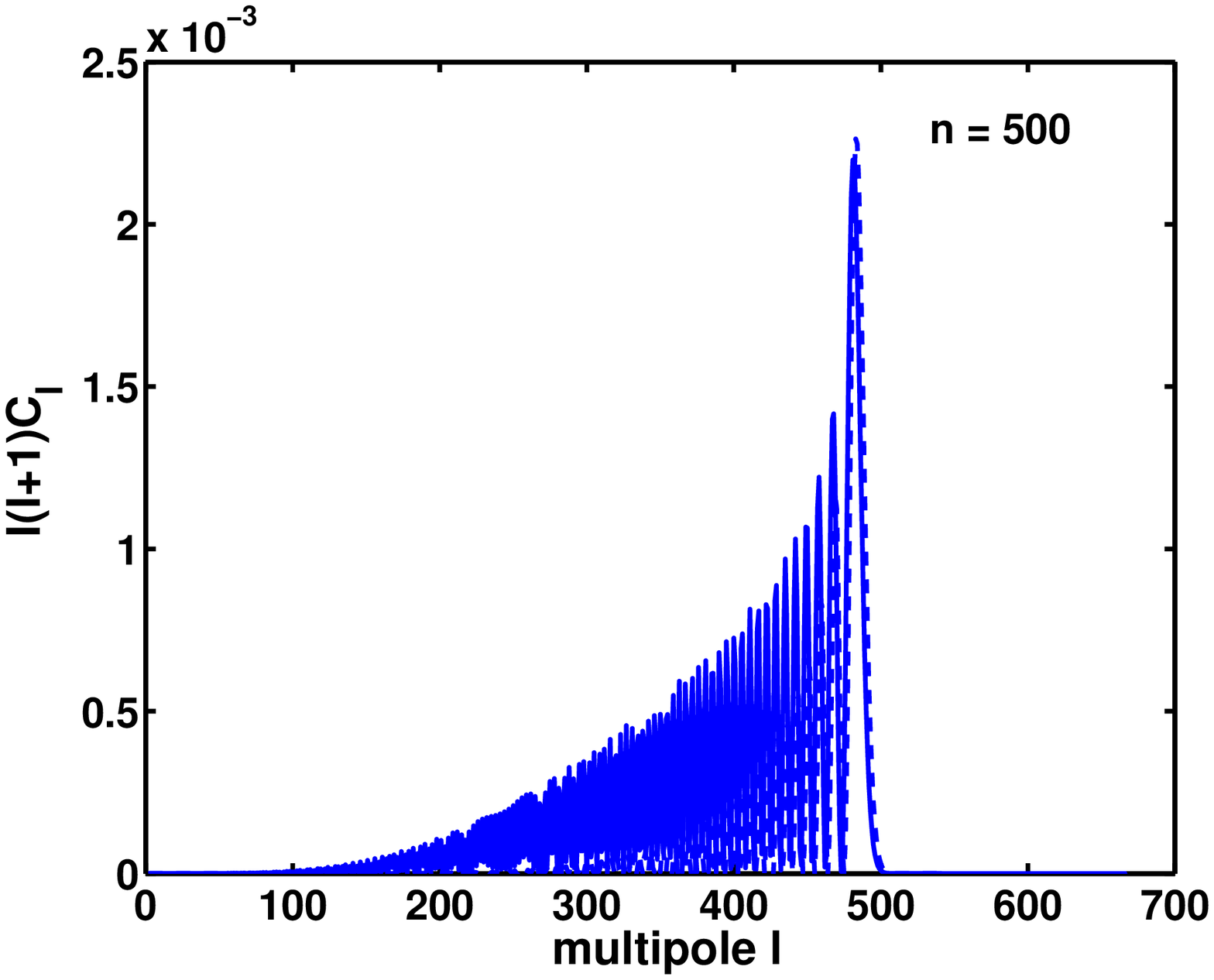}
\end{center}
\caption{The contributions to the power spectrum
$\ell(\ell+1)C_{\ell}^{TT}$ from an individual mode $n$. The solid
line shows the exact result calculated according to
(\ref{a_Tapprox1}), while the dashed line shows the approximation
(\ref{a_Tapprox}). The normalization has been chosen such that
$h_n(\eta_r)=1$.} \label{fig:c_ellsTgw50}
\end{figure}

Neglecting the scattering term, we write
\begin{eqnarray}
a^T_{{\ell}}(n,s) = {\gamma}\sqrt{4\pi(2{\ell}+1)}
\int\limits_{\eta_{dec}}^{\eta_R}d\eta~
\frac{d\stackrel{s}{h}_n}{d\eta} T_{{\ell}}(\zeta).
\label{a_Tapprox1}
\end{eqnarray}
This integral can be taken by parts,
\begin{eqnarray}
a^T_{{\ell}}(n,s) = {\gamma}\sqrt{4\pi(2{\ell}+1)}\left[ -
\stackrel{s}{h}_n(\eta_{dec})T_{\ell}(\zeta_{dec}) +
\int\limits_{\eta_{dec}}^{\eta_R}d\eta ~ n \stackrel{s}{h}_n(\eta)
\frac{dT_{\ell}(\zeta)}{d\zeta} \right].
\nonumber
\end{eqnarray}
The remaining integral contains oscillating functions and its
value is smaller, for sufficiently large $n$'s, than the value of
the integrated term. This is illustrated in Fig.\
\ref{fig:c_ellsTgw50}. Therefore, we have
\begin{eqnarray}
a^T_{{\ell}}(n,s) = - {\gamma}\sqrt{4\pi(2{\ell}+1)}
\stackrel{s}{h}_n(\eta_{dec})T_{\ell}(\zeta_{dec}).
\label{a_Tapprox}
\end{eqnarray}

Finally, we put Eq.\ (\ref{a_Tapprox}) into Eq.\ (\ref{Clxx'}) and
take into account the definition of the metric power spectrum
(\ref{gwpower}). Then, we get
\begin{eqnarray}
C_{\ell}^{TT} = 4\pi\gamma^{2}\int
\frac{dn}{n}h^{2}(n,\eta_{dec})T_{\ell}^{2}(\zeta_{dec}).
\label{C_l^T2}
\end{eqnarray}

The projection factor $T_{\ell}^{2}(\zeta_{dec})$ is given by Eq.\
(\ref{TEBT}). Since the spherical Bessel functions reach maximum
when the argument and the index are approximately equal,
$\zeta_{dec} \approx \ell$, a particular wavenumber $n$ is
predominantly projected onto the multipole $\ell \approx n$:
\begin{eqnarray}
{\ell} \approx \zeta_{dec} = n(\eta_R - \eta_{dec}) \approx n.
\label{projln}
\end{eqnarray}
This can also be seen in Fig.\ \ref{fig:c_ellsTgw50}. Thus, the
oscillatory features of the metric power spectrum
$h^{2}(n,\eta_{dec})$ in the $n$-space are fully responsible for
the oscillatory features of the angular power spectrum
$\ell(\ell+1)C_{\ell}^{TT}$ in the $\ell$-space \cite{Bose2002}.
(We use this opportunity to correct a misprint in Fig.2 of Ref.\
\cite{Bose2002} : the plotted lines are functions $C_l$, not
$l(l+1)C_l$. For the early graphs of $C_l$ see Ref.\ \cite{Allen1994}.)

The g.w.\ metric power spectrum $h^{2}(n,\eta_{dec})$ for the case
$\beta =-2$, and the function ${\ell}(\ell+1)C_{\ell}^{TT}$ caused
by this spectrum, are shown in Fig.\
\ref{figure_recombination4}(a) and Fig.\
\ref{figure_recombination4}(b). The normalization of the metric
power spectrum is such that the function
$\ell(\ell+1)C_{\ell}^{TT}$ at $\ell = 2$ is equal to 1326 $~\mu
\textrm{K} ^{2}$ \cite{Hinshaw2006}. The interval ${\ell}\lesssim
90$ is generated by waves with $n \lesssim 90$. These waves did
not enter the Hubble radius by the time $\eta_{dec}$. Their
amplitudes are approximately equal for all $n$'s in this interval
(compare with Fig.\ \ref{hfigure}). The gradual decrease of the
angular power spectrum at larger $\ell$'s is the reflection of the
gradual decrease of power in shorter gravitational waves whose
amplitudes have been adiabatically decreasing since the earlier
times when the waves entered the Hubble radius.

\subsection{Polarization anisotropy angular power spectrum
\label{sec:polarizationrecombination}}

The decisive function for polarization calculations is
$\Phi_{n,s}(\eta)$. We have approximated this function by
$\Phi_{n,s}^{(0)}$, Eq.\ (\ref{zerothorderapproximation1}), and
compared it with exact result in Fig.\
\ref{figure_recombination2}. For qualitative derivations it is
useful to make further simplifications.

Since $g(\eta)$ is a narrow function, the integral in Eq.\
(\ref{zerothorderapproximation1}) is effective only within a
narrow interval $\Delta\eta_{dec}$. Assuming that the function $d
\stackrel{s}{h}_n(\eta)/d\eta$ does not vary significantly within
this interval, we can take this function from under the integral,
\begin{eqnarray}
\Phi_{n,s}^{(0)}(\eta) \approx - \frac{1}{10}\frac{d
\stackrel{s}{h}_n(\eta)}{d \eta}
\left(g(\eta)\int\limits_0^{\eta}d\eta'~e^{-\frac{3}{10}\tau(\eta,\eta')}\right).
\label{zerothorderapproximation2}
\end{eqnarray}
Clearly, the assumption that the function $d
\stackrel{s}{h}_n(\eta)/d\eta$ is almost constant within the
window $\Delta\eta_{dec}$ gets violated for sufficiently short
waves. With (\ref{zerothorderapproximation2}), we expect
degradation of accuracy for wavenumbers $n$ approaching $n_*$.
This is illustrated by a dotted line in Fig.\
\ref{figure_recombination2}. Nevertheless, the approximation
(\ref{zerothorderapproximation2}) is robust for $n\lesssim n_*$,
and it reveals the importance of first derivatives of metric
perturbations for evaluation of the CMB polarization.

We now introduce the symbol $P$ to denote either $E$ or $B$
components of polarization. In terms of multipoles
$a^P_{\ell}(n,s)$, the angular power spectrum is given by Eq.\
(\ref{Clxx'}). Putting (\ref{zerothorderapproximation2}) into
Eqs.\ (\ref{a_lE}), (\ref{a_lB}) and denoting by
$P_{{\ell}}(\zeta)$ the respective projection functions, we get
\begin{eqnarray}
a^P_{{\ell}}(n,s)&\approx& {\gamma}\sqrt{4\pi(2{\ell}+1)}
\int\limits_{{0}}^{\eta_R}d\eta~\left( -\frac{1}{10} \frac{d
\stackrel{s}{h}_n(\eta)}{d \eta} P_{{\ell}}(\zeta)\right) \left[
g(\eta)\int\limits_0^{\eta}d\eta'
e^{-\frac{3}{10}\tau(\eta,\eta')}\right].
\label{axappr}
\end{eqnarray}
Again referring to the peaked character of $g(\eta)$ and assuming
that the combination $(d \stackrel{s}{h}_n(\eta)/d \eta)
P_{{\ell}}(\zeta)$ does not change significantly within the window
$\Delta\eta_{dec}$, we take this combination from under the
integral,
\begin{eqnarray}
a^P_{{\ell}}(n,s)&\approx& {\gamma}\sqrt{4\pi(2{\ell}+1)}~ D(n)~
\left.\left( -\frac{1}{10} \frac{d \stackrel{s}{h}_n(\eta)}{d
\eta} P_{{\ell}}(\zeta) \right)\right|_{\eta=\eta_{rec}}~\Delta.
\label{a^Xapprox}
\end{eqnarray}
The two new factors in this expression, $D(n)$ and $\Delta$,
require clarification.

The factor $D(n)$ compensates for gradual worsening of our
approximation when the wavenumber $n$ approaches $n_*$. For large
$n$'s, the functions under the integrals change sign within the
window $\Delta \eta_{dec}$, instead of being constant there. This
leads to the decrease of the true value of the integral in
comparison with the approximated one. The evaluation of this
worsening suggests that it can be described by the damping factor
\begin{eqnarray}
D(n)\equiv \left[1 +
\left(\frac{n\Delta\eta_{dec}}{2}\right)^{2}\right]^{-1}.
\nonumber
\end{eqnarray}
We inserted this factor `by hand' in Eq.\ (\ref{a^Xapprox}). Additional
arguments on this damping are given in Ref.\ \cite{Pritchard2005}
(see also \cite{Zhao2006}).

The factor $\Delta$ is the result of the remaining integration
over $\eta$ in Eq.\ (\ref{axappr}) \cite{Basko1980},
\begin{eqnarray}
\Delta \equiv \int\limits_0^{\eta_R}d\eta~
~g(\eta)\left(\int\limits_0^{\eta}d\eta'e^{-\frac{3}{10}\tau(\eta,\eta')}\right)
= \frac{10}{7} \int\limits_0^{\eta_R}d\eta~ \left[
e^{-\frac{3}{10}\tau(\eta)} - e^{-\tau(\eta)} \right].
\nonumber
\end{eqnarray}
Since $e^{-\tau}$ rapidly changes from 0 to 1 around
recombination, the integrand $\left( e^{-(3/10)\tau(\eta)}
-e^{-\tau(\eta)} \right)$ is nonzero only there, and $\Delta$ is
expected to be of the order of $\Delta \eta_{dec}$. To give a
concrete example, we approximate $g(\eta)$ by a gaussian function
\begin{eqnarray}
g(\eta) = \frac{1}{\sqrt{2\pi}(\Delta\eta_{dec}/2)}exp\left(
-\frac{(\eta-\eta_{dec})^{2}}{2(\Delta\eta_{dec}/2)^{2}} \right).
\nonumber
\end{eqnarray}
Then, the quantity $\Delta$ can be found exactly,
\begin{eqnarray}
\Delta =
\frac{5}{7}\Delta\eta_{dec}\int\limits^{+\infty}_{-\infty}dx
\left[ \left(\frac{1}{2} +
\frac{1}{2}erf\left(\frac{x}{\sqrt{2}}\right)\right)^{\frac{3}{10}}
- \left(\frac{1}{2} +
\frac{1}{2}erf\left(\frac{x}{\sqrt{2}}\right)\right)
\right]\approx 0.96 \Delta\eta_{dec}.
\nonumber
\end{eqnarray}
Clearly, factor $\Delta$ in Eq.\ (\ref{a^Xapprox}) demonstrates
the fact that the CMB polarization is generated only during a
short interval of time around recombination.

Finally, substituting (\ref{a^Xapprox}) into (\ref{Clxx'}) and
recalling (\ref{gw_h_shortened}), we obtain the polarization
angular power spectrum:
\begin{eqnarray}
C^{PP}_{\ell} &\approx& 2\pi\gamma^{2}
\frac{1}{100}\Delta^{2}\int\frac{dn}{n}D^{2}(n) \sum_{s=1,2}
\left|\left.{\frac{d\stackrel{s}{h}(n,\eta)}{d\eta}}\right|_{\eta=\eta_{rec}}\right|^{2}
P_{\ell}^{2}(\zeta_{rec}).
\label{approximateC_l3}
\end{eqnarray}

Similarly to the case of temperature anisotropies, the projection
factors $P_{\ell}(\zeta_{dec})$ predominantly translate $n$ into
$\ell$ according to Eq.\ (\ref{projln}). The oscillatory features
of the power spectrum of the first time-derivative of metric
perturbations get translated into the oscillatory features of the
power spectra for $E$ and $B$ components of polarization. This is
illustrated in Fig.\ \ref{figure_recombination4}(c) and Fig.\
\ref{figure_recombination4}(d). The waves with $n \eta_{dec} \ll
\pi$ did not enter the Hubble radius by $\eta = \eta_{dec}$. They
have no power in the spectrum of $dh(n,\eta)/d\eta$ at $n \ll 90$,
and therefore there is no power in polarization at $\ell \ll 90$.
On the other hand, the first gravitational peak at $n \approx 90$
gets reflected in the first polarization peak at $\ell \approx
90$.

\begin{figure}
\begin{center}
\includegraphics[width=6cm]{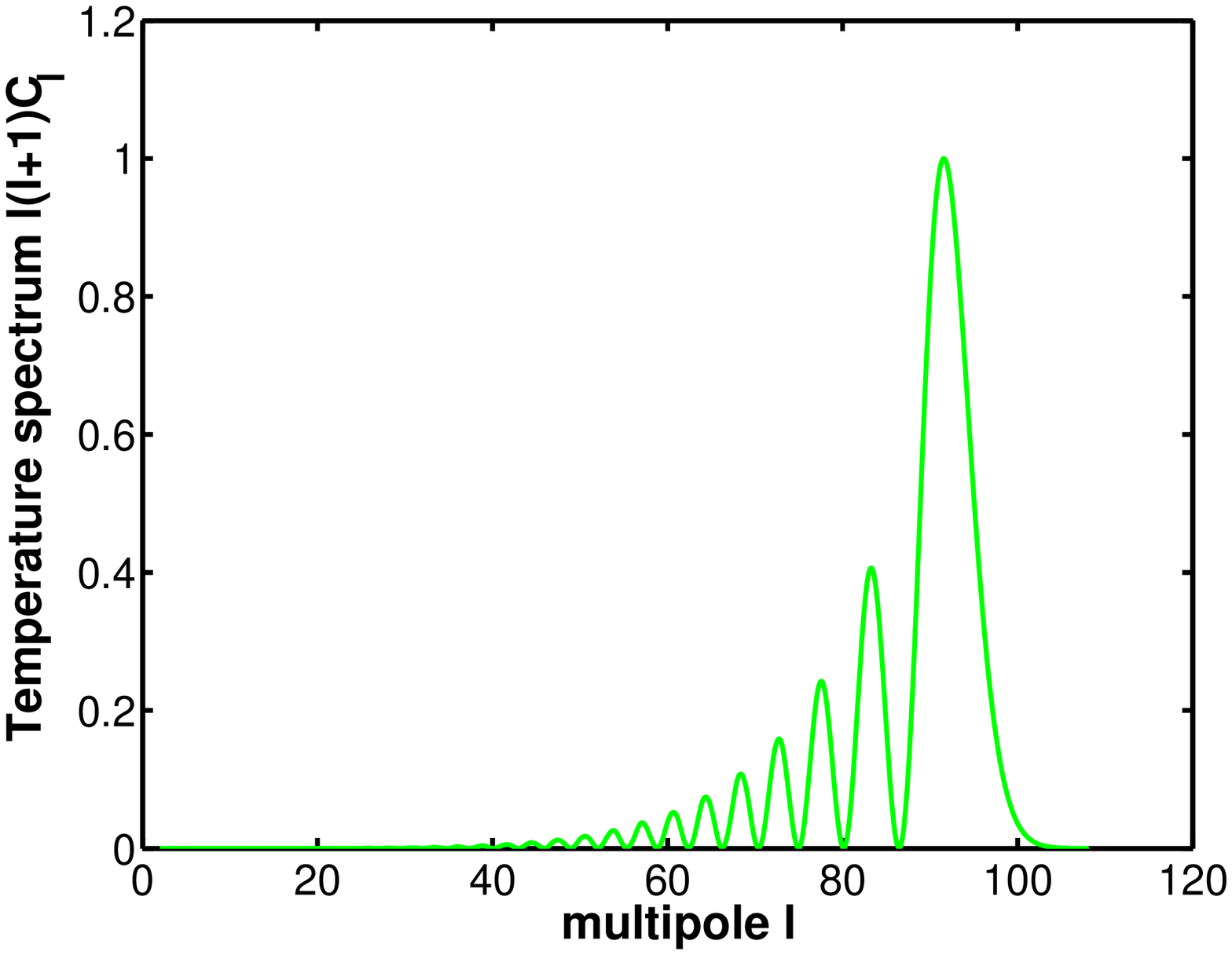}\qquad
\includegraphics[width=6cm]{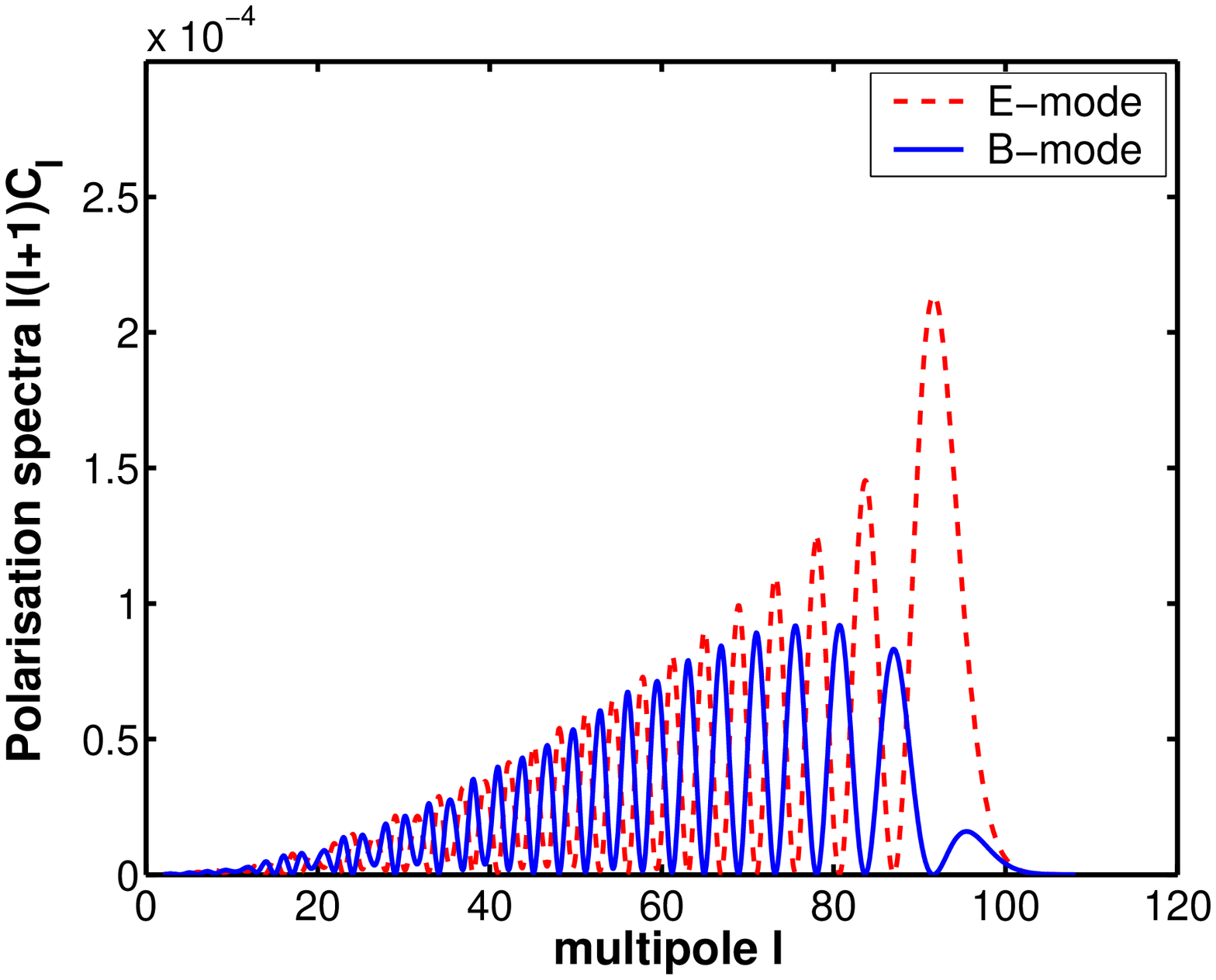}
\end{center}
\caption{Relative contributions of an individual Fourier mode $n =
100$ to various multipoles $\ell$ in the power spectra
$\ell(\ell+1) C_{\ell}$ for temperature and polarization. The
normalization has been chosen arbitrarily but same in both the
graphs.}\label{figure_recombination3a}
\end{figure}

\begin{figure}
\begin{center}
\includegraphics[width=6cm]{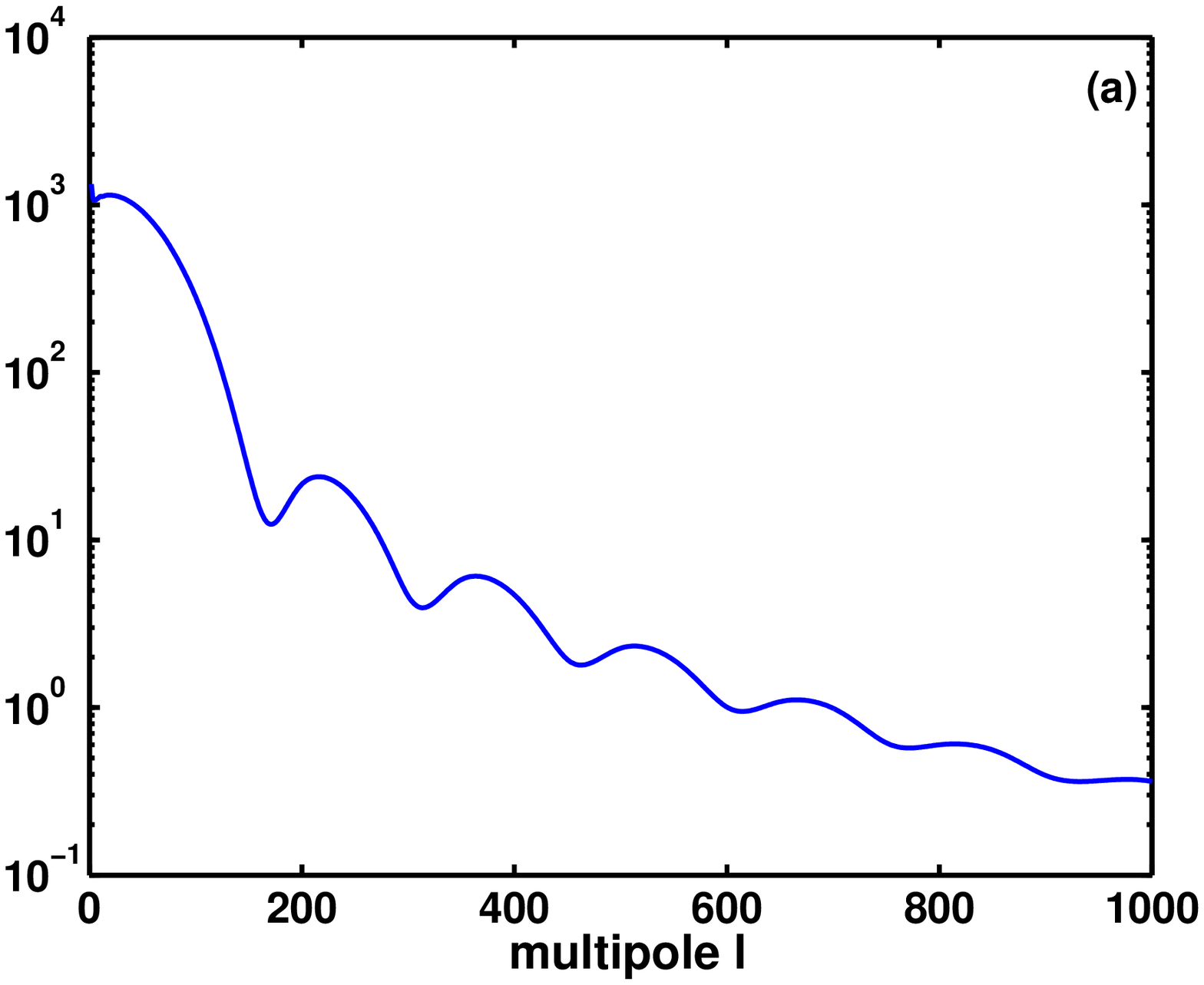}\qquad
\includegraphics[width=6cm]{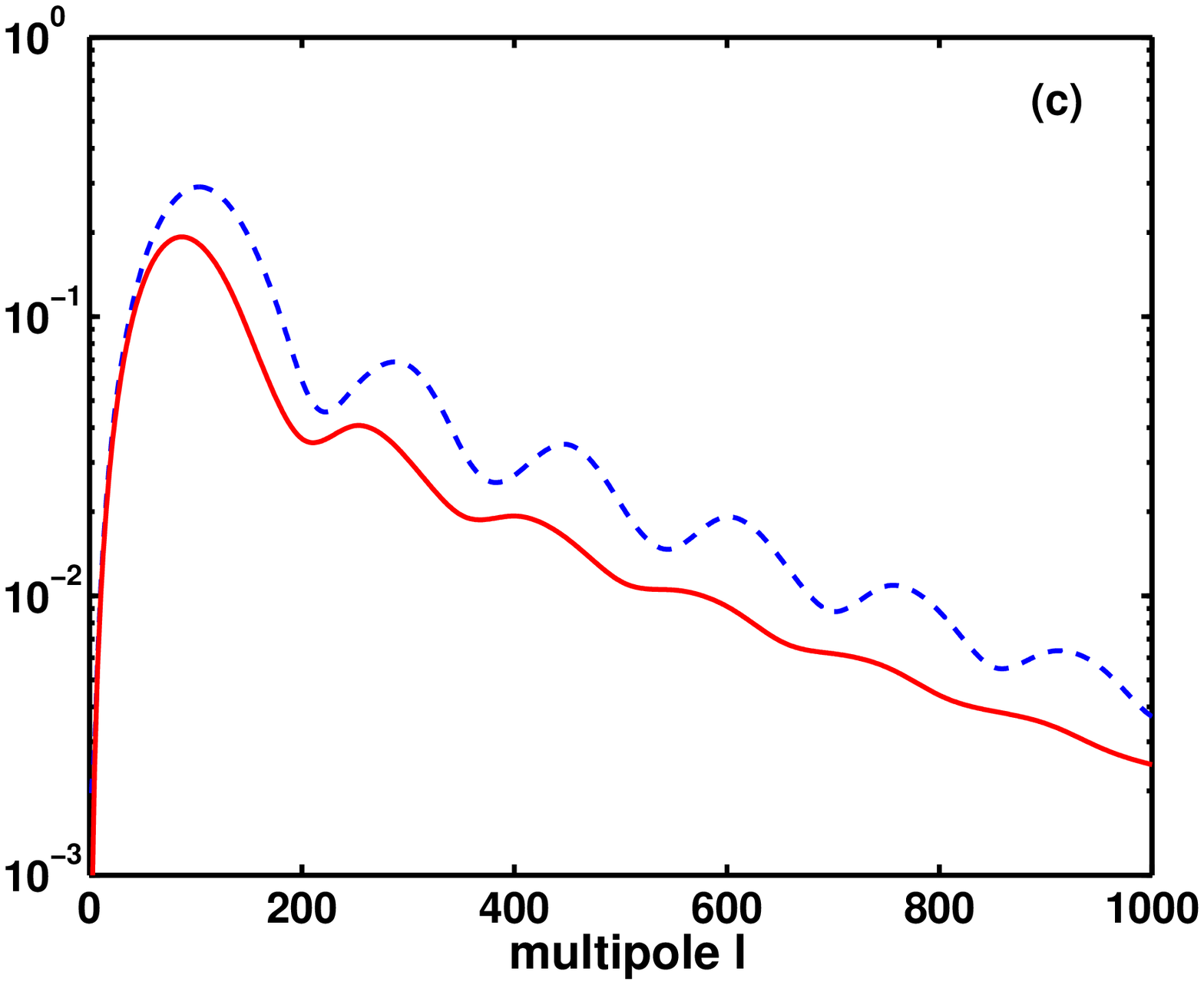}
\\
\includegraphics[width=6cm]{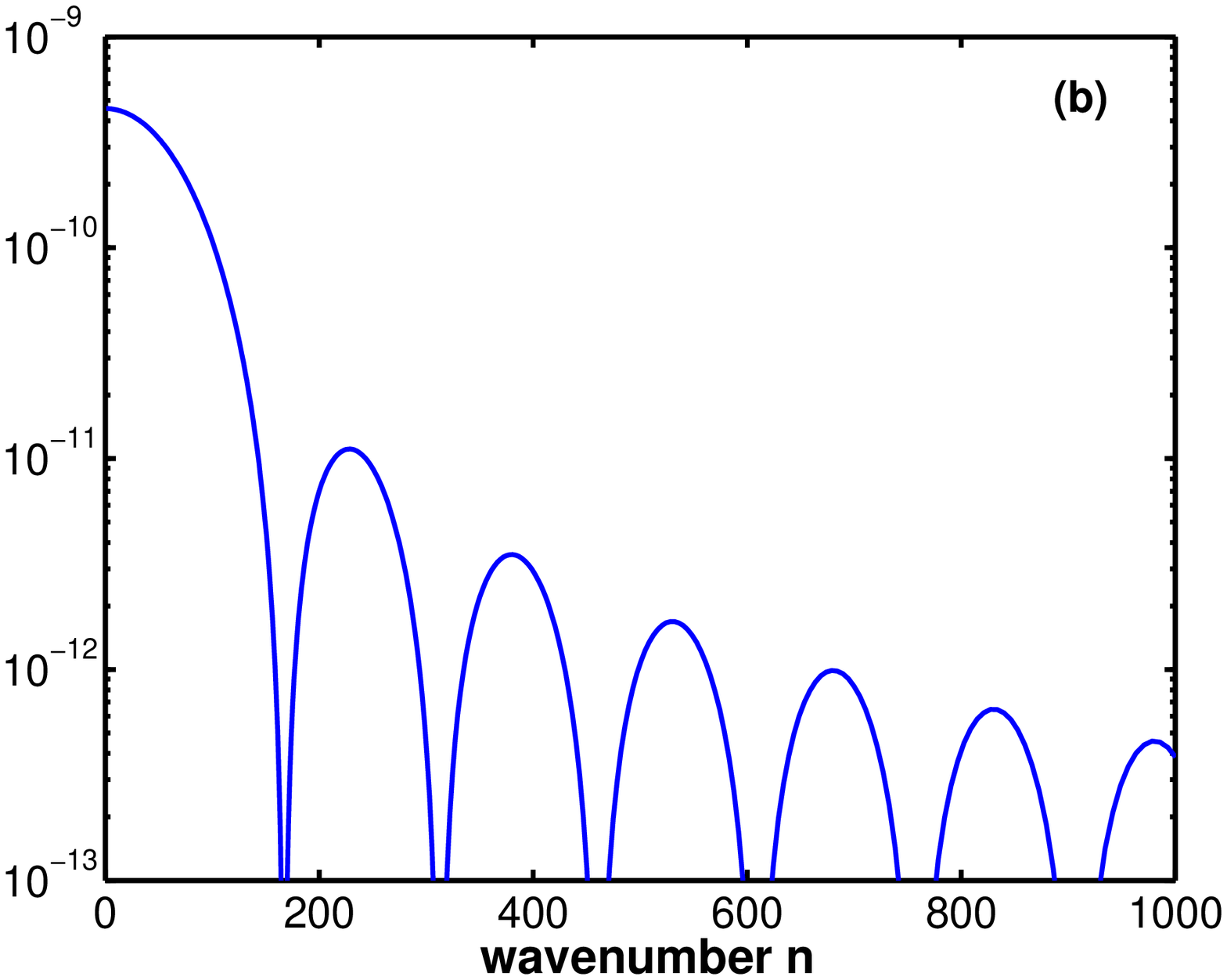}\qquad
\includegraphics[width=6cm]{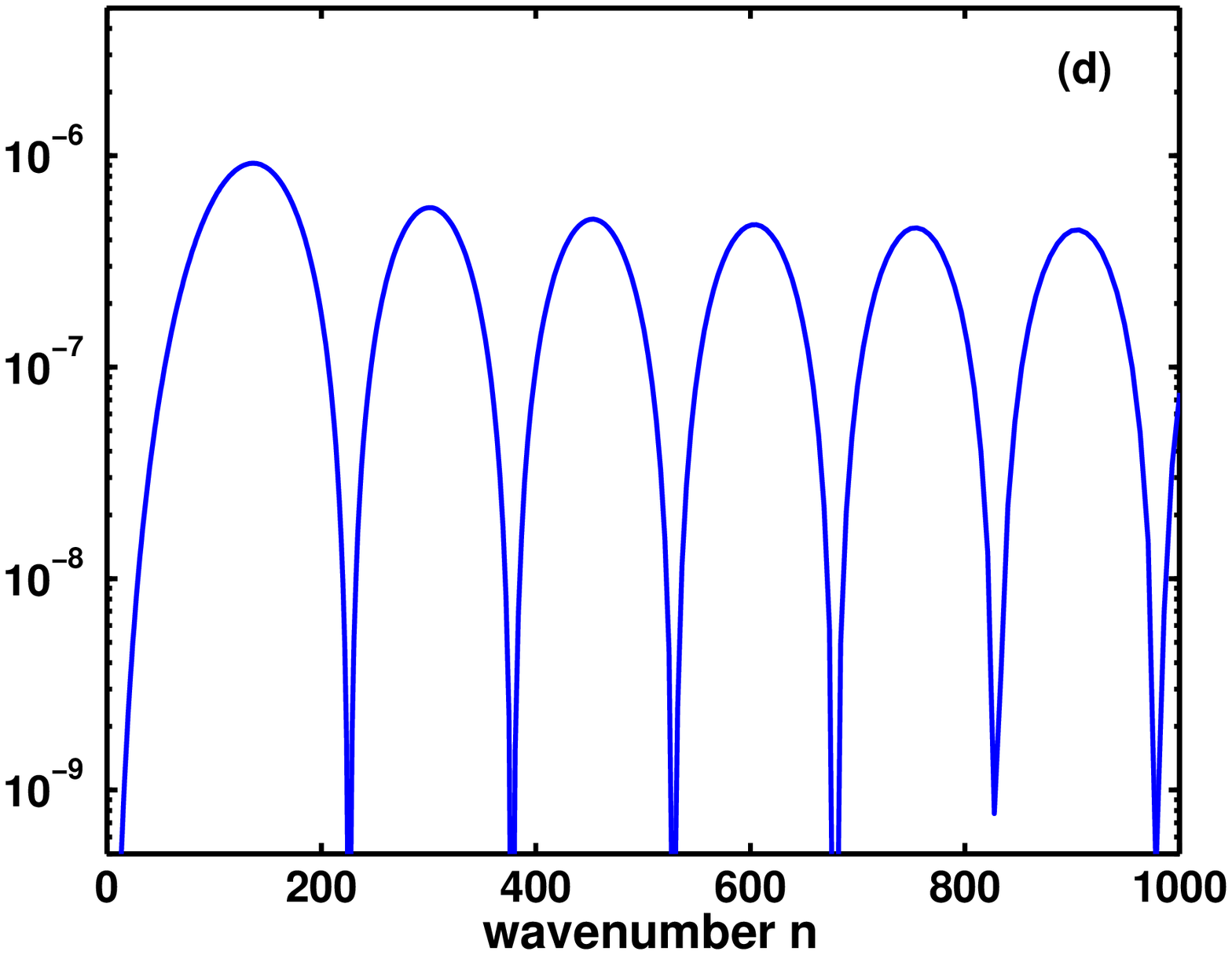}
\end{center}
\caption{The left panel shows (a) the power spectrum of
temperature anisotropies $ \ell (\ell+1)C_{\ell}^{TT}$ (in $~\mu
\textrm{K} ^{2}$) generated by (b) the power spectrum of g.w.\
metric perturbations (\ref{gwpower}), $\beta =-2$. The right panel
shows (c) the power spectra of polarization anisotropies $\ell
(\ell+1)C_{\ell}^{BB}$ (solid line) and $\ell
(\ell+1)C_{\ell}^{EE}$ (dashed line), panel (d) shows the power
spectrum of the first time derivative of the same g.w.\ field,
Eq.\ (\ref{powerder}).} \label{figure_recombination4}
\end{figure}

There is certain difference, however, between the projection
functions $E_{\ell}(\zeta_{dec})$ and $B_{\ell}(\zeta_{dec})$.
This is shown in Fig.\ \ref{figure_recombination3a}. The B-mode
projections are more `smeared' and their maxima are shifted to
somewhat lower $\ell$'s. This explains the visible difference
between $C_{\ell}^{EE}$ and $C_{\ell}^{BB}$ in Fig.\
\ref{figure_recombination4} (see also \cite{Pritchard2005}).
The polarization angular power
spectra, plotted in Fig.\ \ref{figure_recombination4}, were found
from exact numerical calculations.

\subsection{Temperature-Polarization cross correlation
\label{sec:TEtheoretical}}

A very special role belongs to the $TE$ cross-correlation
spectrum. We will show below that the $TE$ correlation at lower
$\ell$'s must be negative for gravitational waves and positive for
density perturbations. This distinctive signature can turn out to
be more valuable for identification of relic gravitational waves
than the presence of the $B$ polarization in the case of
gravitational waves and its absence in the case of density
perturbations. The expected $TE$ signal from gravitational waves
is about two orders of magnitude stronger than the $BB$ signal,
and it is much easier to measure. At lower $\ell$'s, the
contributions to $TE$ from gravitational waves and density
perturbations are comparable in absolute value, so the g.w.\
contribution is not a small effect. The total $TE$
cross-correlation has already been measured at some level
\cite{Kogut2003,Page2006}.

To find the $TE$ power spectrum we have to use the product of
$a^T_{\ell m}(n,s)$ and $a^E_{\ell m}(n,s)$ in Eq.\ (\ref{Clxx'}).
For qualitative analysis we will operate with the approximate
expressions (\ref{a_Tapprox}) and (\ref{a^Xapprox}). Then, the
$TE$ correlation reads
\begin{eqnarray}
C^{TE}_{\ell} &\approx&
\pi\gamma^{2}\left(\frac{\Delta}{10}\right)\int\frac{dn}{n} ~
D(n)\left.\sum_{s=1,2}\left(\stackrel{s}{h^*}(n,
\eta)\frac{d\stackrel{s}{h}(n,\eta)}{d\eta} + \stackrel{s}{h}(n,
\eta)\frac{d\stackrel{s}{h^*}(n,\eta)}{d\eta}
\right)\right|_{\eta=\eta_{dec}}\nonumber \\
&& \qquad \left( T_{\ell}(\zeta_{dec})E_{\ell}(\zeta_{dec})
\frac{}{}\right).
\label{C^TE_l2}
\end{eqnarray}
For a given $\ell$, the projection factor $\left(
T_{\ell}(\zeta_{dec})E_{\ell}(\zeta_{dec}) \frac{}{}\right)$ peaks
at $n \approx \ell$ and is positive there. Therefore, the sign of
$C^{TE}_{\ell}$ is determined by the sign of the term:
\begin{eqnarray}
\left.\frac{1}{2}\sum_{s=1,2}\left(\stackrel{s}{h^*}(n,
\eta)\frac{d\stackrel{s}{h}(n,\eta)}{d\eta} + \stackrel{s}{h}(n,
\eta)\frac{d\stackrel{s}{h^*}(n,\eta)}{d\eta}
\right)\right|_{\eta=\eta_{dec}}=
\left.\left(\frac{dh^{2}(n,\eta)}{d\eta}\right)\right|_{\eta=\eta_{dec}}
\label{tederiv}
\end{eqnarray}

\begin{figure}
\begin{center}
\includegraphics[width=6cm]{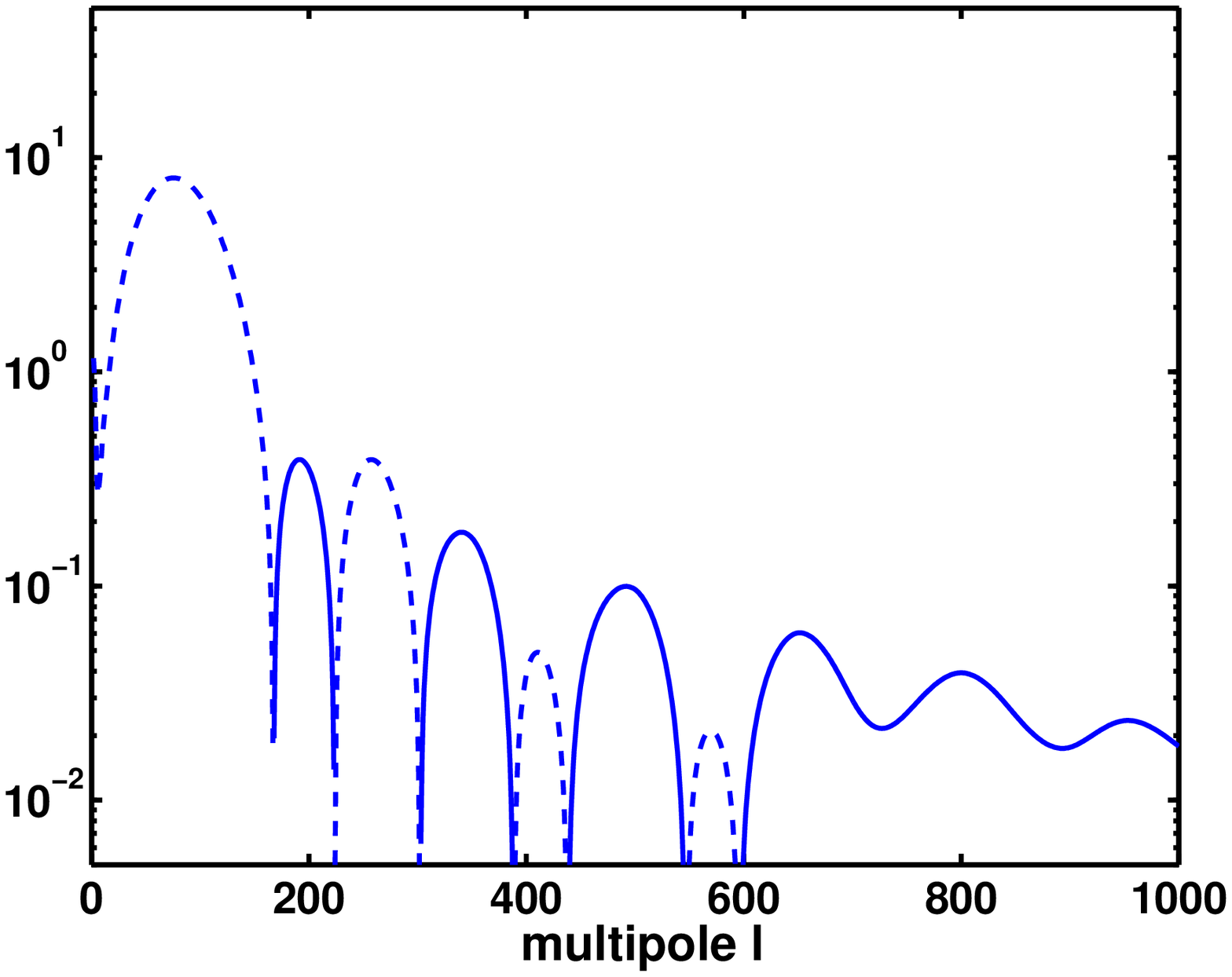}
\\
\includegraphics[width=6cm]{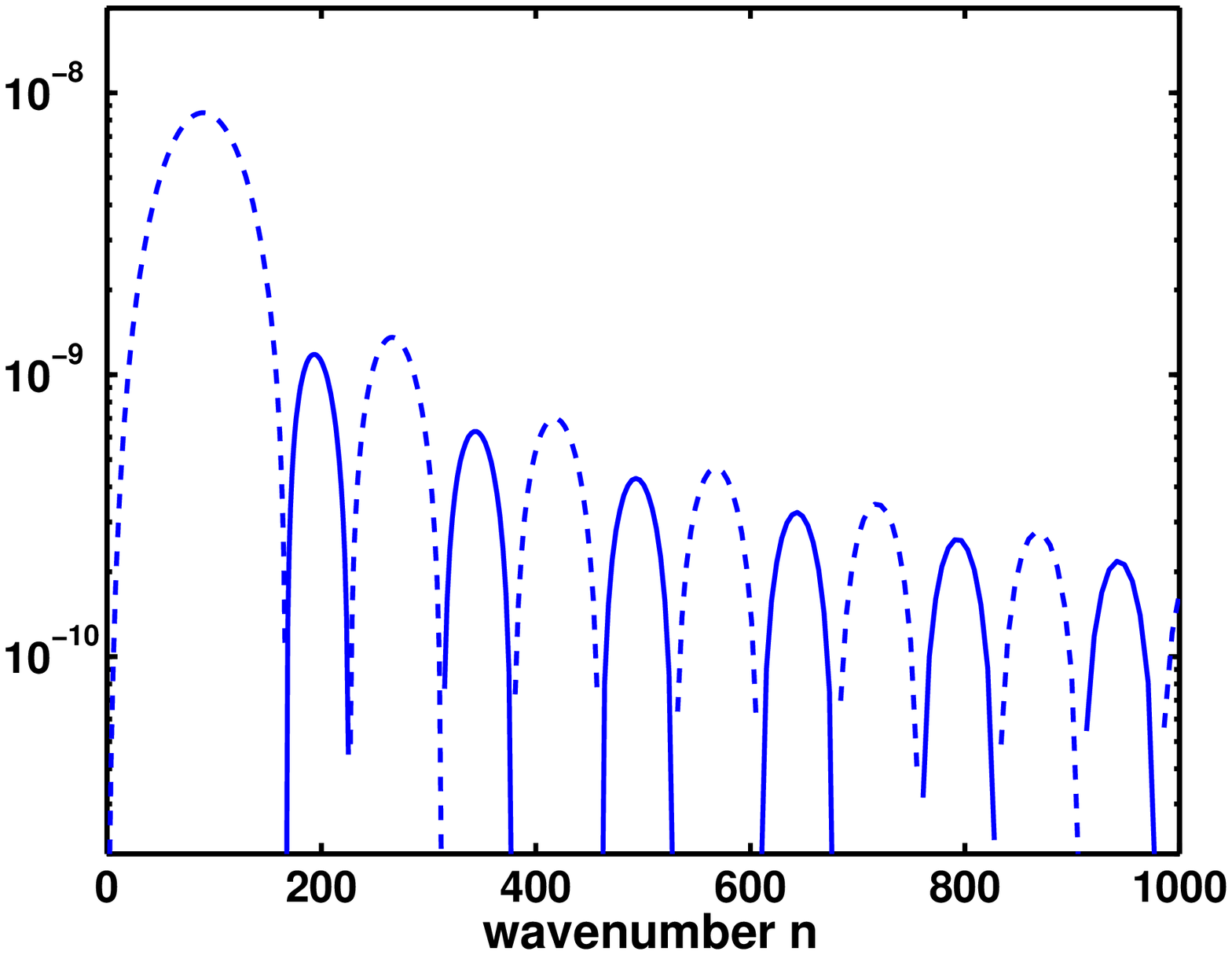}
\end{center}
\caption{The bottom panel shows the spectrum (\ref{tederiv}) of
gravitational waves, whereas the top panel shows the angular power
spectrum $\ell(\ell+1) C^{TE}_{\ell}$ caused by these waves. The
negative values of these functions are depicted by broken lines.}
\label{powcorrTE}
\end{figure}

The adiabatic decrease of the g.w.\ amplitude upon entering the
Hubble radius is preceded by the monotonic decrease of the g.w.\
mode function $\stackrel{s}{h}_{n}(\eta)$ as a function of $\eta$.
This behaviour is illustrated in Fig.\ \ref{hfigure}. It is clear
from the graph that for $n \lesssim 100$ the quantity
(\ref{tederiv}) is negative, because the first derivative of
$\stackrel{s}{h}_n(\eta)$ is negative. Therefore, for $\ell
\lesssim 90$ the correlation $C^{TE}_{\ell}$ must be negative. For
larger $\ell$'s the $TE$ correlation goes through zero, changes
sign and oscillates reflecting the oscillations of the function
(\ref{tederiv}) in the $n$-space. An exact numerical graph for
$TE$ correlation caused by gravitational waves is depicted in
Fig.\ \ref{powcorrTE}. The graph clearly shows how the sign and
features of the spectrum (\ref{tederiv}) get translated into the
sign and features of the $TE$ correlation.

It is shown in Appendix \ref{app:densityperturbations} that in the
case of density perturbations the $TE$ correlation must be
positive at lower $\ell$'s. This is because the relevant metric
perturbations associated with density perturbations are growing in
time and therefore the first time-derivative of metric
perturbations is positive. Because of other contributions, the
$TE$ correlation is expected to change sign at $\ell\approx 70$
\cite{Kogut2003}, \cite{LAMBDA}. The region of intermediate
multipoles $15\leq\ell\leq 90$ should be of a particular interest.
On one hand, the multipoles $\ell > 15$ are not affected by the
reionization era and its uncertain details, except the overall
suppression by $e^{-2\tau_{reion}}$. On the other hand, at $\ell <
90$ the g.w.\ contribution to $TE$ is not much smaller numerically
than the contribution from density perturbations. The lower
multipoles $\ell \lesssim 15$ are affected by reionization, and we
shall study reionization in the next Section.


\section{Effects of Reionization Era \label{reionera}}

The reionization of the intergalactic medium by first sources of
light has occurred relatively late, at $z\sim 30 - 7$ (see, for
example, \cite{Page2006, Keating2006}). In contrast to the
recombination part of the visibility function $g(\eta)$, which is
narrow and high, the reionization part of $g(\eta)$ is broad and
much lower (see Fig.\ \ref{figure_ionization_history}). For a
crude qualitative analysis, one can still apply analytical
formulas derived for recombination. One has to replace
$\eta_{dec}$ with $\eta_{reion}$ and $\Delta \eta_{dec}$ with
$\Delta \eta_{reion}$. The waves that start entering the Hubble
radius at $\eta_{reion}$, i.e. waves with wavenumbers $n
\eta_{reion} \sim \pi$, provide the most of power to the spectrum
of first time derivative of metric perturbations. Therefore, these
waves, with $n \sim 12$, produce a `bump' in polarization spectra
at the projected $\ell \sim 6$. However, these wavenumbers are
comparable with the wavenumber $n_{**}$, $n_{**} \Delta
\eta_{reion} \sim 2 \pi$, that characterizes the width of the
visibility function at reionization. The assumption $n \ll n_{**}$
is not well satisfied, the analytical approximation becomes crude,
and one has to rely mostly on numerics for more accurate answers.

\begin{figure}
\begin{center}
\includegraphics[width=6cm]{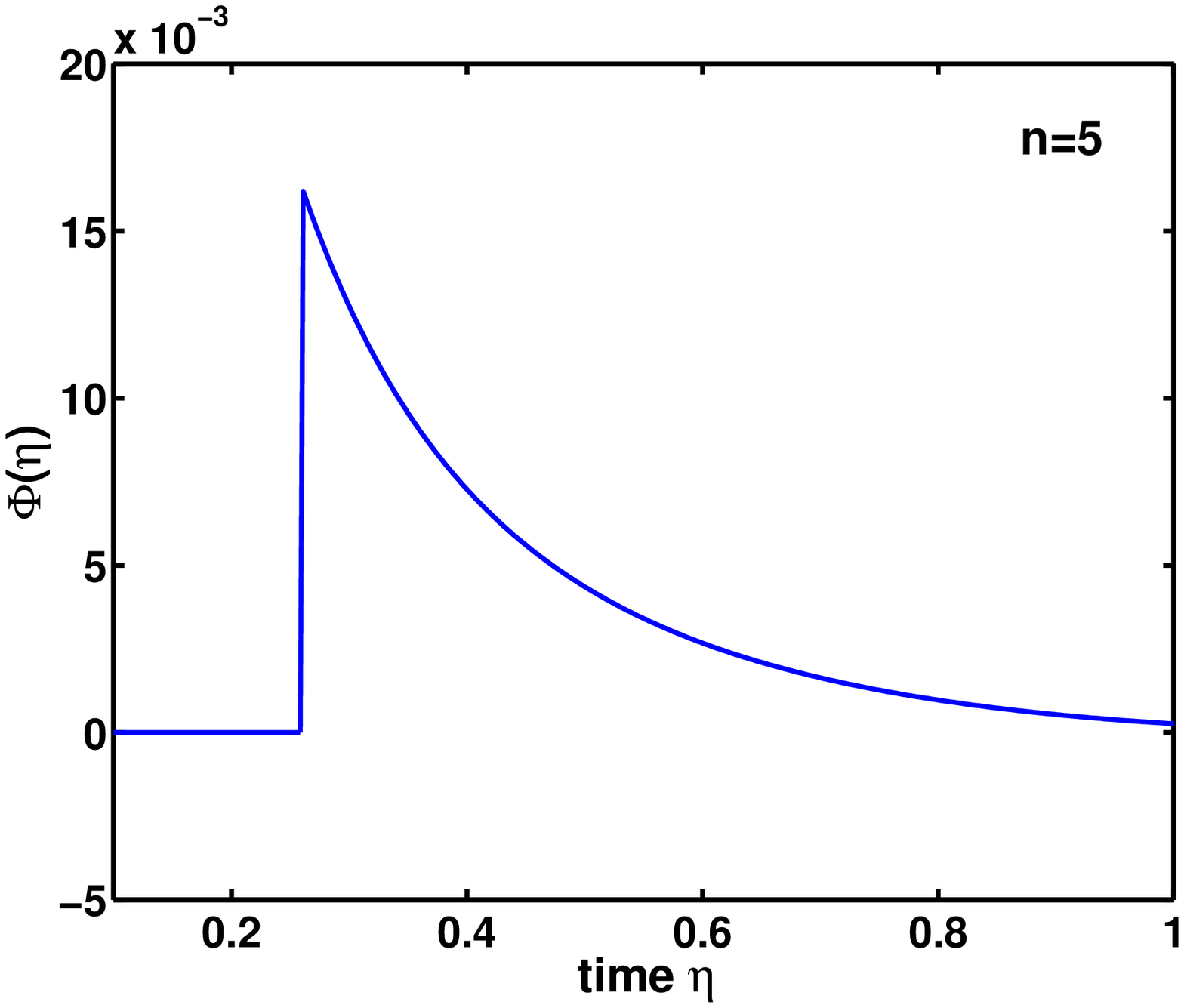}\qquad
\includegraphics[width=6cm]{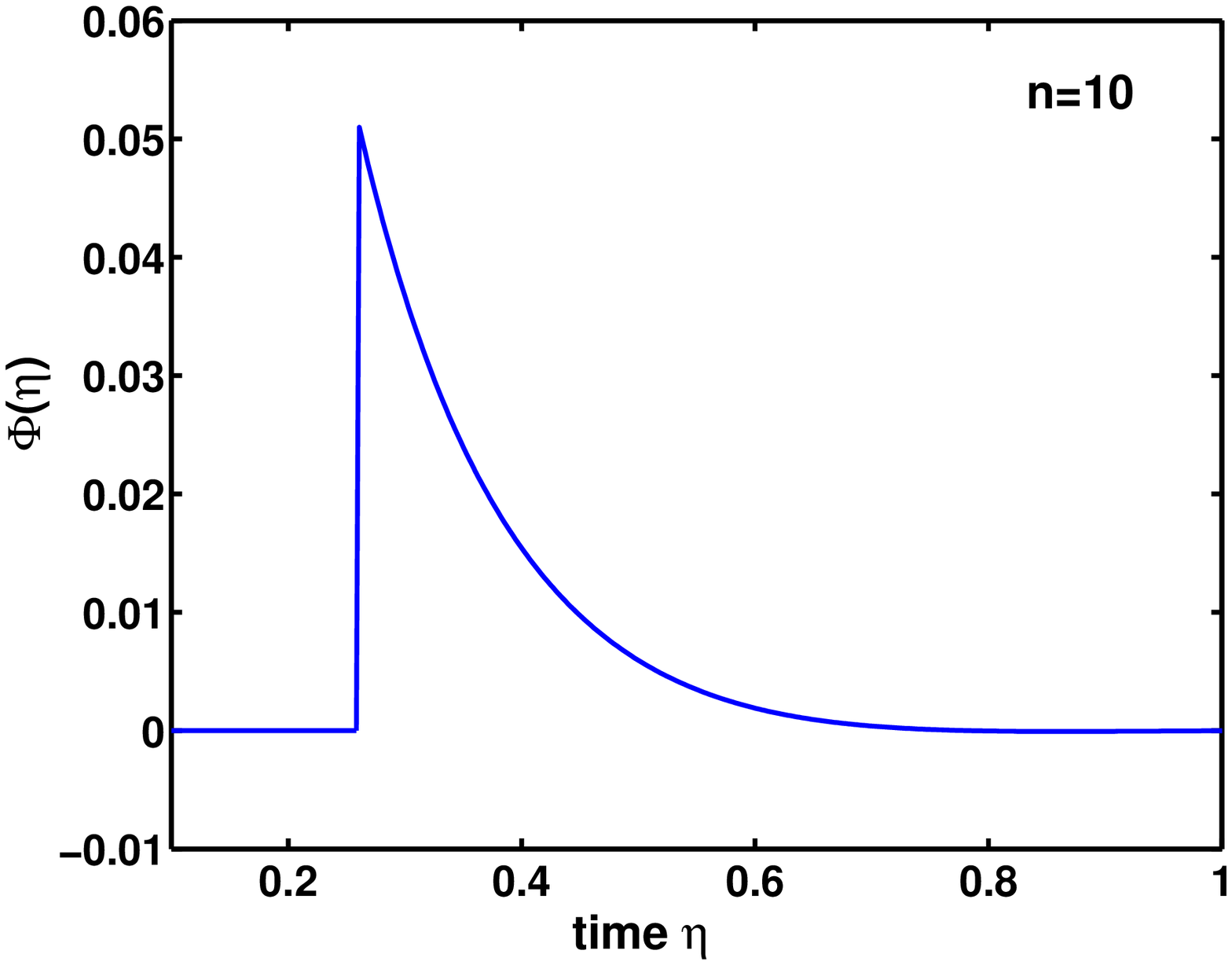}
\\
\includegraphics[width=6cm]{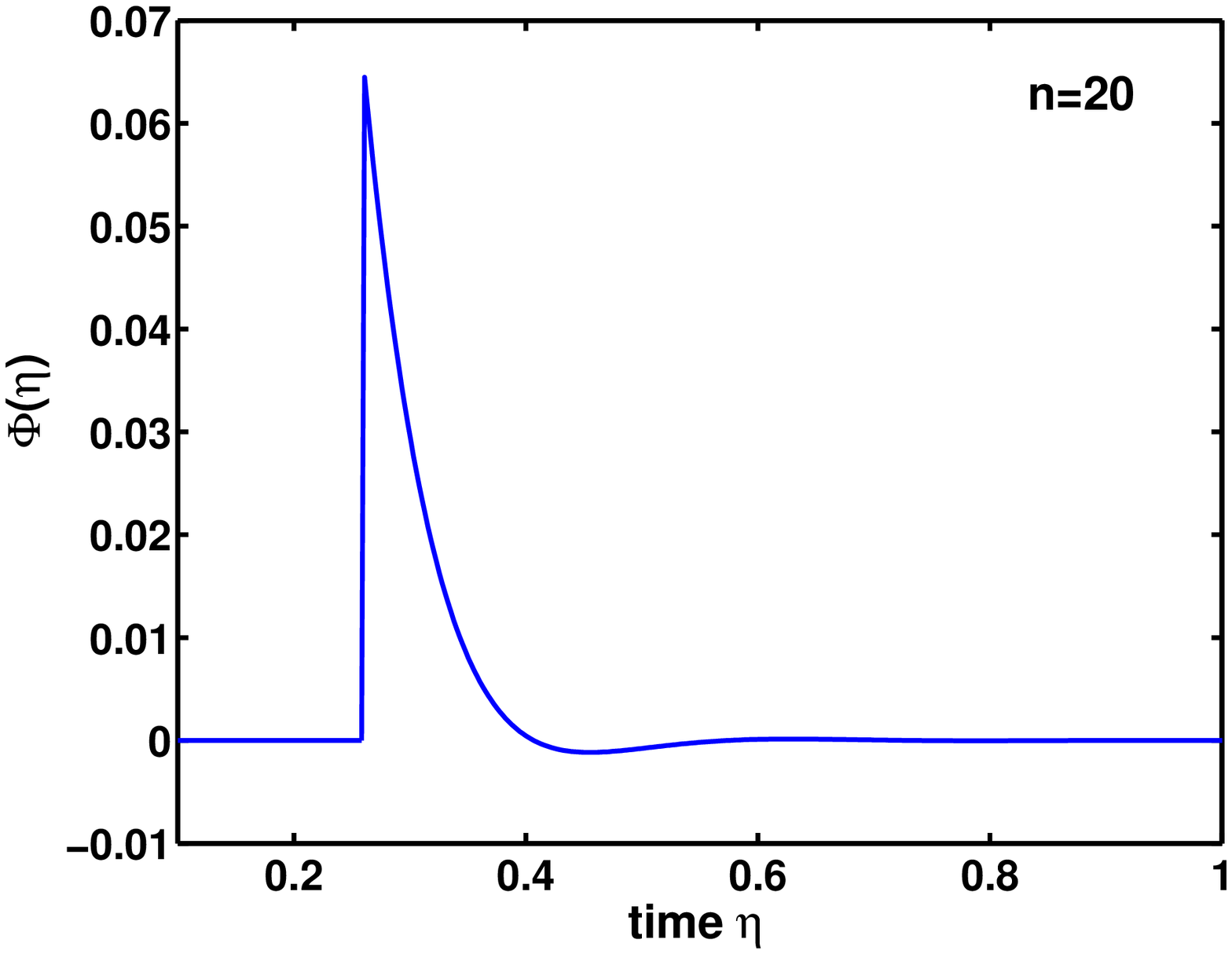}\qquad
\includegraphics[width=6cm]{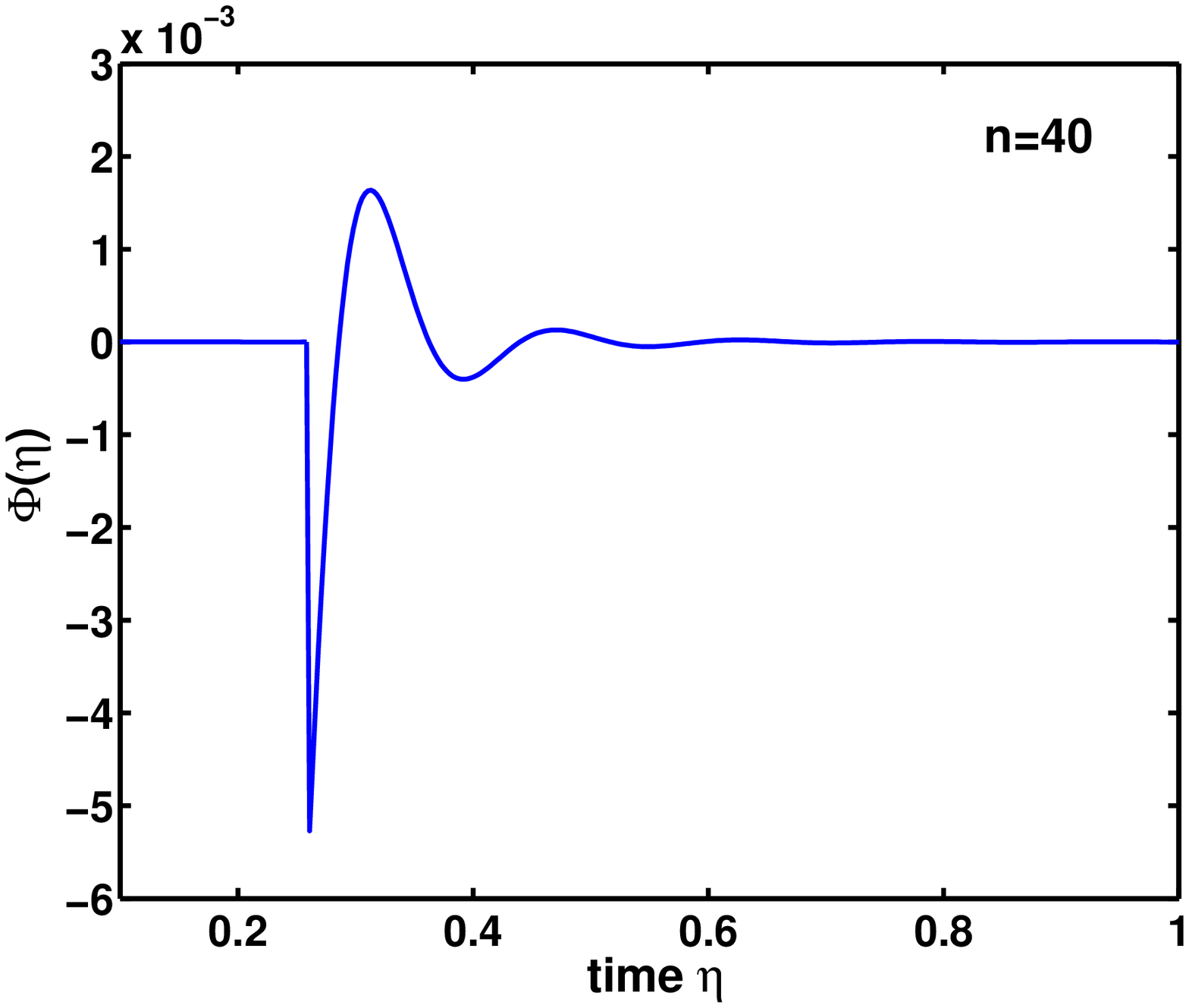}
\end{center}
\caption{The source function $\Phi_n(\eta)$ in reionization era
for different wavenumbers.}\label{figure_reion2}
\end{figure}

\begin{figure}
\begin{center}
\includegraphics[width=6cm]{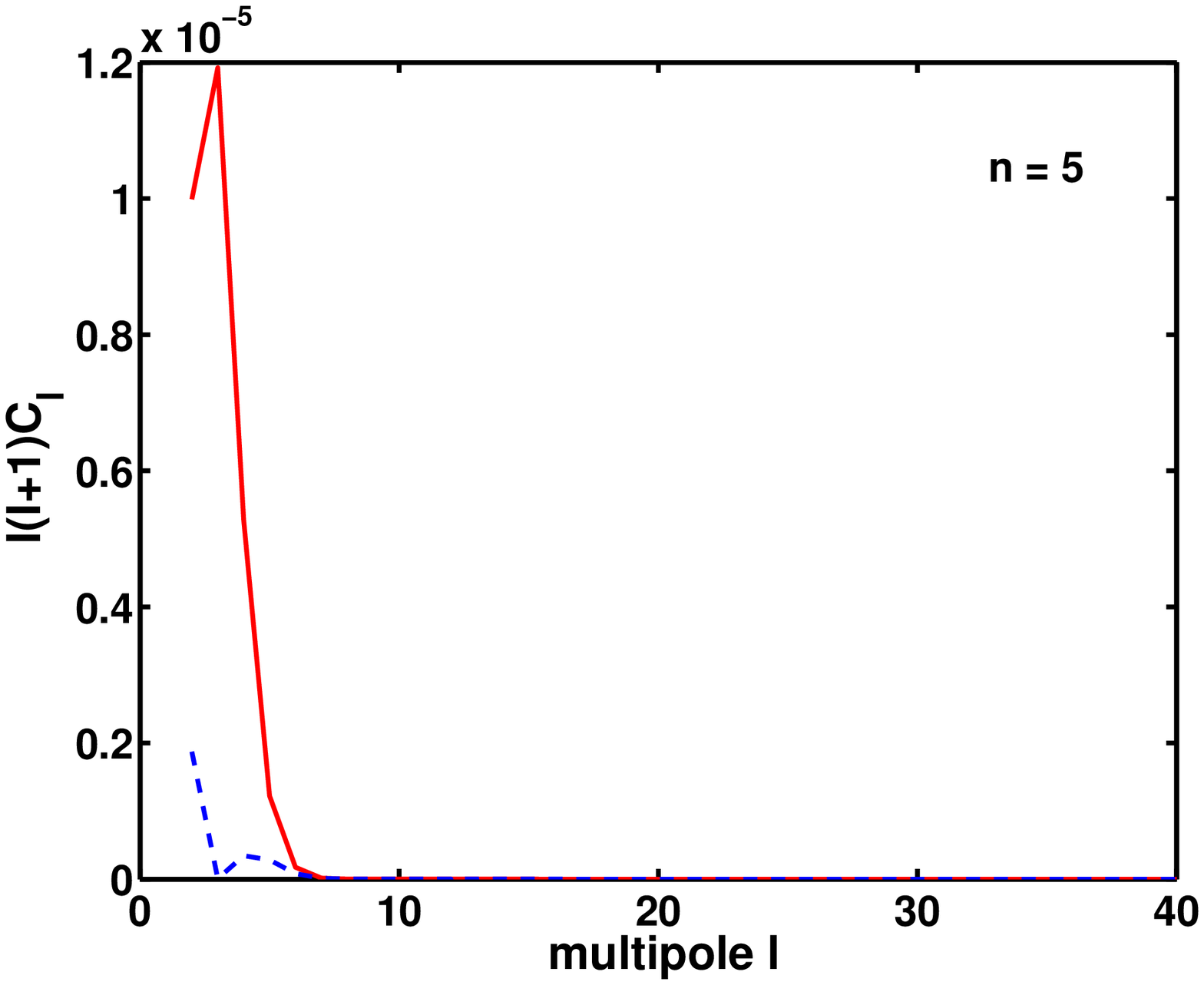}\qquad
\includegraphics[width=6cm]{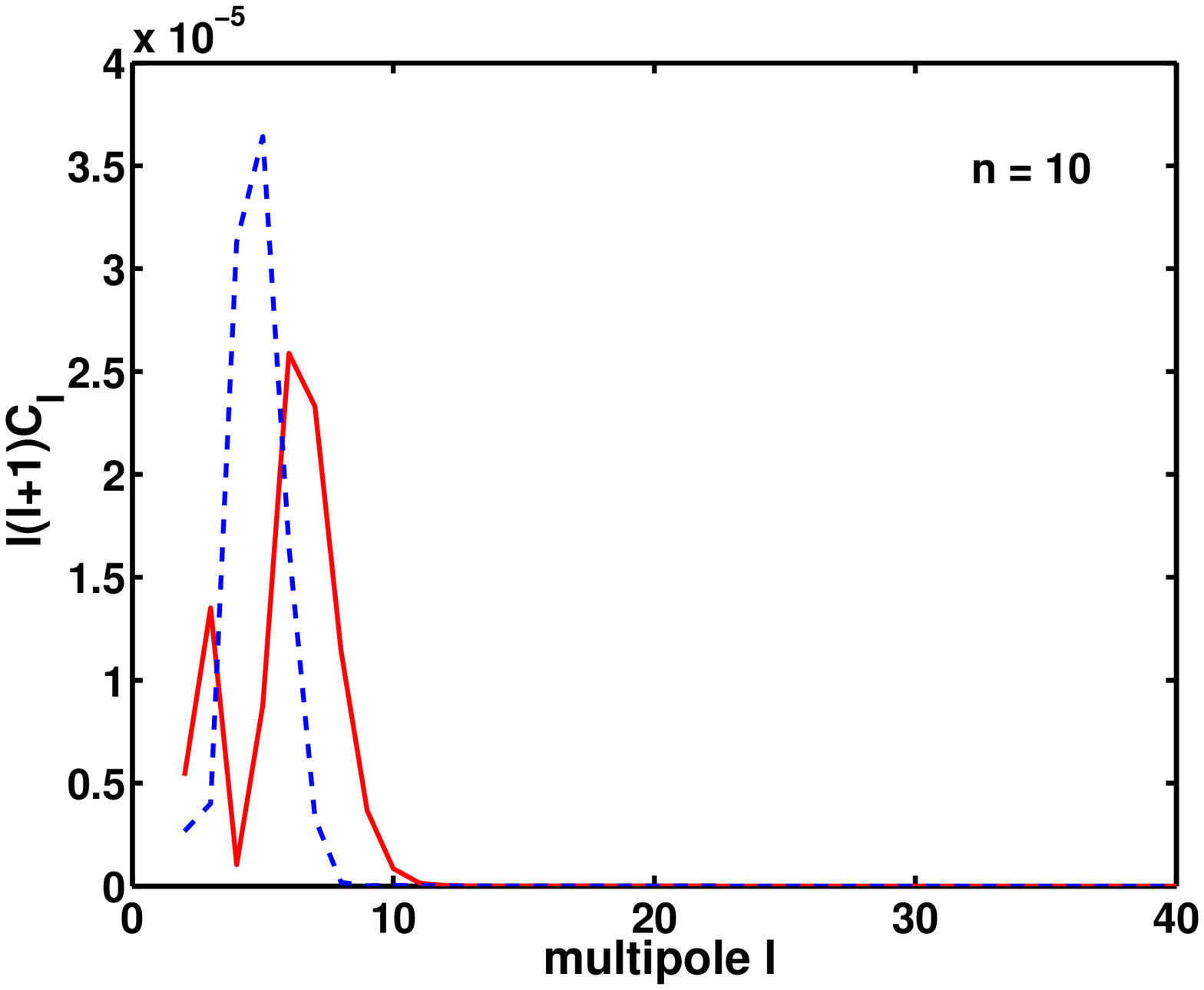}
\\
\includegraphics[width=6cm]{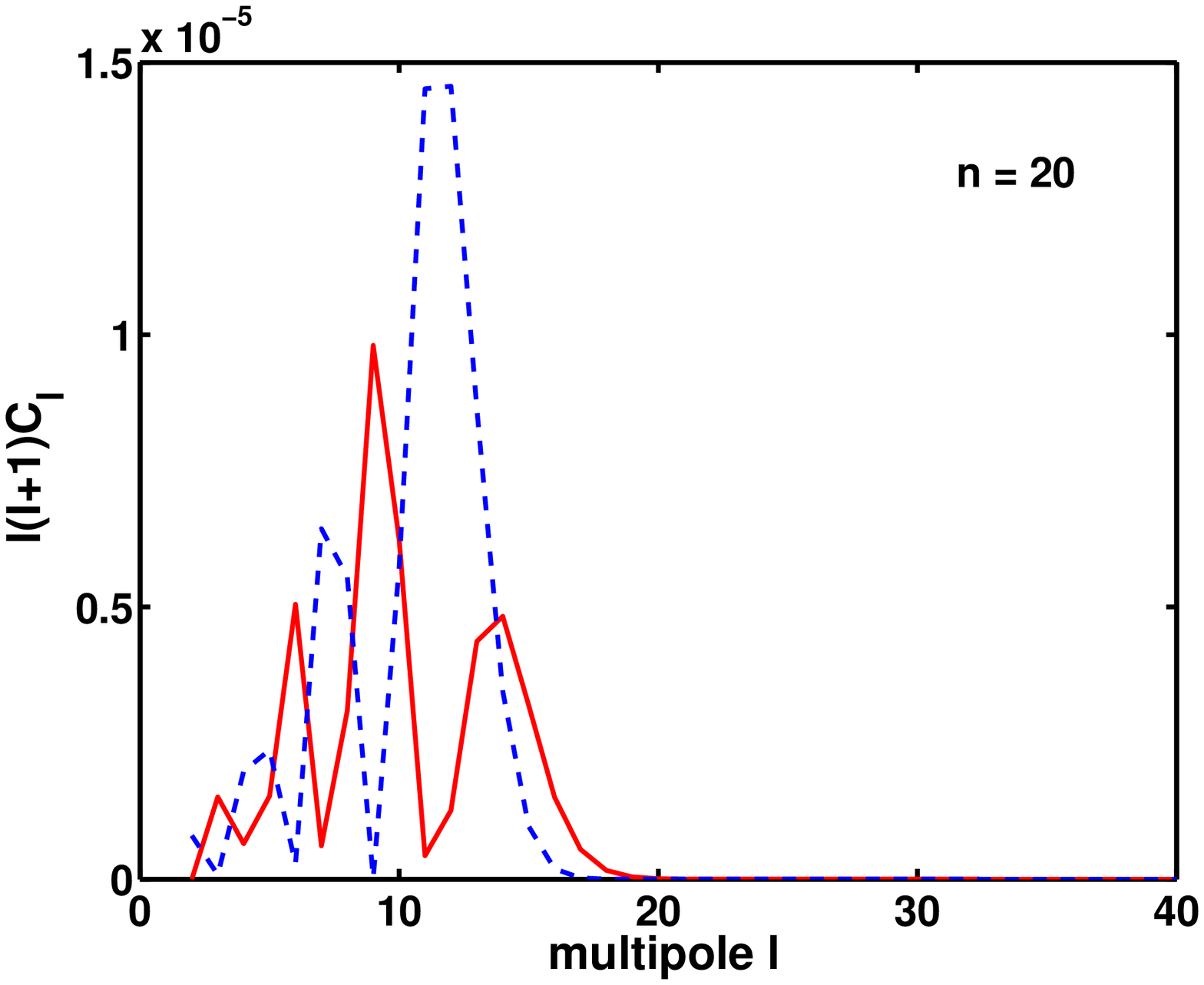}\qquad
\includegraphics[width=6cm]{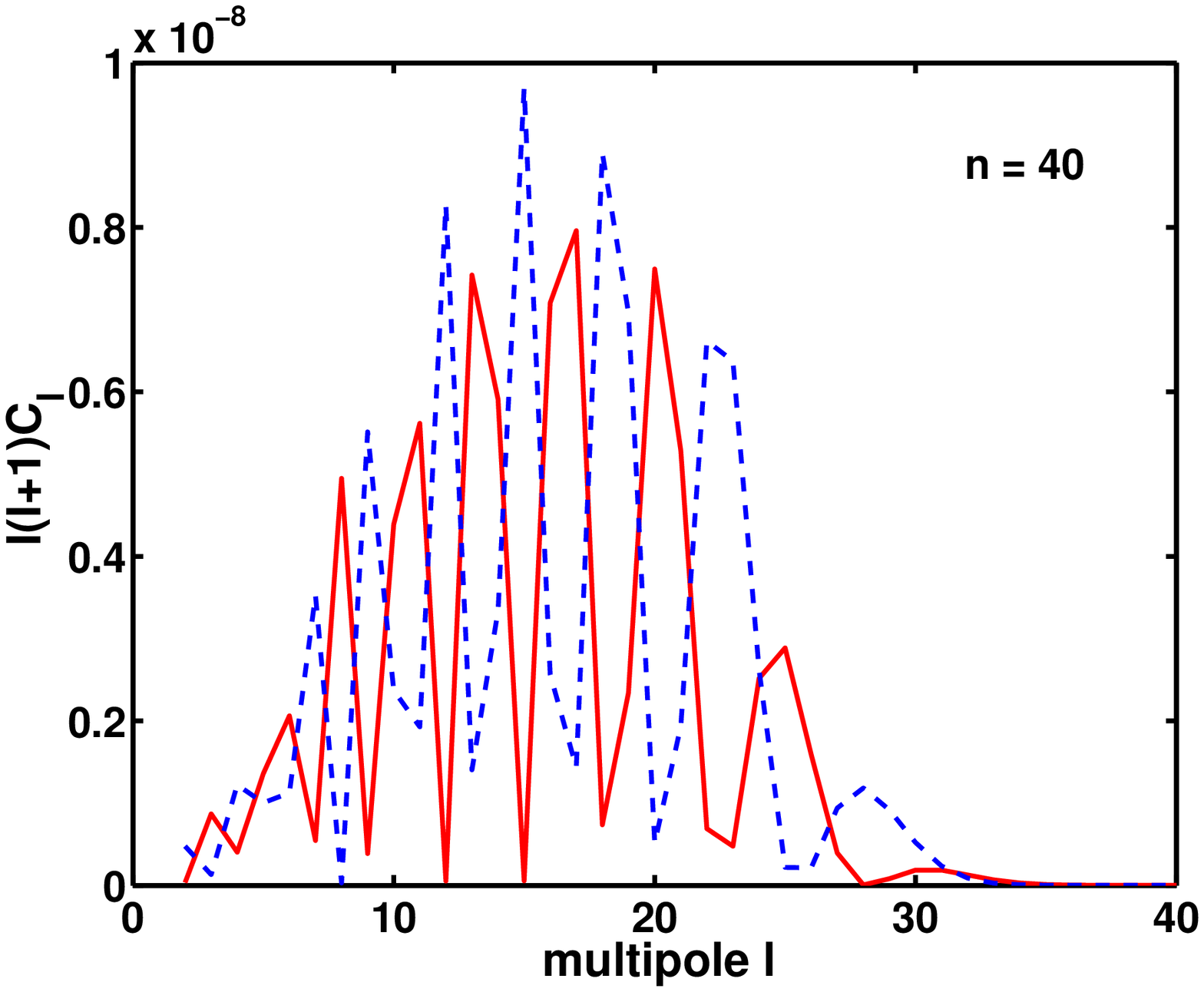}
\end{center}
\caption{The contributions to the angular power spectra from an
individual mode $n$. The dashed line is for
$\ell(\ell+1)C_{\ell}^{EE}$ and the solid line is for
$\ell(\ell+1)C_{\ell}^{BB}$. }\label{figure_reion3}
\end{figure}

The numerical solution to the integral equation
(\ref{integralequation}) in the reionization era is shown in Fig.\
\ref{figure_reion2}. One can see that the damping effect is
expected to commence from $n \gtrsim 20$, as the polarization
source function $\Phi_n(\eta)$ begins to show an oscillatory
behaviour. The projection relationships are also far away from the
almost one-to-one correspondence $\ell \approx n$ that was typical
for recombination era. In Fig.\ \ref{figure_reion3} we show the
contributions of a given $n$ to various $\ell$'s in the
polarization power spectra $\ell(\ell+1) C_{\ell}$. One can see
that a considerable portion of power from a given $n$ is
distributed over many lower-order $\ell$'s.

\begin{figure}
\begin{center}
\includegraphics[width=8cm]{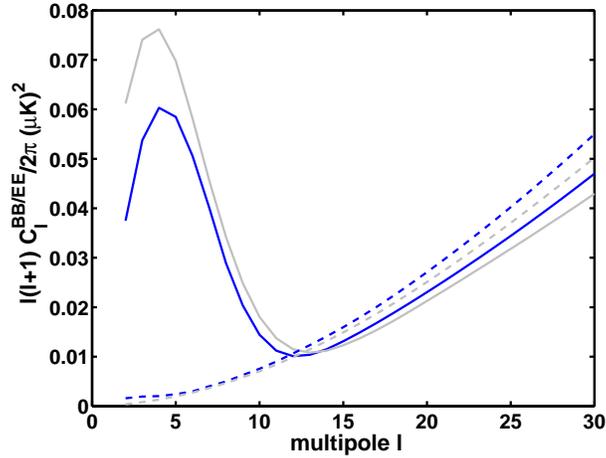}
\end{center}
\caption{The reionization `bump' in the polarization power
spectra. The dashed lines show the spectra when the reionization
era is ignored. The solid lines include both eras. The darker
lines are for $B$-mode, and the lighter lines for $E$-mode.
}\label{figure_reion4}
\end{figure}

The total effect of the reionization era is shown in Fig.\
\ref{figure_reion4}. This numerical result was based on our
simplified model of homogeneous reionization, as described in
Appendix \ref{app:Astrophysics}. The reionization `bumps' at lower
multipoles for $E$ and $B$ components of polarization are similar
in shape and numerical value.

\section{Comparison with Available Observations; Signatures of Relic
Gravitational Waves \label{comparison}}

The theory that we are using here is applicable to any primordial
spectral index ${\rm n}$. The initial conditions for gravitational
waves, Eqs.\ (\ref{initcond}), (\ref{Bnorm}), as well as analogous
initial conditions for density perturbations, hold $\beta$, and
hence ${\rm n}$, as a free constant parameter. The spectral index
${\rm n}$ can be larger, equal, or smaller than 1. However, from
the theoretical point of view, the `red' primordial spectra ${\rm
n} < 1$ ($\beta < -2$) seem to be unacceptable, or at least
questionable. If $\beta <-2$, the mean square fluctuations of
gravitational field, Eq.\ (\ref{meansq}), are power-law divergent
in the limit of very small wavenumbers $n$. One could argue that
the extrapolation of the primordial spectrum to the `infrared'
region of very small $n$'s is uncertain, and for some reason the
shape of the spectrum bends in the `infrared' region making the
integral (\ref{meansq}) convergent at the lower limit. We prefer
not to hide behind this possibility. If the shape of the
primordial spectrum is allowed to be varied, then practically
anything in the CMB data can be explained by the properly adjusted
primordial spectrum. Therefore, our theoretical preference (unless
the data will enforce us to change this preference) is a constant
primordial spectral index ${\rm n} > 1$ ($\beta > -2$). Obviously,
such primordial spectra entail no difficulty in the `ultraviolet'
region of very large $n$'s, because such short-wavelength
fluctuations (today's wavelength $\sim$ 3 cm.) did not satisfy the
requirements of superadiabatic amplification and simply have not
been generated.

\begin{figure}
\begin{center}
\includegraphics[width=8cm]{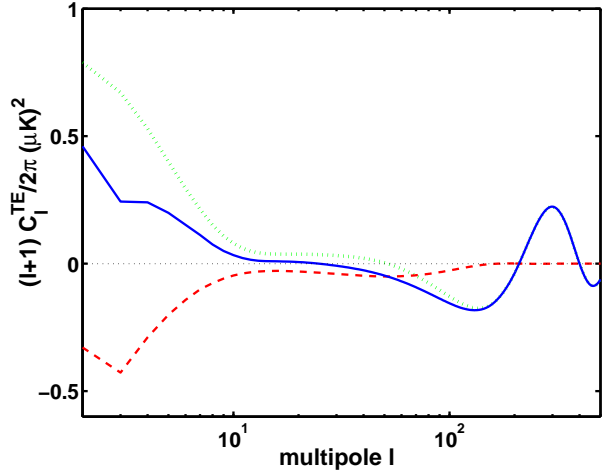}
\end{center}
\caption{ The dotted line shows the contribution of density
perturbations alone, and the dashed line shows the contribution of
gravitational waves alone. The solid line is the sum of these
contributions. It is seen from the graph that the inclusion of
g.w.\ makes the total curve to be always below the d.p.\ curve.}
\label{te_just_adding_gw}
\end{figure}

We now return to the $TE$ correlation. In Fig.\
\ref{te_just_adding_gw} we show the contributions to the $TE$
correlation function from relic gravitational waves (gw) and
primordial density perturbations (dp). (In order to enhance lowest
$\ell$'s we use the combination $(\ell+1)C_{\ell}$ rather than
$\ell(\ell+1)C_{\ell}$.) For this illustration, we choose a flat
primordial spectrum ${\rm n} =1$ ($\beta=-2$) and assume equal
contributions from (gw) and (dp) to the temperature quadrupole:
$R=1$, where
\[
R \equiv C_{\ell =2}^{TT}(gw)/C_{\ell =2}^{TT}(dp).
\]
We include the effects of reionization according to the model with
$\tau_{reion} =0.09$. The (gw) contribution is numerically
calculated from the solution to the integral equation
(\ref{integralequation}), whereas the (dp) contribution is plotted
according to the CMBfast code \cite{Zaldarriaga1997,LAMBDA}. One
can see from the graph that the negative $TE$ correlation function
at lower $\ell$'s is only possible if there is a significant
amount of primordial gravitational waves. One can also see from
the graph that a mis-interpretation of the total $TE$ effect as
being caused by density perturbations alone, could lead to a
serious mis-estimation of $\tau_{reion}$.

The $TE$ correlation at lower $\ell$'s measured by the WMAP
mission \cite{Kogut2003,Page2006} shows clusters of data points,
including the negative ones, that lie systematically below the
theoretical curve based on density perturbations alone. It is true
that the data points in the interval $10 \lesssim \ell \lesssim
70$ are concentrated near a zero level, the error bars are still
large, and the measured $TE$ correlation can be appreciably
different from the theoretical statistically averaged $TE$
correlation. However, the recent paper \cite{Page2006}, page 26,
explicitly emphasizes the detection of the $TE$ anticorrelation by
WMAP: ``The detection of the TE anticorrelation near $\ell \approx
30$ is a fundamental measurement of the physics of the formation
of cosmological perturbations...". As we have already stated
several times, the $TE$ anticorrelation at lower $\ell$'s, such as
$\ell \approx 30$, can only take place (within the framework of
all other common assumptions) when a significant amount of relic
gravitational waves is present.

Our theoretical position, as explained in this and previous
papers, is such that we are asking not `if' relic gravitational
waves exist, but `where' they are hiding in the presently
available data. We shall now discuss some models that fit the CMB
data and contain significant amounts of relic gravitational waves.
More accurate observations with WMAP, and especially Planck,
should firmly settle on the issue of the sign and value of the
$TE$ correlation at lower multipoles. Hopefully, these
observations will establish the presence of relic gravitational
waves beyond reasonable doubts.

\begin{figure}
\begin{center}
\includegraphics[width=6cm]{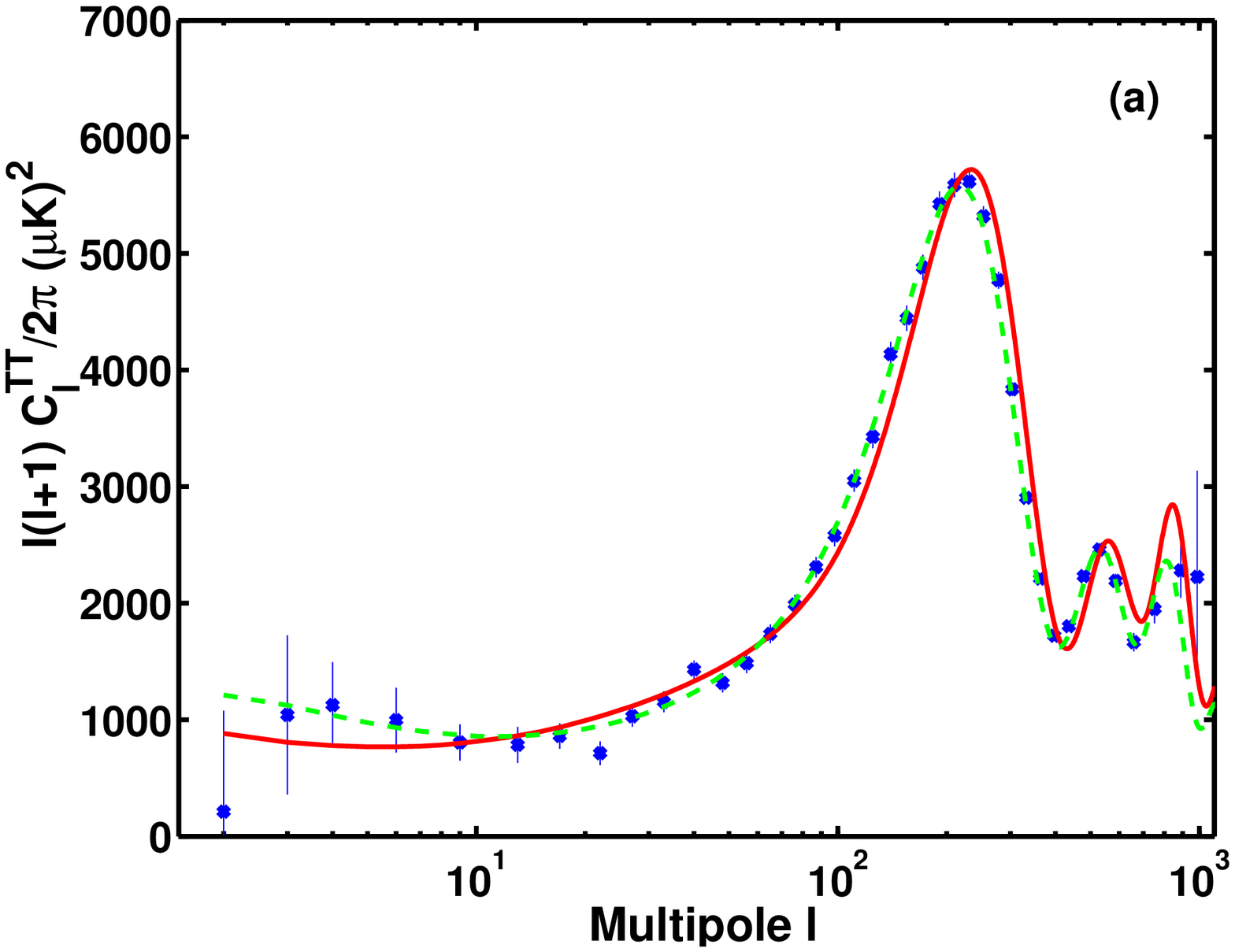}\qquad
\includegraphics[width=6cm]{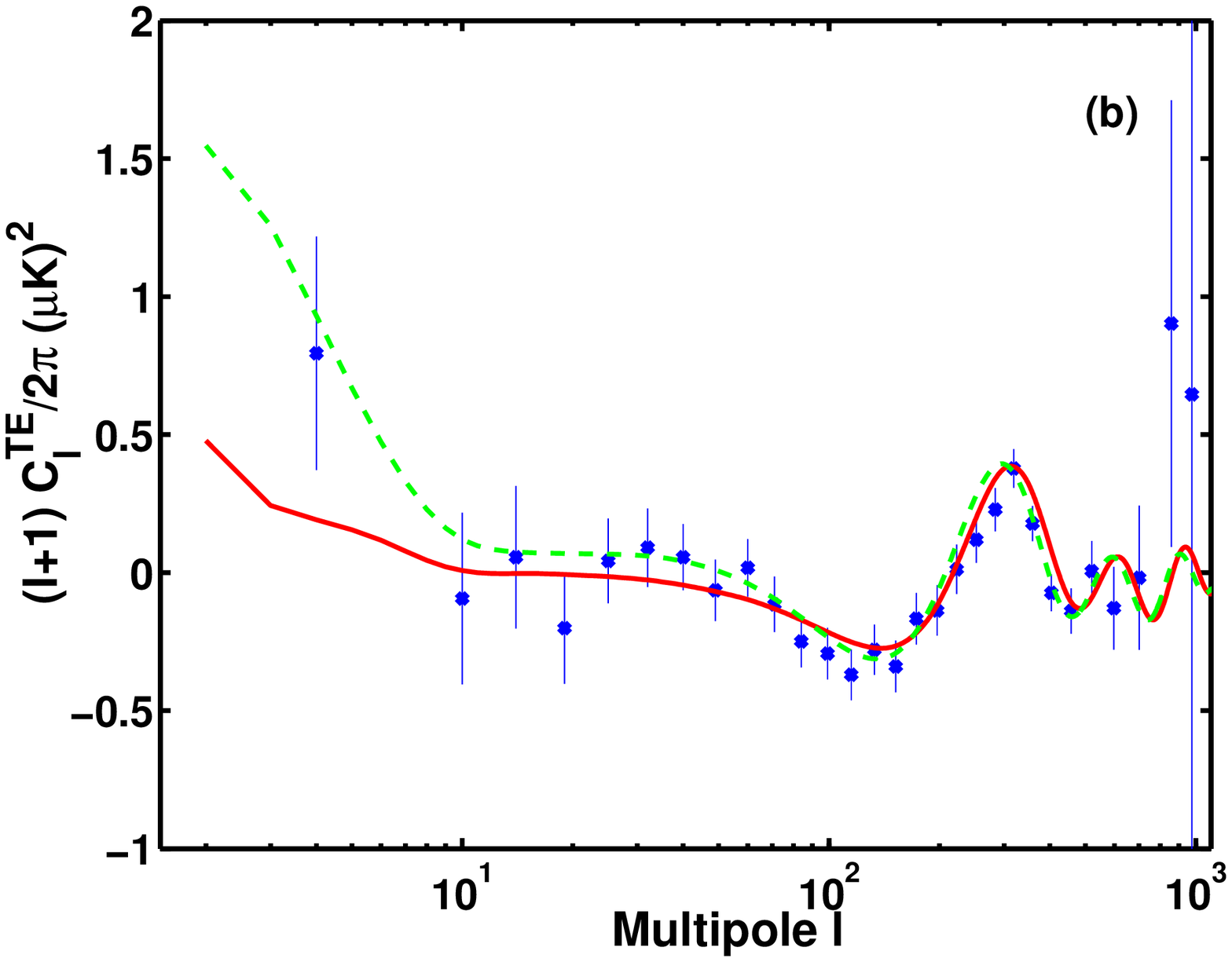}
\end{center}
\caption{The dashed line shows the best fit $\Lambda$CDM model
without gravitational waves. The solid line shows a model with
spectral index $n=1.2$ and gravitational waves $R=1$.}
\label{figure_tt_and_te_fit}
\end{figure}

To sharpen the discussion of allowed parameters, we take the model
with ${\rm n}=1.2$, $\tau_{reion} =0.09$ and $R= 1$. We take the
values $\Omega_m h^{2}=0.03$, $\Omega_b h^{2}=0.12$, $h=0.75$, and
we normalize the g.w.\ contribution to
$\frac{\ell(\ell+1)}{2\pi}C^{TT}_{\ell} = 440 ~\mu \textrm{K}
^{2}$ at $\ell=2$. In Fig.\ \ref{figure_tt_and_te_fit}a we show
our calculation of the $TT$ correlation function in comparison
with WMAP data and the best fit $\Lambda$CDM model
\cite{Spergel2006}, \cite{LAMBDA}. (If it comes to the necessity
of explaining `dark energy', natural modifications of general
relativity will be superior to unnatural modifications of the
matter sector \cite{Babak2003}.) One can see from the graph that
even this model (which lies, arguably, on a somewhat extreme end)
is consistent with the $TT$ data at all $\ell$'s and significantly
alleviates the much discussed tension between theory and
experiment at $\ell=2$. In Fig.\ \ref{figure_tt_and_te_fit}b we
show the $TE$ correlation for exactly the same model. One can see
that the inclusion of relic gravitational waves makes more
plausible the negative data points at lower $\ell$'s. Since the
relative contribution of gravitational waves becomes small at
$\ell \gtrsim 90$, the higher $\ell$ portion of the graph is
governed by density perturbations alone. Obviously, models with a
little smaller ${\rm n}$ or $R$ reach the same goals.

\begin{figure}
\begin{center}
\includegraphics[width=10cm]{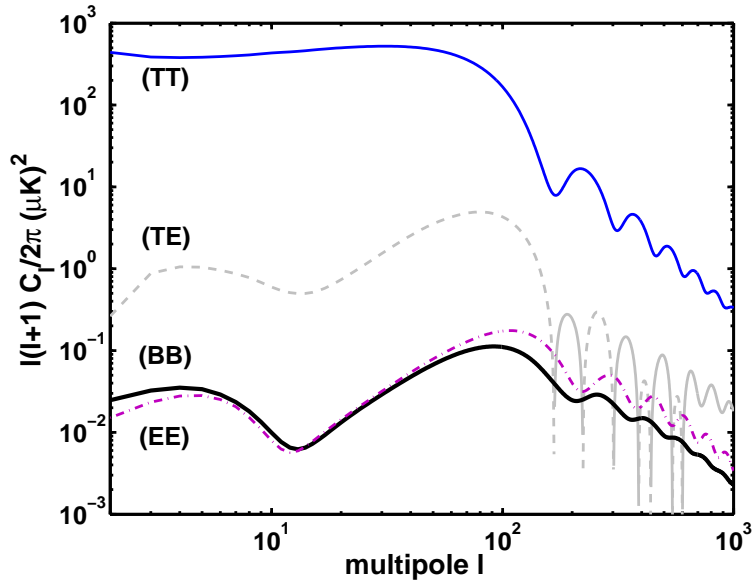}
\end{center}
\caption{The summary of CMB temperature and polarization
anisotropies due to relic gravitational waves with $n=1.2$ and
$R=1$.} \label{figureCMBgw}
\end{figure}

Finally, in Fig.\ \ref{figureCMBgw} we combine together all
correlation functions induced by relic gravitational waves. The
graphs are based on the discussed model with ${\rm n}=1.2$, $R=1$.
The future detection of the $BB$ correlation will probably be the
cleanest proof of presence of relic gravitational waves. For the
discussed model, the predicted level of the $BB$ signal is $
\left.\frac{\ell(\ell+1)}{2\pi}C_{\ell}^{BB} \right|_{\ell\approx
90} \sim 0.1 ~\mu \textrm{K} ^{2}$ in the region near the first
peak at $\ell \sim 90$. This level of $B$-mode polarization should
be detectable by the experiments currently in preparation, such as
CLOVER \cite{Maffei2005}, BICEP \cite{BICEP2003} and others (see
also \cite{Keating2006a}). Obviously, in this paper, we ignore
many complications, including astrophysical foregrounds. More
information can be found in Ref.\ \cite{Keating1998, Sazhin2004,
Naselsky2006, Lewis2006}.

\section{Conclusions}

In this paper, we summarized the properties of relic gravitational
waves that are directly relevant to CMB calculations. We explained
the reasons why we are working with g.w.\ backgrounds possessing
specific statistical properties, amplitudes, spectral indices,
etc. Then, we worked out, essentially from first principles, a
theory of CMB anisotropies induced by relic gravitational waves.
Some parts of this theory are rederivations and confirmations of
previous studies, some parts are new. The important advantage of
our approach, as we see it, is a transparent physical picture. We
believe we have demonstrated in the paper that every detail of the
derived correlation functions is under full analytical control.
Clear understanding of the participating physical processes has
led us to the conclusion that the $TE$ correlation in CMB can be a
valuable probe of relic gravitational waves. We compared our
theoretical findings with the WMAP data. We believe that the $TE$
anticorrelation detected by WMAP at $\ell \approx 30$ is certain
evidence for relic gravitational waves in the already available
data. We propose more accurate observations of the $TE$
correlation at lower $\ell$'s and believe that these observations
have the potential of providing a firm positive result.



\appendix


\section{Polarization states and randomness of gravitational waves
\label{detailsgw}}

As stated above, the quantities $|\stackrel{s}{h}_n(\eta)|^{2}$
describe the magnitude of the mean-square fluctuations of the
g.w.\ field in the corresponding polarization states $s$. We
consider a particular mode ${\bf n}$ of the field:
\begin{eqnarray}
h_{ij}({\bf n}, \eta, {\bf x}) = \sum_{s=1,2}
\stackrel{s}{p}_{ij}({\bf n}) \left(\stackrel{s}{h}_n(\eta)
e^{i{\bf n} \cdot{\bf x}} \stackrel{s}{c}_{\bf
n}+{\stackrel{s}{h}_n}^{*}(\eta) e^{-i{\bf n} \cdot{\bf
x}}\stackrel{s}{c}_{\bf n}^{*}\right).
\label{hmode}
\end{eqnarray}
The lowest-order independent g.w.\ correlation functions amount to
\begin{eqnarray}
\langle h_{ij}({\bf n}, \eta, {\bf x}) \stackrel{s}{p}{}^{ij}({\bf
n})~ h_{kl}({\bf n'}, \eta, {\bf x}') \stackrel{s'}{p}{}^{kl}({\bf
n'})\rangle= 8 \left|\stackrel{s}{h}_n(\eta) \right|^{2}
\delta_{ss'} \delta^{(3)}({\bf n} -{\bf n'}).
\label{corf}
\end{eqnarray}

The action of the random g.w.\ field (\ref{hmode}) on free
particles leads to their relative oscillatory motion. We refer
this motion to a local inertial frame. Let a ring of free
particles to lie in the (${\bf l}, {\bf m}$)-plane; the ring
encircles the axis ${\bf n}$. Then, the mean-square amplitude of
oscillations in the '+' polarisation state is determined by
$|\stackrel{1}{h}_n|^{2}$, whereas the mean-square amplitude of
oscillations in the '$\times$' polarisation state is determined by
$|\stackrel{2}{h}_n|^{2}$. In general, the random
gravitational-wave field can be such that the oscillation
amplitudes $|\stackrel{1}{h}_n|^{2}$ and $|\stackrel{2}{h}_n|^{2}$
are different. But if $|\stackrel{1}{h}_n|^{2} \neq
|\stackrel{2}{h}_n|^{2}$, then the (averaged) observed picture of
oscillations is not symmetric with respect to rotations around
${\bf n}$.

Formally, the correlation functions of the field with
$|\stackrel{1}{h}_n|^{2} \neq |\stackrel{2}{h}_n|^{2}$ do not have
symmetry with respect to the change of polarisation basis. Indeed,
the transition to the primed basis according to Eq.\
(\ref{ptensors2}) brings the gravitational wave mode (\ref{hmode})
to the form
\begin{eqnarray}
h_{ij}({\bf n}, \eta, {\bf x}) = \sum_{s=1,2}
\stackrel{s}{p'}_{ij}({\bf n}) \left(\stackrel{s}{b}_{\bf n}
(\eta) e^{i{\bf n}\cdot{\bf x}} + {\stackrel{s}{b}_{\bf
n}}^{*}(\eta) e^{-i{\bf n}\cdot{\bf x}} \right),
\nonumber
\end{eqnarray}
where
\begin{eqnarray}
\begin{array}{l}
\stackrel{1}{b}_{\bf n} (\eta) = \cos {2\psi}
\stackrel{1}{h}_n(\eta) \stackrel{1}{c}_{\bf n} + \sin{2\psi}
\stackrel{2}{h}_n(\eta) \stackrel{2}{c}_{\bf n},\\
\stackrel{2}{b}_{\bf n} (\eta) = - \sin{2\psi}
\stackrel{1}{h}_n(\eta) \stackrel{1}{c}_{\bf n} + \cos{2\psi}
\stackrel{2}{h}_n(\eta) \stackrel{2}{c}_{\bf n}.
\end{array}\nonumber
\end{eqnarray}

Taking into account the relationships (\ref{statCs}), we can now
derive the correlation functions for the new polarisation
components:
\begin{eqnarray}
\begin{array}{l}
\langle \stackrel{1}{b}_{\bf n} \stackrel{1}{b^*}_{\bf n'}\rangle
= \left(|\stackrel{1}{h}_n|^{2} \cos^{2}{2\psi} +
|\stackrel{2}{h}_n|^{2}
\sin^{2}{2\psi}\right) \delta^{(3)}({\bf n} -{\bf n'}), \\
\langle \stackrel{2}{b}_{\bf n} \stackrel{2}{b^*}_{\bf n'}\rangle
= \left(|\stackrel{1}{h}_n|^{2} \sin^{2}{2\psi} +
|\stackrel{2}{h}_n|^{2}
\cos^{2}{2\psi}\right) \delta^{(3)}({\bf n} -{\bf n'}), \\
\langle \stackrel{1}{b}_{\bf n} \stackrel{2}{b^*}_{\bf n'}\rangle
= -\left(|\stackrel{1}{h}_n|^{2} - |\stackrel{2}{h}_n|^{2} \right)
\sin{2\psi} \cos{2\psi}~ \delta^{(3)}({\bf n} -{\bf n'}).
\end{array}
\label{newstat}
\end{eqnarray}
It is seen from (\ref{newstat}) that, in general, the
$\psi$-dependence survives, and the assumption of statistical
independence of polarisation components in one basis is not
equivalent to this assumption in another basis. However, one
recovers the original correlation functions (\ref{corf}) from
(\ref{newstat}) if the conditions (\ref{eqval}) are fulfilled.

Similar properties hold true for circular polarizations. A g.w.\
mode $h_{ij}({\bf n},\eta, {\bf x})$ expanded over circular
polarisation states is given by
\begin{eqnarray}
h_{ij}({\bf n},\eta, {\bf x}) = {\bar h}_{ij}({\bf n}, \eta, {\bf
x}) + {{\bar h}_{ij}}^{*}({\bf n}, \eta, {\bf x}), \label{hmodec}
\nonumber\end{eqnarray} where
\begin{eqnarray}
{\bar h}_{ij}({\bf n}, \eta, {\bf x})= \sum_{s=L, R}
\stackrel{s}{p}_{ij}({\bf n}) \stackrel{s}{h}_n(\eta) e^{i{\bf n}
\cdot{\bf x}} \stackrel{s}{c}_{\bf n}. \label{hmodec2}
\end{eqnarray}
We assume that the complex random coefficients
$\stackrel{s}{c}_{\bf n}$ $(s= L, R)$, satisfy the statistical
conditions (\ref{statCs}).

The relevant independent correlation functions are calculated to
be
\begin{eqnarray}
\langle \bar{h}_{ij}({\bf n},\eta, {\bf x})
{\stackrel{s}{p}{}^{ij*}}({\bf n})~ {{\bar h}_{kl}}^{*}({\bf
n'},\eta, {\bf x}') \stackrel{s'}{p}{}^{kl}({\bf n'})\rangle= 4
\left|\stackrel{s}{h}_n(\eta) \right|^{2} \delta_{ss'}
\delta^{(3)}({\bf n} -{\bf n'}), ~~~(s = L, R). \label{corfc}
\end{eqnarray}
If the observer views the motion of test particles from the $-{\bf
n}$ direction, i.e. against the direction of the incoming
gravitational wave, the function $\left|\stackrel{R}{h}_n(\eta)
\right|^{2}$ is responsible for the mean-square amplitude of the
right-handed (clockwise) rotations of individual particles. The
function $\left|\stackrel{L}{h}_n(\eta) \right|^{2}$ is
responsible for the left-handed (anti-clockwise) rotations. (For
more details about the motion of free particles in the field of
gravitational waves, see \cite{Baskaran2004}.)

Expansion (\ref{hmodec2}) preserves its form under transformations
(\ref{pctensors2}), if one makes the replacements:
$\stackrel{L}{h}{'}_n=\stackrel{L}{h}_n e^{i2\psi}$,
$\stackrel{R}{h}{'}_n=\stackrel{R}{h}_n e^{-i2\psi}$. Therefore,
the correlation functions (\ref{corfc}) do not change, regardless
the value of the amplitudes $\left|\stackrel{s}{h}_n(\eta)
\right|^{2}$ ($s= L,R$). On the other hand, discrete
transformations (\ref{pctensors3}) generate the replacements:
$\stackrel{L}{h}{'}_n=\stackrel{R}{h}_n$,
$\stackrel{R}{h}{'}_n=\stackrel{L}{h}_n$. Therefore, the sense of
correlation functions (\ref{corfc}) changes from $L$ to $R$ and
viceversa. The symmetry between left and right is violated, unless
the the conditions (\ref{eqval2}) are fulfilled.


\section{\label{app:Astrophysics}Astrophysical prerequisites}

(i) Ionization History

The ionization history of the Universe enters our equations
through the density of free electrons $N_e(\eta)$, i.e. electrons
available for Thompson scattering. Specifically, we operate with
the quantity $q(\eta)$:
\begin{eqnarray}
q(\eta) = \sigma_Ta(\eta)N_e(\eta). \label{q}
\end{eqnarray}
The optical depth $\tau$ between some instant of time $\eta'$ and
a later instant $\eta$ is defined by the integral
\begin{eqnarray}
\tau(\eta,\eta') = \int\limits_{\eta'}^{\eta}d\eta''q(\eta'').
\nonumber
\end{eqnarray}
The optical depth from some $\eta$ to the present time $\eta_R$ is
denoted $\tau(\eta)$ and is given by
\begin{eqnarray}
\tau(\eta) \equiv \tau(\eta_R,\eta) =
\int\limits_{\eta}^{\eta_R}d\eta'q(\eta'). \nonumber
\end{eqnarray}
It follows from the above definitions that $\tau(\eta, \eta') =
\tau(\eta') - \tau(\eta)$.

A key role in our discussion is played by the quantity $g(\eta)$
called the visibility function:
\begin{eqnarray}
g(\eta) = q(\eta)e^{-\tau(\eta)} =
\frac{d}{d\eta}e^{-\tau(\eta)},~~~~~~ \int_0^{\eta_R} g(\eta) d
\eta =1 \nonumber
\end{eqnarray}

\begin{figure}
\begin{center}
\includegraphics[width=6cm]{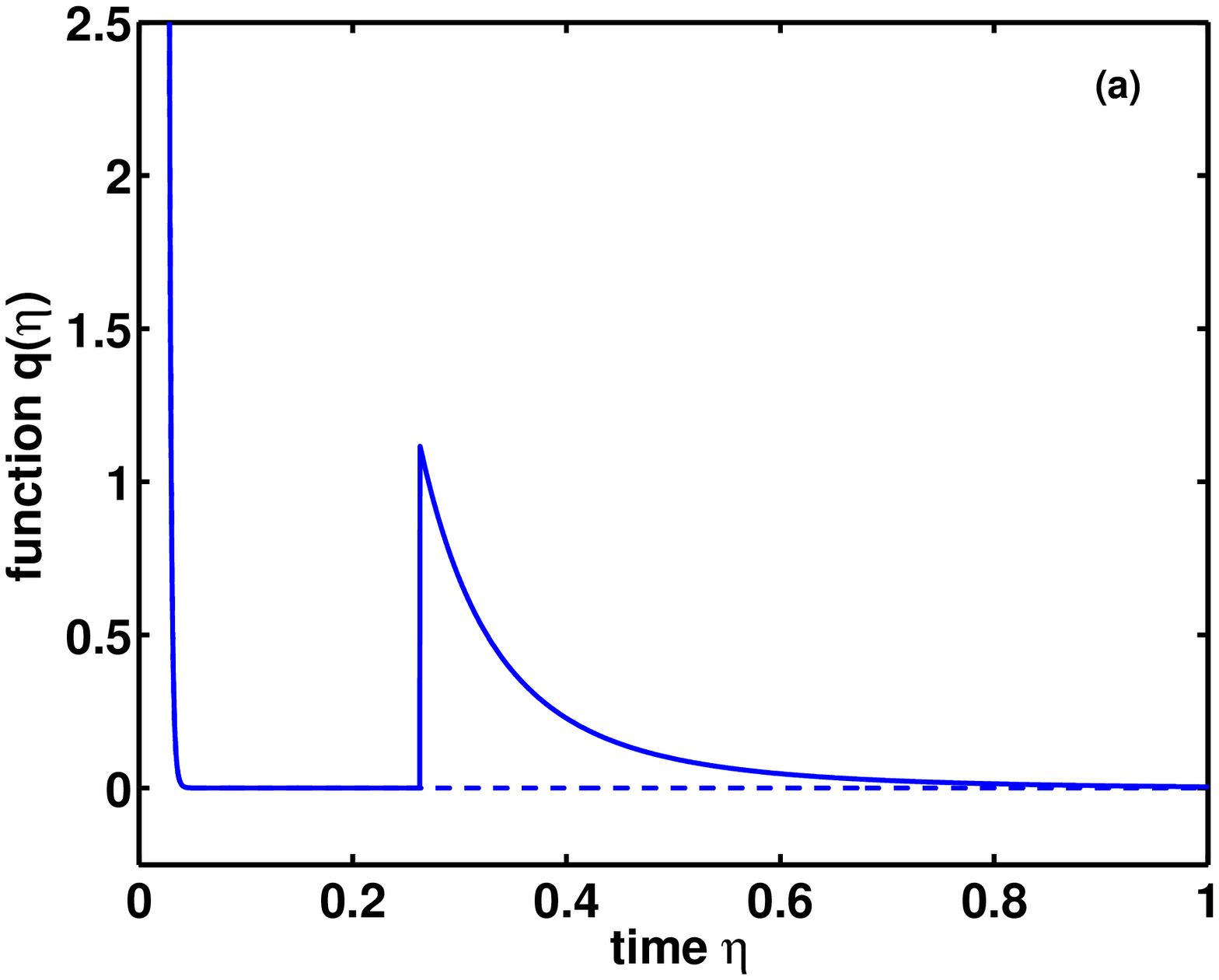}\qquad
\includegraphics[width=6cm]{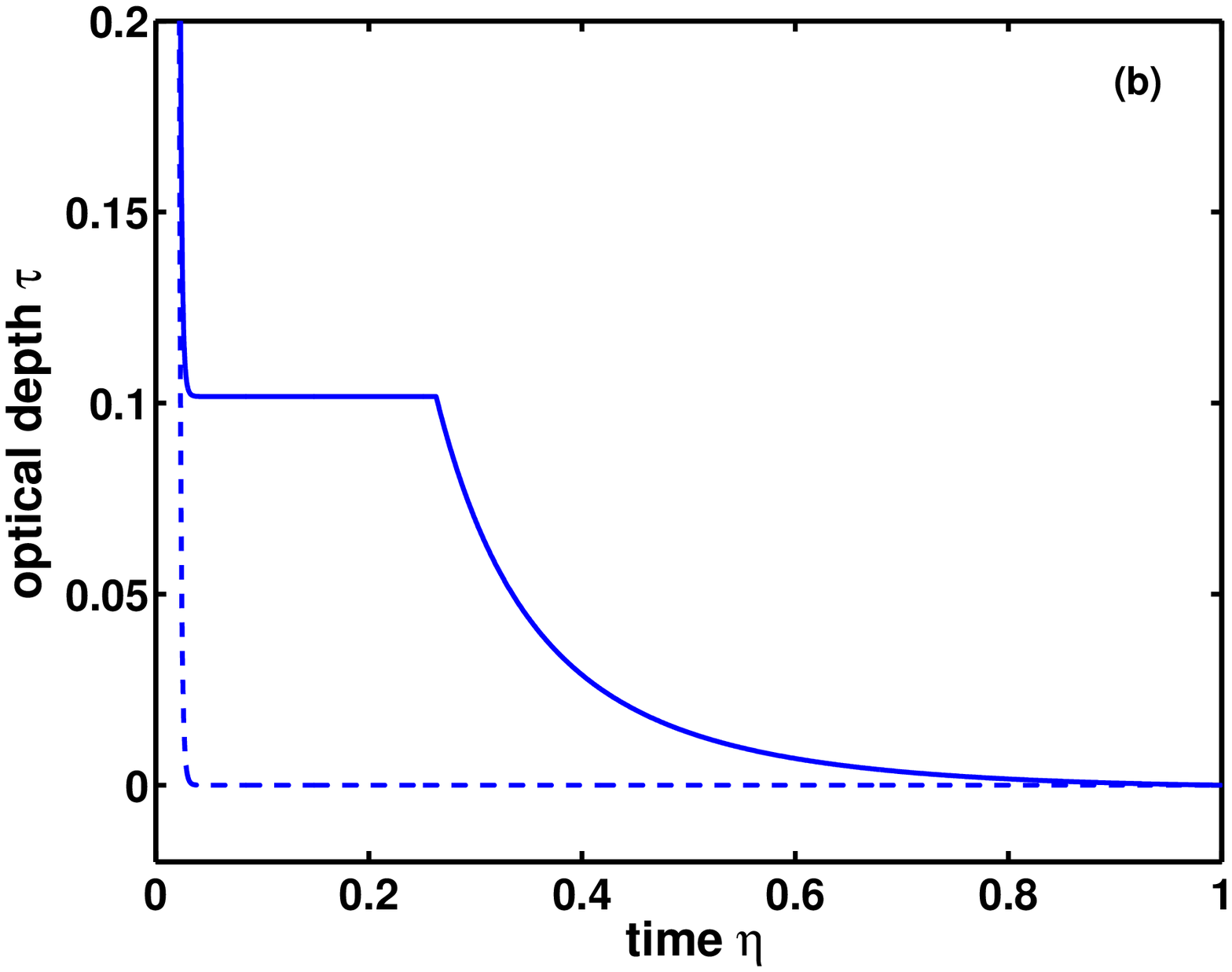}\\
\includegraphics[width=6cm]{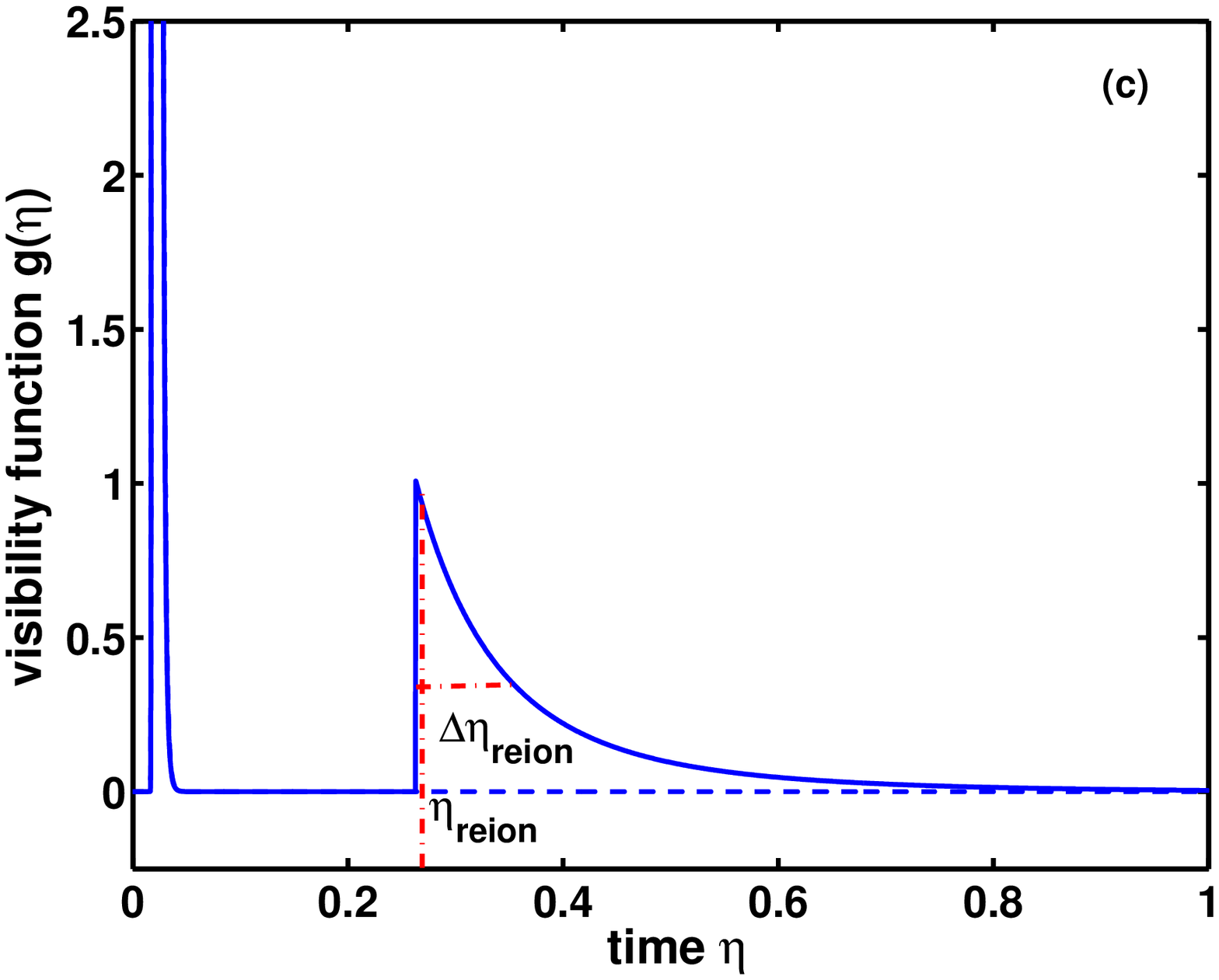}\qquad
\includegraphics[width=6cm]{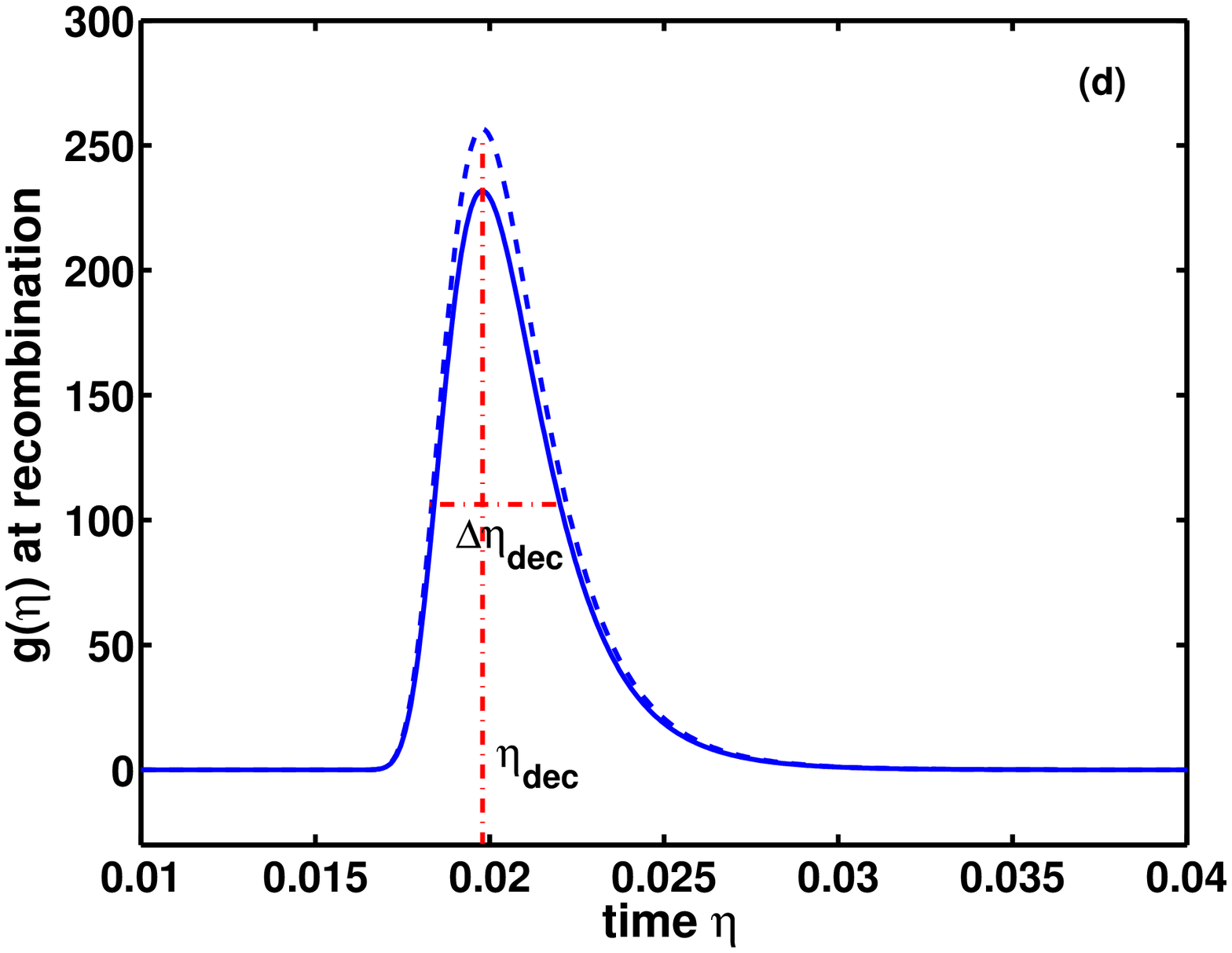}
\end{center}
\caption{The graphs (a), (b), (c) show the quantities $q(\eta)$,
$\tau(\eta)$, $g(\eta)$. The graph (d) is a zooming of $g(\eta)$
in the region of recombination. The dashed line shows a model
without reionization. The solid line takes reionization into
account, with $\tau_{reion} = 0.1$.}
\label{figure_ionization_history}
\end{figure}

The state of ionization is determined by microphysical processes
during all the evolution of the Universe
\cite{Naselskii1987,Doroshkevich2003}. For our purposes it is
sufficient to focus on two eras: early recombination and late
reionization. The recombination of primordial plasma into atomic
hydrogen and helium is accompanied by decoupling of CMB radiation
from the rest of matter. (For simplicity, we do not distinguish
here the notions of recombination and decoupling.) This relatively
quick process has happened at redshifts around $z_{dec}\approx
1100$. Much later, at redshifts around $z_{reion} \sim 10$
\cite{Page2006} the intergalactic medium has become ionized again,
presumably, by ionizing radiation of first condensed objects.

The density of free electrons is modelled \cite{Peebles1993} by
the expression
\begin{eqnarray}
N_e(\eta) = \left(1-\frac{Y_p}{2}\right)\frac{X_e(\eta) \Omega_b
\rho_c}{m_p}\left(\frac{a(\eta_R)}{a(\eta)}\right)^{3},\nonumber
\end{eqnarray}
where $Y_p \approx 0.23$ is the primordial helium mass fraction,
$X_e(\eta)$ is the fraction of ionized electrons, $\Omega_b$ is
the baryon content, and $m_p$ is the mass of a proton. In the
framework of linear perturbation theory it is sufficient to regard
the electron density as homogeneous, i.e. depending only on
$\eta$. For $X_e(\eta)$ we use the fitting formula \cite{Hu1994}:
\begin{eqnarray}
X_e(\eta) =
\left(1-\frac{Y_p}{2}\right)^{-1}\left(\frac{c_2}{1000}\right)
\left(\frac{m_p}{2\sigma_T l_H \rho_c}\right)\Omega_b^{c_1-1}
\left(\frac{z}{1000}\right)^{c_2-1}
\left(\frac{a'}{a}\right)(1+z)^{-1}, \label{hu94}
\end{eqnarray}
where $c_1 = 0.43$, $c_2 = 16+1.8\ln\Omega_B$, and $z$ is a
redshift.

As for the reionization, we assume that it was practically
instantaneous and happened at a redshift $z_{reion}\approx 16$.
The function $X_e$ is determined by Eq.\ (\ref{hu94}) for
$z>z_{reion}$ and $X_e = 1$ for $z \leq z_{reion}$.

To plot the graphs for $q(\eta)$, $\tau(\eta)$ and $g(\eta)$ in
Fig.\ \ref{figure_ionization_history}, we use the above-mentioned
parameters and $\Omega_b=0.046$ \cite{Spergel2006,LAMBDA}. It is
seen from Fig.\ \ref{figure_ionization_history} that the
visibility function $g(\eta)$ is sharply peaked at the era of
recombination. The peak can be characterized by the position of
its maximum $\eta_{dec}$, and the characteristic width
$\Delta\eta_{dec}$. A similar, but less pronounced, peak is
present also at the era of reionization.

(ii) Frequency dependence of the Stokes parameters

As is seen from Eq.\ (\ref{firstorderradiativetransfer}), the
frequency dependence of both, temperature and polarization,
anisotropies is governed by the function ${f(\tilde{\nu})}$
\cite{Basko1980}. We assume that the unperturbed radiation field
has a black-body spectrum,
\begin{eqnarray}
n_0(\tilde{\nu}) =
\frac{1}{exp\left(h\tilde{\nu}/k\tilde{T}\right) - 1}, \nonumber
\end{eqnarray}
where $\tilde{T} = T a(\eta)$ and the present-day value of $T$ is
$T(\eta_R) \approx 2.73K$.

It follows from these formulae that $f(\tilde{\nu})$ is
approximately 1 in the Reyleigh-Jeans part of the spectrum, and
$f(\tilde{\nu})$ varies as $h\tilde{\nu}/k\tilde{T}$ in the Wein
part. In practice, we are usually interested in the total
intensity, i.e. Stokes parameters integrated over all photon
frequencies. This integration produces the factor $\gamma$:
\begin{eqnarray}
\gamma \equiv \int
d\tilde{\nu}~\frac{h\tilde{\nu}^{3}}{c^{2}}n_0(\tilde{\nu})f(\tilde{\nu})
= -4I_{0}, \label{gamma}
\end{eqnarray}
where $I_{0} = \int d\tilde{\nu} ~\frac{h\tilde{\nu}^{3}}{c^{2}}
n_{0}(\tilde{\nu}) $ is the total intensity of the unperturbed
radiation field. The factor $\gamma$ often appears in the main
text of the paper.


\section{\label{app:poln} Two essential variables for temperature
and polarization}

It is seen from Eq.\ (\ref{fourierfirstorderradiativetransfer})
that the frequency dependence of ${\bf \hat{n}^{(1)}}$ is
determined by the factor $f(\tilde{\nu}) n_{0}(\tilde{\nu})$.
Therefore, we can single out this factor and write
\begin{eqnarray}
{\bf \hat{n}^{(1)}}(\eta,\tilde{\nu},\mu,\phi) =
\frac{1}{2}f(\tilde{\nu}) n_{0}(\tilde{\nu}){\bf
\hat{n}^{(1)}}(\eta,\mu,\phi).
\nonumber
\end{eqnarray}
The equation that follows from
(\ref{fourierfirstorderradiativetransfer}) and (\ref{eep}) reads
\begin{eqnarray}
&& \left[ \frac{\partial }{\partial \eta} + q(\eta) + in\mu
\right]
{\bf \hat{n}^{(1)}}(\eta,\mu,\phi) = \nonumber \\
&& = (1-\mu^{2})e^{\pm 2i\phi}\frac{d h_{n}}{d \eta} {\bf \hat{u}}
+
\frac{q(\eta)}{4\pi}\int\limits_{-1}^{+1}\int\limits_{0}^{2\pi}d\mu'
d\phi' ~ {\bf \hat{P}}(\mu,\phi; \mu',\phi'){\bf
\hat{n}^{(1)}}(\eta,\mu',\phi').
\label{A2}
\end{eqnarray}

The quantity ${\bf \hat{n}^{(1)}}(\eta,\mu,\phi)$ as a function of
$\phi$ can be expanded in a series
\begin{eqnarray}
{\bf \hat{n}^{(1)}}(\eta,\mu,\phi) = \sum_{m=-\infty}^{+\infty}
{\bf \hat{n}^{(1)}}_{m}(\eta,\mu)e^{im\phi}. \label{nphidec}
\end{eqnarray}
The explicit structure of the Chandrasekhar matrix ${\bf
\hat{P}}(\mu,\phi;\mu',\phi')$ is given by the expression
\cite{Chandrasekhar1960} (equation(220) on p.42):
\begin{eqnarray}
{\bf \hat{P}}(\mu,\phi;\mu',\phi')={\bf Q}\left[{\bf P^{(0)}}
+(1-\mu^{2})^{\frac{1}{2}}(1-\mu'^{2})^{\frac{1}{2}}{\bf
P^{(1)}+{\bf P^{(2)}}}\right], \nonumber
\end{eqnarray}
where matrices ${\bf Q}$,~ ${\bf P^{(0)}}(\mu;\mu')$,~ ${\bf
P^{(1)}}(\mu,\phi;\mu',\phi')$, ~${\bf
P^{(2)}}(\mu,\phi;\mu',\phi')$ read
\begin{eqnarray}
{\bf Q}
=\left(\begin{array}{ccc}1&0&0\\0&1&0\\0&0&2\end{array}\right),
\nonumber
\end{eqnarray}

\begin{eqnarray}
{\bf P^{(0)}}= \frac{3}{4}\left(\begin{array}{ccc}
2\left(1-\mu^{2}\right)\left(1-\mu'^{2}\right)+\mu^{2}\mu'^{2}&\mu^{2} & 0\\
\mu'^{2}&1&0\\
0&0&0
\end{array}\right)\nonumber,
\end{eqnarray}

\begin{eqnarray}
{\bf P^{(1)}} = \frac{3}{4}\left(\begin{array}{ccc}
4\mu\mu'\cos{\left(\phi'-\phi\right)}&0&-2\mu\sin{\left(\phi'-\phi\right)}\\
0&0&0\\
2\mu'\sin{\left(\phi'-\phi\right)}&0&\cos{\left(\phi'-\phi\right)}\\
\end{array}\right)\nonumber,
\end{eqnarray}

\begin{eqnarray}
{\bf P^{(2)}} = \frac{3}{4}\left(\begin{array}{ccc}
\mu^{2}\mu'^{2}\cos{2\left(\phi'-\phi\right)}&-\mu^{2}\cos{2\left(\phi'-\phi\right)}
&-\mu^{2}\mu'\sin{2\left(\phi'-\phi\right)}\\
-\mu'^{2}\cos{2\left(\phi'-\phi\right)}&\cos{2\left(\phi'-\phi\right)}
&\mu'\sin{2\left(\phi'-\phi\right)}\\
\mu\mu'^{2}\sin{2\left(\phi'-\phi\right)}&-\mu\sin{2\left(\phi'-\phi\right)}
&\mu\mu'\cos{2\left(\phi'-\phi\right)}
\end{array}\right).
\nonumber
\end{eqnarray}

The structure of the Chandrasekhar matrix ${\bf \hat{P}}(\mu,\phi;
\mu',\phi')$ is such that it does not mix the $m\phi$-dependence,
that is,
\begin{eqnarray}
\frac{1}{4\pi}\int\limits_{-1}^{+1}\int\limits_{0}^{2\pi}d\mu'
d\phi' ~ {\bf \hat{P}}(\mu,\phi; \mu',\phi'){\bf
\hat{n}^{(1)}}_{m}(\eta,\mu')e^{im\phi'} \sim e^{im\phi}.
\nonumber
\end{eqnarray}
Moreover, this integral vanishes for all $|m| > 2$. This means
that Eq.\ (\ref{A2}) is a homogeneous differential equation for
all ${\bf \hat{n}^{(1)}}_{m}$ with $m\neq \pm2$. Assuming zero
initial conditions at some initial $\eta$, we obtain ${\bf
\hat{n}^{(1)}}_{m}= 0$ for all $m\neq \pm2$. Hence, we are left
with three functions of $(\eta, \mu)$:
\begin{eqnarray}
{\bf \hat{n}^{(1)}}(\eta,\mu,\phi) = {\bf \hat{n}^{(1)}}
(\eta,\mu)e^{\pm2i\phi}.
\label{varmuphi}
\end{eqnarray}

We are now able to show that only two out of the three functions
${\bf \hat{n}^{(1)}}(\eta,\mu)$ are independent. Indeed, using
(\ref{varmuphi}) in Eq.\ (\ref{A2}) we arrive at a system of three
linear equations for the components of ${\bf
\hat{n}^{(1)}}(\eta,\mu)$:
\begin{eqnarray}
\left( \frac{\partial}{\partial\eta} + q(\eta) + in\mu
\right)\left(\begin{array}{c} \hat{n}_1(\eta,\mu) \\ \hat{n}_2(\eta,\mu) \\
\hat{n}_3(\eta,\mu)
\end{array}\right) = (1-\mu^{2})\frac{dh_n}{d\eta}\left(\begin{array}{r}
1\\1\\0
\end{array}\right) + \frac{3}{8}q(\eta)\left(\begin{array}{c}
\mu^{2} \\ -1 \\ \pm 2i\mu
\end{array}\right)\mathcal{I}(\eta),
\label{A4}
\end{eqnarray}
where $\mathcal{I}(\eta)$ is the remaining integral that was left
over from the last term in Eq.\ (\ref{A2}):
\begin{eqnarray}
\mathcal{I}(\eta) =
\frac{1}{2}\int\limits_{-1}^{+1}d\mu'\left[\frac{}{}
\mu'^{2}\hat{n}_1(\eta,\mu')-\hat{n}_2(\eta,\mu')\pm
i\mu'\hat{n}_3(\eta,\mu')\right]. \label{scrii}
\end{eqnarray}

Despite the complicated appearance, only two of the three
equations (\ref{A4}) are really coupled. Indeed, making linear
combinations of equations (\ref{A4}) it is easy to show that the
combination $2i\mu(\hat{n}_1-\hat{n}_2) \mp(1+\mu^{2})\hat{n}_3$
satisfies a homogeneous differential equation. Assuming zero
initial conditions, we derive
\begin{eqnarray}
2i\mu(\hat{n}_1-\hat{n}_2) \mp(1+\mu^{2})\hat{n}_3 = 0.
\nonumber\end{eqnarray} As two independent and essential variables
we choose
\begin{eqnarray}
&&\alpha(\eta,\mu) =
\frac{\hat{n}_1(\eta,\mu)+\hat{n}_2(\eta,\mu)}{(1-\mu^{2})},
~~~~~~\beta(\eta,\mu) =
\frac{\hat{n}_1(\eta,\mu)-\hat{n}_2(\eta,\mu)} {(1+\mu^{2})} = \pm
\frac{ \hat{n}_3(\eta,\mu)}{2i\mu}. \nonumber
\end{eqnarray}
In terms of $\alpha$ and $\beta$ equations of radiative transfer
reduce to Eq.\ (\ref{eqbeta}) and Eq.\ (\ref{eqxi}), and the
definition (\ref{scrii}) for $\mathcal{I}(\eta)$ takes the form of
Eq.\ (\ref{I}).

Although we have considered a gravitational wave perturbation, the
existence of only two essential variables is a general statement
and it applies to density and rotational perturbations as well. In
general, the perturbed radiative transfer equation contains an
arbitrary function $f(\eta, \mu, \phi)$ in front of ${\bf
\hat{u}}$, rather than a specific combination $(1-\mu^{2})e^{\pm
2i\phi}{d h_{n}}/{d \eta}$ quoted in Eq.\ (\ref{A2}). Function
$f(\eta, \mu, \phi)$ can be expanded in a series similar to Eq.\
(\ref{nphidec}):
\begin{eqnarray}
f(\eta,\mu,\phi) = \sum_{m=-\infty}^{+\infty}
f_{m}(\eta,\mu)e^{im\phi}.
\nonumber
\end{eqnarray}
Since the scattering integral (second term in the r.h.s of Eq.\
(\ref{A2})) vanishes for all $|m|>2$, the functions ${\bf
\hat{n}^{(1)}}_{m}(\eta,\mu)$ are fully determined by
$f_{m}(\eta,\mu)$ ($|m|>2$) and describe the temperature
variations only. To discuss polarization, we have to consider
three remaining cases $m = 0, \pm 1, \pm 2$.

The explicit identification of the two essential variables
$\alpha(\eta, \mu)$ and $\beta(\eta, \mu)$ for the case $m= \pm 2$
has been given above. Specifically for gravitational waves, $f_{2}
= (1-\mu^{2}) {d\stackrel{L}{h}_{n}(\eta)}/{d\eta}$,~ $f_{-2} =
(1-\mu^{2}) {d\stackrel{R}{h}_{n}(\eta)}/{d\eta}$.

The identification of the two essential variables for the case
$m=0$ proceeds in a similar manner. Having calculated the
scattering integral for $m=0$ one can show that the equation,
analogous to Eq.\ (\ref{A4}), will now read
\begin{eqnarray}
\left( \frac{\partial}{\partial\eta} + q(\eta) + in\mu
\right)\left(\begin{array}{c} \hat{n}_1(\eta,\mu) \\ \hat{n}_2(\eta,\mu) \\
\hat{n}_3(\eta,\mu)
\end{array}\right) = \left(f_0(\eta, \mu)+ q(\eta) I_0^{(1)}(\eta)\right)
\left(\begin{array}{r} 1\\1\\0
\end{array}\right) + \frac{3}{8}q(\eta)\left(\begin{array}{c}
3\mu^{2}-2 \\ 1 \\ 0
\end{array}\right)\mathcal{J}(\eta),\nonumber
\end{eqnarray}
where
\begin{eqnarray}
I_0^{(1)}(\eta)
=\frac{1}{4}\int\limits_{-1}^{+1}d\mu'\left[\frac{}{}
\hat{n}_1(\eta,\mu')+\hat{n}_2(\eta,\mu')\right],
\nonumber
\end{eqnarray}
\begin{eqnarray}
\mathcal{J}(\eta)=
\frac{1}{3}\int\limits_{-1}^{+1}d\mu'\left[\frac{}{} (3\mu'^{2}
-2)\hat{n}_1(\eta,\mu')+\hat{n}_2(\eta,\mu')\right].
\nonumber
\end{eqnarray}
Obviously, $\hat{n}_3$ satisfies a homogeneous equation and can be
put to zero. Specifically for density perturbations (see Appendix
\ref{app:densityperturbations}), function $f_0(\eta, \mu)$
consists of terms representing gravitational field perturbations
and the Doppler term arising due to the baryon velocity:
\begin{eqnarray}
f_0(\eta, \mu) = \frac{1}{2} \left( \frac{d h}{d \eta} - \mu^{2}
\frac{d h_l}{d \eta} + iq(\eta) \mu v_b(\eta) \right).
\nonumber
\end{eqnarray}

We have checked that in the case $m=\pm 1$ the problem also
reduces to only two essential variables. The combination
$\hat{n}_1 - \hat{n}_2 \mp i\mu \hat{n}_3$ vanishes at the zero
initial data.


\section{\label{app:densityperturbations}Temperature and
polarization anisotropies caused by density perturbations}

To discover relic gravitational waves in the CMB data we have to
distinguish their effects from the effects of density
perturbations. The theory of temperature and polarization
anisotropies caused by primordial density perturbations is very
much similar to the theory of relic gravitational waves. We start
from the metric Fourier expansion (\ref{fourierh}) with the
polarisation tensors (\ref{ptensors4}). Having derived and solved
integral equations of radiative transfer in the presence of
density perturbations, we arrive at our final goal of
distinguishing the $TE$ cross-correlations.

{\bf{(i) Radiative transfer equations}}

The equations of radiative transfer in the presence of a single
mode $n$ of density perturbations are similar to Eq.\
(\ref{fourierfirstorderradiativetransfer}) and read
\begin{eqnarray}
&&\left[ \frac{\partial }{\partial \eta} + q(\eta) + ie^in_i
\right]
{\bf {\hat{n}}^{(1)}}_{\bf n}(\eta,\tilde{\nu},e^i) = \nonumber \\
&& = \frac{f(\tilde{\nu})n_0(\tilde{\nu})}{2}
\left[e^ie^j\stackrel{s}{p}_{ij}({\bf
n})\frac{d\stackrel{s}{h}_{n}(\eta)}{d\eta} - q(\eta)e^i v_i
\right]{\bf \hat{u}} + \frac{q(\eta)}{4\pi}\int d\Omega' ~ {\bf
\hat{P}}(e^i ; {e'^j}){\bf {\hat{n}}^{(1)}}_{\bf
n}(\eta,\tilde{\nu},e'^j),\nonumber \\
\label{fourierfirstorderradiativetransferDP}
\end{eqnarray}
where the extra term $e^i v_i$ takes care of the movement of
scattering electrons with respect to the chosen synchronous
coordinate system \cite{Kaiser1983,Bond1984,Ma1995}.

For technical reasons, it is convenient to work with the `scalar'
$h(\eta)$ and `longitudinal' $h_l(\eta)$ polarization mode
functions, instead of the original $\stackrel{s}{h}_n(\eta)$
\cite{Grishchuk1994}. The relationship between them is
\begin{eqnarray}
\stackrel{1}{h}_n(\eta) = \sqrt{\frac{3}{2}}\left( h(\eta) -
h_l(\eta) \right), ~~~ \stackrel{2}{h}_n(\eta) =
\sqrt{\frac{1}{3}}h_l(\eta),
\nonumber
\end{eqnarray}
where the wavenumber index $n$ on $h(\eta)$ and $h_l(\eta)$ is
implicit. Both polarization components of metric perturbations
participate in Eq.\ (\ref{fourierfirstorderradiativetransferDP}).
In the frame associated with the density wave, i.e. for ${\bf n}/n
= (0, 0, 1)$, the structures $e^ie^j\stackrel{s}{p}_{ij}$ (see
Eq.\ \ref{ptensors4}) and $e^iv_i=-i\mu v_b$ depend on $\mu=\cos
\theta$, but not on the azimuthal angle $\phi$.

By argumentation similar to that in Appendix \ref{app:poln}, one
can show that a solution (not vanishing on zero initial data) to
Eq.\ (\ref{fourierfirstorderradiativetransferDP}) must have the
form
\begin{eqnarray}
{\bf {\hat{n}}}^{(1)}_{n}(\eta,\tilde{\nu},\mu) =
\frac{f(\tilde{\nu}) n_0(\tilde{\nu})}{2}\left
[\alpha_{n}(\eta,\mu)\left(
\begin{array}{c}1 \\ 1 \\0
\end{array} \right)
+ \beta_{n}(\eta,\mu) \left( \begin{array}{c}1 \\ -1 \\ 0
\end{array} \right)\right ].
\label{nintermsofalphabetaDP}
\end{eqnarray}
Substituting Eq.\ (\ref{nintermsofalphabetaDP}) into Eq.\
(\ref{fourierfirstorderradiativetransferDP}) we arrive at a system
of coupled equations for $\alpha$ and $\beta$
\begin{subequations}
\begin{eqnarray}
\left[\frac{{\partial}}{\partial\eta}+q(\eta)+in\mu \right]
\alpha_n(\eta,\mu) &=&
\frac{1}{2}\left(\frac{dh}{d\eta}-\mu^{2}\frac{dh_l}{d\eta}\right)
+ q(\eta)\left(\mathcal{I}_1 + i\mu v_b -
\frac{1}{2}P_2(\mu)\mathcal{I}_2\right),\label{eqalphaDP}\\
\left[\frac{{\partial}}{\partial\eta}+q(\eta)+in\mu \right]
\beta_n(\eta,\mu) &=&
\frac{1}{2}q(\eta)\left(1-\frac{}{}P_2(\mu)\right)\mathcal{I}_2,\label{eqbetaDP}
\end{eqnarray}
\end{subequations}
where
\begin{subequations}
\begin{eqnarray}
\mathcal{I}_1(\eta) &=&
\frac{1}{2}\int\limits_{-1}^{+1}d\mu~\alpha_n(\eta,\mu),
\label{I_1}\\
\mathcal{I}_2(\eta) &=& \frac{1}{2}\int\limits_{-1}^{+1}
d\mu~\left[ \left(1-P_2(\mu)\right)\beta_n(\eta,\mu) -\frac{}{}
P_2(\mu)\alpha_n(\eta,\mu) \right] \label{I_2}
\end{eqnarray}
\end{subequations}
The quantity $\mathcal{I}_1$ is the monopole component of the
perturbed radiation field, whereas $\mathcal{I}_2$ is the
quadrupole ($\ell = 2$) component, responsible for the generation
of polarization.

To make contact with previous work we note that the variables
$\alpha$ and $\beta$ are closely related to the variables
$\Delta_T$ and $\Delta_P$ from Ref.\ \cite{Zaldarriaga1997}.
Assuming a black body unperturbed radiation field, we have
$\Delta_T = -\alpha$, $\Delta_P=-\beta$. We also note that the
mode functions $h$ and $h_l$ are related to the mode functions
$\textrm{h}$ and $\eta$ used in \cite{Ma1995,Zaldarriaga1997} by
$h= - 2\eta$, $h_l = -(\textrm{h}+6\eta)$. Keeping in mind the
difference in notations, one can verify that equations
(\ref{eqalphaDP}) and (\ref{eqbetaDP}) are equivalent to equations
(11) in Ref.\ \cite{Zaldarriaga1997}.

{\bf{(ii) Integral equations and their solutions}}

A formal solution to equations (\ref{eqalphaDP}) and
(\ref{eqbetaDP}) can be written as
\begin{subequations}
\begin{eqnarray}
\alpha_n(\eta, \mu) &=& \int\limits_{0}^{\eta}d\eta'
e^{-\tau(\eta,\eta')-in\mu(\eta-\eta')}\left[
\frac{1}{2}\left(\frac{dh}{d\eta}-\mu^{2}\frac{dh_l}{d\eta}\right)
+ q\left(\mathcal{I}_1 + i\mu v_b -
\frac{1}{2}P_2(\mu)\mathcal{I}_2\right) \right] ,\nonumber\\
&&\label{alphaDP}
\\ \beta_n(\eta, \mu) &=&
\frac{1}{2}\left(1-\frac{}{}P_2(\mu)\right)\int\limits_{0}^{\eta}d\eta'
e^{-\tau(\eta,\eta')-in\mu(\eta-\eta')}\mathcal{I}_2.
\label{betaDP}
\end{eqnarray}
\end{subequations}
Proceeding in a manner similar to that in Sec.\
\ref{sec:integralequations}, we substitute (\ref{alphaDP}) and
(\ref{betaDP}) into (\ref{I_1}) and (\ref{I_2}). After certain
rearrangements we arrive at two coupled integral equations for
$\mathcal{I}_1$ and $\mathcal{I}_2$:
\begin{subequations}
\begin{eqnarray}
\mathcal{I}_1(\eta) &=& \int\limits_0^{\eta}d\eta'q(\eta')
e^{-\tau(\eta,\eta')}\left[ \mathcal{I}_1(\eta')j_0(x) +
\frac{1}{2}\mathcal{I}_2(\eta')j_2(x) \right] \nonumber\\
&&+ \int\limits_0^{\eta}d\eta' e^{-\tau(\eta,\eta')}\left[
\frac{1}{2}\left(\frac{dh}{d\eta'}j_0(x)+\frac{dh_l}{d\eta'}
\frac{d^{2}j_0}{dx^{2}}\right) + q(\eta')v_b(\eta')j_1(x)
\right],\label{equationI_1}
\end{eqnarray}
\begin{eqnarray}
\mathcal{I}_2(\eta) &=& \int\limits_0^{\eta}d\eta'q(\eta')
e^{-\tau(\eta,\eta')}\left[ \mathcal{I}_1(\eta')j_2(x) +
\mathcal{I}_2(\eta')
\left(\left(\frac{18}{x^{2}}-\frac{1}{2}\right)j_2(x)-
\frac{3}{2x}j_1(x)\right) \right] \nonumber\\
&&+ \int\limits_0^{\eta}d\eta' e^{-\tau(\eta,\eta')}\left[
\frac{1}{2}\left(\frac{dh}{d\eta'}j_2(x)+\frac{dh_l}{d\eta'}\left(j_2(x)
+ 3\frac{d^{2}j_0}{dx^{2}}\right)\right) +
q(\eta')v_b(\eta')\frac{dj_2}{dx} \right],\nonumber \\ &&
\label{equationI_2}
\end{eqnarray}
\end{subequations}
where the argument of spherical Bessel functions $j_{\ell}(x)$ is
$x=n(\eta-\eta')$.

Equations (\ref{equationI_1}) and (\ref{equationI_2}) together
with the continuity equations for matter perturbations and
Einstein equations for metric perturbations (see for example
\cite{Ma1995,Grishchuk1994}), form a closed system of coupled
integro-differential equations. In previous treatments
\cite{Bond1984,Ma1995,Zaldarriaga1997}, the radiative transfer
equations were presented as an infinite series of coupled ordinary
differential equations.

Similarly to what was done in Sec.\
\ref{sec:analyticalsolutioninteq}, we can analyze Eqs.\
(\ref{equationI_1}), (\ref{equationI_2}) in terms of expansions in
powers of $n$,
\begin{eqnarray}
\mathcal{I}_1 = \sum_{k=0}^{\infty}\mathcal{I}_1^{(k)}
n^{2k},~~~~~~ \mathcal{I}_2 =
\sum_{k=0}^{\infty}\mathcal{I}_2^{(k)} n^{2k}. \nonumber
\end{eqnarray}
The kernels of Eqs.\ (\ref{equationI_1}), (\ref{equationI_2}) are
expanded in powers of $x$. In the long-wavelength approximation,
i.e. for $x \ll 1$, we can limit ourselves by first non-vanishing
terms. The zero order approximation brings us to the equations
\begin{subequations}
\begin{eqnarray}
\mathcal{I}^{(0)}_1(\eta) = \int\limits_0^{\eta}d\eta'q(\eta')
e^{-\tau(\eta,\eta')} \mathcal{I}^{(0)}_1(\eta') +
\int\limits_0^{\eta}d\eta' e^{-\tau(\eta,\eta')}\left[
\frac{1}{2}\left(\frac{dh}{d\eta'}-\frac{1}{3}\frac{dh_l}{d\eta'}\right)
+ \frac{1}{3}q(\eta')v_b(\eta')x \right],
\nonumber
\end{eqnarray}
\begin{eqnarray}
\mathcal{I}^{(0)}_2(\eta) =
\frac{7}{10}\int\limits_{0}^{\eta}d\eta'
q(\eta')e^{-\tau(\eta,\eta')}\mathcal{I}^{(0)}_2(\eta') +
\frac{1}{15}\int\limits_{0}^{\eta}d\eta'e^{-\tau(\eta,\eta')}
\left[ -\frac{dh_l}{d\eta'} + 2q(\eta')v_b(\eta')x \right].
\nonumber
\end{eqnarray}
\end{subequations}
The solution to these equations is given by
\begin{subequations}
\begin{eqnarray}
\mathcal{I}^{(0)}_1(\eta) = \int\limits_0^{\eta}d\eta'\left[
\frac{1}{2}\left(\frac{dh}{d\eta'}-\frac{1}{3}\frac{dh_l}{d\eta'}\right)
+ \frac{1}{3}n
\int\limits_{0}^{\eta'}d\eta''q(\eta'')e^{-\tau(\eta',\eta'')}v_b(\eta'')
\right],\label{solutionI_1zero}
\end{eqnarray}
\begin{eqnarray}
\mathcal{I}^{(0)}_2(\eta)
=\frac{1}{15}\int\limits_{0}^{\eta}d\eta'e^{-\frac{3}{10}\tau(\eta,\eta')}
\left[ -\frac{dh_l}{d\eta'} +
2n\int\limits_{0}^{\eta'}d\eta''q(\eta'')e^{-\tau(\eta',\eta'')}v_b(\eta'')
\right]. \label{solutionI_2zero}
\end{eqnarray}
\end{subequations}

{\bf{(iii) Multipole coefficients}}

We are mostly interested in the present-day values of $\alpha$ and
$\beta$ and, hence, we put $\eta=\eta_R$ in (\ref{alphaDP}) and
(\ref{betaDP}). Irrespective of approximation in which the
functions $\mathcal{I}_1(\eta)$, $\mathcal{I}_2(\eta)$ are known,
the multipole coefficients for the radiation field can be found in
a way similar to that in Sec.\ \ref{sectionmultipoleexpansion}:
\begin{eqnarray}
\begin{array}{c}
a_{\ell m}^T(n) = (-i)^{\ell}\gamma\sqrt{4\pi(2\ell+1)}\delta_{m
0} a^T_{\ell}(n), \\ a_{\ell m}^E(n)
=(-i)^{\ell}\gamma\sqrt{4\pi(2\ell+1)}\delta_{m 0} a^E_{\ell}(n),
\\ a_{\ell m}^B(n) = 0,
\end{array}
\nonumber
\end{eqnarray}
where $a^T_{\ell}(n)$, $a^E_{\ell}(n)$ are given by
\begin{subequations}
\begin{eqnarray}
a^T_{\ell}(n) &=& \int\limits_{0}^{\eta_R}d\eta
\left[e^{-\tau}\left( \frac{1}{2}\frac{dh}{d\eta} +
\frac{1}{2}\frac{dh_l}{d\eta}\frac{d^{2}}{d\zeta^{2}} \right) +
g(\eta)\left(\mathcal{I}_1 + v_b\frac{d}{d\zeta} -
\frac{3}{4}\mathcal{I}_2\left(1+\frac{d^{2}}{d\zeta^{2}}\right)\right)
\right]j_{\ell}(\zeta) ,\nonumber \\ && \label{a_lTDP} \\
a^E_{\ell}(n) &=&\frac{3}{4}\left( \frac{({\ell}+2)!}{({\ell}-2)!}
\right)^{\frac{1}{2}}\int\limits_{0}^{\eta_R}d\eta
~g(\eta)\mathcal{I}_2\frac{j_{\ell}(\zeta)}{\zeta^{2}},\label{a_lEDP}
\end{eqnarray}
\end{subequations}
and $\zeta=n(\eta_R-\eta)$. As expected, in the case of density
perturbations all $a_{\ell m} = 0$ for $m\neq0$ and $a_{\ell
m}^B(n) = 0$ \cite{Seljak1997a},
\cite{Zaldarriaga1997,Kamionkowski1997,Hu1997b}. The formal reason
for this is that all quantities in (\ref{nintermsofalphabetaDP})
do not depend on $\phi$ (A more detailed exposition can be found
in \cite{Zaldarriaga1997,Kamionkowski1997}).

Let us start from the temperature multipoles. For illustration, we
consider an instanteneous recombination. We replace $e^{-\tau}$
with the step function, $g(\eta)$ with the $\delta$-function, and
we neglect the $\mathcal{I}_2$ term. Then, we get
\begin{eqnarray}
a^T_{\ell}(n) &=& \left.\left[\left(\mathcal{I}_1 +
v_b\frac{d}{d\zeta}\right)j_{\ell}(\zeta)\right]
\right|_{\eta=\eta_{dec}} +
\frac{1}{2}\int\limits_{\eta_{dec}}^{\eta_R}d\eta \left(
\frac{dh}{d\eta} + \frac{dh_l}{d\eta}\frac{d^{2}}{d\zeta^{2}}
\right)j_{\ell}(\zeta).\nonumber
\end{eqnarray}
This expression can be further simplified by taking the remaining
integral by parts. After some rearrangements we arrive at the
final expression
\begin{eqnarray}
a^T_{\ell}(n) = &&
\left.\left(\frac{1}{2n^{2}}\frac{d^{2}h_l}{d\eta^{2}}\delta_{\ell
0} + \frac{1}{6n}\frac{dh_l}{d\eta}\delta_{\ell
1}\right)\right|_{\eta=\eta_R} \nonumber \\ && +
\left.\left[\left(-\frac{1}{2n^{2}}\frac{d^{2}h_l}{d\eta^{2}}+
\mathcal{I}_1 + \left(-\frac{1}{2n}\frac{dh_l}{d\eta}+
v_b\right)\frac{d}{d\zeta}\right)j_{\ell}(\zeta)\right]
\right|_{\eta =\eta_{dec}} \nonumber \\ && + \frac{1}{2}
\int\limits_{\eta_{dec}}^{\eta_R}d\eta \left( \frac{dh}{d\eta} +
\frac{1}{n^{2}}\frac{d^{3}h_l}{d\eta^{3}} \right)j_{\ell}(\zeta).
\label{a_lTDP1}
\end{eqnarray}

For the growing mode of density perturbations in the
matter-dominated era, the integrand of the remaining integral
vanishes. If, in addition, the intrinsic temperature perturbation
$\mathcal{I}_1$ and the plasma velocity $v_b$ are zero at $\eta=
\eta_{dec}$, we recover from Eq.\ (\ref{a_lTDP1}) the four terms
of the full Sachs-Wolfe formula (43) in Ref.\ \cite{Sachs1967}. We
would like to note in passing that the origin of the often used,
including this paper, combination $\ell(\ell +1) C_{\ell}$ is in
fact a historical accident, when a wrong motivation leads to a
convenient notation. This combination arises in the essentially
incorrect formula $\ell(\ell+1)C_{\ell} =const$ that can be
derived after the unjustified neglect of three out of four terms
in the original Sachs-Wolfe formula (43) in Ref.\
\cite{Sachs1967}. This clarification is important not only by
itself, but also for correct physical interpretation of reasons
for the rise toward the `first peak' at $\ell \approx 220$ of the
actually observed function $\ell(\ell+1)C_{\ell}$. For more
details, see \cite{Dimitropoulos2002}.

We now turn to polarization anisotropies. We use the zero order
approximation (\ref{solutionI_2zero}) in (\ref{a_lEDP}),
\begin{eqnarray}
a^E_{\ell} &=&\frac{1}{20}\left( \frac{({\ell}+2)!}{({\ell}-2)!}
\right)^{\frac{1}{2}}\int\limits_{0}^{\eta_R}d\eta
~qe^{-\frac{7}{10}\tau}\frac{j_{\ell}(\zeta)}{\zeta^{2}}\int
\limits_{0}^{\eta}d\eta'e^{-\frac{3}{10}\tau(\eta')}
\left(-\frac{dh_l}{d\eta'} +
2n\int\limits_{0}^{\eta'}d\eta''q(\eta'')e^{-\tau(\eta',\eta'')}v_b(\eta'')
\right).\nonumber
\end{eqnarray}
As $g(\eta)$ is a narrow function, the integrand is localized near
$\eta =\eta_{dec}$. By the same argumentation that was used in
Sec.\ \ref{sec:polarizationrecombination}, the above expression
can be reduced to quantities evaluated at $\eta=\eta_{dec}$,
\begin{eqnarray}
a^E_{\ell}(n) \approx\frac{1}{20}\left(
\frac{({\ell}+2)!}{({\ell}-2)!}
\right)^{\frac{1}{2}}\left[\frac{j_{\ell}(\zeta)}{\zeta2}\left.\left(
-\frac{dh_l}{d\eta}\Delta + 2nv_b\tilde{\Delta}\right)
\right]\right|_{\eta=\eta_{dec}},
\label{a_lEDP1}
\end{eqnarray}
where the factors $\Delta$ and $\tilde{\Delta}$ are of the order
of the width of $g(\eta)$ in the recombination era. Explicitly,
they are given by
\begin{eqnarray}
\Delta = \int\limits_{0}^{\eta_R}d\eta
~g(\eta)\int\limits_{0}^{\eta}d\eta'~e^{-\frac{3}{10}\tau(\eta,\eta')},
~~~\tilde{\Delta}= \int\limits_{0}^{\eta_R}d\eta
~g(\eta)\int\limits_{0}^{\eta}d\eta'~e^{-\frac{3}{10}\tau(\eta,\eta')}
\int\limits_{0}^{\eta'}d\eta''~q(\eta'')e^{-\tau(\eta',\eta'')}.\nonumber
\end{eqnarray}

The angular power spectra in the case of density perturbations are
described by formulas analogous to (\ref{Clxx'}). By the steps
similar to those in Section \ref{powsp} it can be shown that the
power spectra are given by
\begin{eqnarray}
C_{\ell}^{XX'} = \frac{\mathcal{C}^{2}}{\pi}\int ndn a^X_{\ell}(n)
a^{X'}_{\ell}(n), \label{C_ellDP}
\end{eqnarray}
where $\mathcal{C}$ for density perturbations is
$\mathcal{C} =\sqrt{24\pi}l_{Pl}$.

{\bf{(iiii) Temperature-polarization correlations}}

We focus on the $TE$ correlation at the relatively small
multipoles, $\ell \lesssim 70$. The dominant contribution to these
multipoles comes from density perturbations that did not enter the
Hubble radius at recombination, $n \lesssim 70$. It is sufficient
to consider the early time evolution of these perturbations in the
matter-dominated era.

Restricting our analysis to the growing mode, we write for the
metric and matter perturbations (see, for example,
\cite{Grishchuk1994}):
\begin{eqnarray}
h(\eta) = B_n =const,~~~h_l(\eta) = B_{n} (n\eta)^{2},~~~~
\mathcal{I}^{(0)}_1 = \frac{\delta T}{T}= -\frac{2}{3}B_{n}
(n\eta)^{2}, ~~~ v_b = -\frac{2}{9}B_{n} (n\eta)^{3}.\nonumber
\end{eqnarray}
Substituting this solution into (\ref{a_lTDP1}), (\ref{a_lEDP1})
and taking into account only the lowest-order terms in $n$ (the
long-wavelength approximation) we arrive at
\begin{subequations}
\begin{eqnarray}
a^T_{\ell} &\approx& -B_n\left.\frac{}{}
j_{\ell}(\zeta)\right|_{\eta=\eta_{dec}},
\label{a_lTDP2} \\
a^E_{\ell} & \approx& -\frac{1}{10}n^{2}B_n \eta_{dec} \Delta
\left(\frac{(\ell+2)!}{(\ell-2)!}\right)^{\frac{1}{2}}\left.
\left(\frac{j_{\ell}(\zeta)}{\zeta^{2}}\right)\right|_{\eta=\eta_{rec}}.
\label{a_lEDP2}
\end{eqnarray}
\end{subequations}
Finally, from Eq.\ (\ref{C_ellDP}) we obtain for the $TE$
correlation:
\begin{eqnarray}
C_{\ell}^{TE} &\approx& \frac{\mathcal{C}^{2}}{\pi}
\frac{\eta_{dec} \Delta}{10}
\left(\frac{(\ell+2)!}{(\ell-2)!}\right)^{\frac{1}{2}} \int dn
~n^{3}B_n^{2} \left.\left(\frac{j_{\ell}(\zeta)}{\zeta}
\right)^{2}\right|_{\eta=\eta_{dec}}.
\end{eqnarray}

All terms in the above expression are strictly positive.
Therefore, the $TE$ correlation caused by density perturbations,
in contrast to the case of gravitational waves, must be positive
at lower multipoles $\ell \lesssim 50$ where our approximations
are still valid. As mentioned before, this difference between
gravitational waves and density perturbations boils down to the
difference in the sign of first time-derivative of the associated
metric perturbations.



%
%


\end{document}